\documentclass[preprint,aps,tightenlines,showkeys,nofootinbib,superscriptaddress]{revtex4}
\usepackage{color}
\usepackage{enumerate}
\usepackage{latexsym}
\usepackage[dvips]{graphicx}
\def\be{\begin{equation}}
\def\ee{\end{equation}}
\def\bear{\begin{eqnarray}}
\def\eear{\end{eqnarray}}
\newcommand{\beq}{\begin{eqnarray}}
\newcommand{\eeq}{\end{eqnarray}}

\newcommand{\ci}{\cite}
\newcommand{\ga}{{\gamma}}

\newcommand{\de}{{\delta}}

\newcommand{\La}{{\Lambda}}

\newcommand{\om}{{\omega}}

\newcommand{\no}{{\nonumber}}
\newcommand{\f}{\frac}

\newcommand{\ra}{\rightarrow}

\newcommand{\Sch}{Schwarzschild }

\begin{document}

\preprint{arXiv:2004.12185v3 [hep-th]}
\today\\

\title{Quasi-Normal Modes and Stability of Einstein-Born-Infeld \\
Black Holes in de Sitter Space}

\author{Chong Oh Lee \footnote{E-mail address: cohlee@gmail.com} }
\author{Jin Young Kim \footnote{E-mail address: jykim@kunsan.ac.kr, Corresponding author} }
\affiliation{Department of Physics, Kunsan National University,
Kunsan 54150, Korea}

\author{Mu-In Park \footnote{E-mail address: muinpark@gmail.com}}
\affiliation{Center for Quantum Spacetime,
Sogang University, Seoul, 121-742, Korea}

\begin{abstract}
We study gravitational perturbations of electrically charged black holes in
(3+1)-dimensional Einstein-Born-Infeld gravity with a positive cosmological constant.
For the {\it axial} perturbations, we obtain a set of decoupled
Schr\"{o}dinger-type equations,
whose formal expressions, in terms of metric functions, are the same as those without
cosmological constant, corresponding to the Regge-Wheeler equation in the proper limit. We compute the quasi-normal modes (QNMs) of the decoupled
perturbations using the Schutz-Iyer-Will's WKB method. We discuss the stability of the charged
black holes by investigating the dependence of quasi-normal frequencies
on the
parameters of the theory,
correcting some errors in the literature. It is
found that all the axial perturbations are {\it stable}
for the cases where
the WKB method applies.
 There are cases where
the conventional WKB method does not apply, like the three-turning-points
problem,
so that a more generalized formalism is necessary for studying their QNMs and stabilities.
We find that, for the degenerate horizons with the ``point-like" horizons at the origin,
the QNMs  are quite {\it long-lived}, close to
the quasi-resonance modes,
in addition to the ``frozen" QNMs
for the Nariai-type horizons and the usual (short-lived) QNMs
for the extremal black hole horizons. This is a genuine effect of the
branch
which does not have the general relativity
limit. We also study the exact solution near the (charged) Nariai limit
and find good agreements even far beyond the limit for the
imaginary frequency parts.
\end{abstract}


\maketitle

\newpage

\section{Introduction}

Recent detections
of gravitational waves from the merger of compact astronomical objects
confirmed directly the existence of black holes \cite{prl116,prl118,prl119}.
Gravitational waves from the mergers of various compact objects, like black holes
or neutron stars, are common in our universe and expected to be observed
frequently in the near future. Also the recent observation of the shadow images
from a supermassive black hole, along with the observation of gravitational waves,
opened a new era to test Einstein's general relativity (GR) under extreme
conditions, like
black holes \cite{EHTelescope}.

The non-linear generalization of Maxwell's electromagnetism by Born and Infeld (BI)
\cite{borninfeld1,borninfeld2} was originally developed to obtain a finite self-energy of a point charge by modifying Maxwell's theory at the short distance. The BI-type action attracted renewed interests as an effective action of D-branes in string theory \cite{Leig:1989}.
When combined with Einstein action, the short distance behaviors of the metric and black hole solutions are different from those of Einstein-Maxwell theory.
On the large scale, natural existence of
electrically charged black holes is questionable because our universe seems to be
charge neutral. However, on the small scale which may be described by
Einstein-Born-Infeld (EBI) theory, the charged black holes may exist,
as charged elementary particles do at the high energy.
It may be also interesting as a preliminary to study the magnetically
charged black holes, which may exist from the collapse of
neutron stars with a
strong magnetic field, known as magnetars
\cite{magnetar}.

Quasi-normal modes (QNMs) are the oscillations with complex frequencies in open
systems with energy dissipations. Perturbations of a black hole are described
by QNMs and the complex frequencies of QNMs carry the characteristic properties
of the black hole like mass, charge, and angular momentum, independent of the
initial perturbations. QNMs of black holes have attracted much interest in
relation with the gravitational waves since these can be used as a key tool
for the test of GR. Theoretically, this also provides a (simple)
criterion of the stability, under linear perturbations, of a black hole by
looking at the sign of the imaginary part of quasi-normal frequency.

In this paper, we consider electrically charged black holes in
EBI
gravity with a positive cosmological constant and
compute QNMs of gravitational perturbations of EBI black holes. We adopt the
Schutz-Iyer-Will's WKB method \cite{Schutz:1985a, iyerwill, iyer} for
numerical computations of QNMs and consider the {\it axial} perturbations
for simplicity.
We obtain a set of decoupled Schr\"{o}dinger-type equations, corresponding to
Regge-Wheeler equation, whose formal expressions, in terms of metric functions, are the same as those without cosmological constant. We discuss the stability of the charged black holes by investigating the dependence of quasi-normal frequencies on
the
parameters of the
theory,
{\it correcting some errors in the literature}. It is
found that all the axial perturbations are {\it stable}
for the cases where
the WKB method applies.
There are cases where
the conventional WKB method does not apply, like the three-turning-points problem,
so that a more generalized formalism is necessary for studying their QNMs and stabilities.
We find that, for the degenerate horizons with the ``point-like" horizons at
the origin,
the QNMs  are quite {\it long-lived}, close to
the quasi-resonance modes (QRMs), in addition to the ``frozen" QNMs
with the vanishing frequency, $\om \approx 0$, for the Nariai-type horizons and the usual (short-lived) QNMs
for the extremal black hole horizons. This is a genuine effect of the
{\it non-GR} branch of $\beta Q = 1/2$, {which does not have the GR limit $\beta \ra \infty$ with the BI parameter $\beta$ and electric charge $Q$.}
{We study the exact solution near the (charged) Nariai limit
and find good agreements even far beyond the limit for the
imaginary frequency parts.}

The organization of this paper is as follow. In Sec. 2, we review the electrically charged black hole solution in EBI gravity with a positive cosmological constant.
In Sec. 3, we consider the axial perturbations from the spherically symmetric EBI black hole background and obtain a set of decoupled Regge-Wheeler equations.
In Sec. 4, we use the Shutz-Iyer-Will's third-order WKB method to compute the QNMs numerically and discuss the stabilities of the EBI black holes.
In Sec. 5, we study the exact solution near the (charged) Nariai limit and compare it with the numerical results in Sec. 4.
In Sec. 6, we conclude with several discussions. Throughout this paper, we use the conventional units for the speed of light $c$, the electric and magnetic constants $\epsilon_0, \mu_0$, and the Boltzman constant $k_B$, $c =4 \mu_0=4/\epsilon_0= k_B = 1$, but keep the Newton constant $G$ and the Planck constant $\hbar$ unless stated otherwise.

\section{Background Solutions}

In this section, we briefly review the black hole solution in EBI gravity with a positive cosmological constant $\La$ in $(3+1)$ dimensions, from which the
gravitational perturbations will be considered.
The EBI action is given by
\be
 S = \int d^{4} x \sqrt{-g} \left [ \frac{(R - 2\Lambda)}{16 \pi G}
  + L(F)                            \right ],
 \label{ebiaction}
\ee
where the BI Lagrangian density $L(F)$ is given by
 \be
 L(F) = 4 \beta^2 \left( 1 - \sqrt{ 1 + \frac{ F_{\mu\nu} F^{\mu\nu} }{2 \beta^2} }
                  \right) .
 \label{BI}
 \ee
Here, 
{$\beta~$} is
the BI's
coupling constant with dimensions $[{\rm length}]^{-2}$ \cite{borninfeld1,borninfeld2}.
In the weak field or equivalently strong coupling limit, $| F_{\mu\nu} F^{\mu\nu} | \ll 2 \beta^2$, $L(F)$ may be expanded as
\beq
L(F)= -{F}^2+\f{2 F^4}{\beta^2}+{\cal O}\left(\f{F^6}{\beta^4} \right),
\eeq
where $F^2 \equiv F_{\mu\nu} F^{\mu\nu}$ and the usual Maxwell's electrodynamics
is recovered at the lowest order.
The correction terms,
when $F$ is comparable to $\beta$, represent the effect of the non-linear BI fields at the short distance $\sim \beta^{-1}$ and the possible electromagnetic field strength is bounded by $-F^2 \leq 2 \beta^2$.

Taking $16 \pi G = 1$ for simplicity, the equations of motion are
obtained as
\be
\nabla_{\mu} \left ( \frac{F^{\mu\nu}}{\sqrt{ 1 + \frac{ F^2 }{2 \beta^2} }}
             \right ) = 0 ,
\label{eomforBI}
\ee
\be
 R_{\mu\nu} - \frac{1}{2} R g_{\mu\nu} + \Lambda g_{\mu \nu}=  \frac{1}{2}T_{\mu\nu},
 \label{eomforgrav}
 \ee
where the energy-momentum tensor for BI fields is given by
 \be
  T_{\mu\nu}= g_{\mu\nu} L(F) +\frac{4 F_{\rho\mu} {F^\rho}_{\nu} }{\sqrt{ 1 + \frac{ F^2 }{2 \beta^2} }}  .
   \label{Tmunu}
   \ee
Let us now consider a static and spherically symmetric solution with
the metric ansatz
 \be
 ds^2 = -N^2(r) dt^2 + \frac{1}{f(r)} dr^2 + r^2 (d \theta^2 + \sin^2 \theta d \phi^2).
 \label{ansatz}
 \ee
For the static, electrically-charged case where the only non-vanishing component of the
field strength tensor is {$F_{rt} \equiv E_r $}, the general solutions for the metric and the BI
electric field are obtained
 as
 \beq
 N^2(r)
&=&f(r)=
1 - \frac{2 M}{r} - \frac{\Lambda}{3} r^2
 + \frac{2 }{3} \beta^2 r^2 \left ( 1 - \sqrt{ 1+ \frac{Q^2} { \beta^2 r^{4} } } \right )
  + \frac{4}{3} \frac{Q^2} {r^{2}}~ {_2 F}_1 \left ( \frac{1}{2} , \frac{1}{4}; \frac{5}{4};
  \frac{-Q^2}{\beta^2 r^{4}} \right ),
   \label{f:sol2} \\
   E_r(r) &=& \frac{Q}{\sqrt{ r^{4} + \frac{Q^2}{\beta^2}    } } ,
 \label{Esol}
\eeq
in terms of the hypergeometric function \cite{Garc:1984,Li:2016}. Here $Q$ is the electric charge \footnote{The charge convention differs from the usual one, like that of Rasheed in \cite{Garc:1984}. One can convert to the usual convention by replacing $Q \ra Q/16 \pi$.} and $M$ is the ADM mass which is composed of the intrinsic mass $C$ and (finite) self energy of a point charge $M_0$, defined by
\beq
M &= &C+ M_{0}, \\
M_{0} &=& \frac{2}{3} \sqrt{ \frac{\beta Q^3}{\pi} } \Gamma\left( \frac{1}{4} \right) \Gamma\left( \frac{5}{4} \right).
\label{M0}
 \eeq
In the large-distance limit $r \gg \sqrt{Q/\beta}$, the solutions become
\beq
 N^2&=&f = 1 - \frac{\Lambda}{3} r^2 - \frac{2 M }{r}
 +  \frac{Q^2} {r^{2}} -\frac{Q^4}{20 \beta^2 r^6} + {\cal O}(r^{-10}), \\
 \label{f:larger}
E_r &=& \frac{Q}{r^2} -\frac{\beta^2 Q^3}{r^{6}} +{\cal O}(r^{-10}),
 \label{E:larger}
 \eeq
which show the usual Reissner-Nordstrom-(Anti) de Sitter ({RN-(A)dS}) black hole at the leading order and the short-distance corrections at the sub-leading orders. On the other hand, in the short-distance limit $r \ll \sqrt{Q/\beta}$, we have
\beq
 N^2&=&f = 1 - 2 \beta Q - \frac{2C}{r} + \frac{1}{3} (2 \beta^2 - \Lambda) r^2
 -\frac{\beta^3}{5 Q}r^4+{\cal O}(r^6), \\
 \label{f:smaller}
 E_r &=&\f{Q}{|Q/\beta|}-\f{Q}{2 |Q/\beta|^3}  r^4 +{\cal O}(r^8),
 \label{E:smaller}
 \eeq
which show the {\it regularized} solutions near the origin, with the {\it milder} (curvature) singularity of the metric and the {\it finite} electric field at the origin \cite{Li:2016,Kim:2016pky}
due to the non-linear BI fields for a finite  coupling $\beta$.

\begin{figure}
\includegraphics[width=7cm,keepaspectratio]{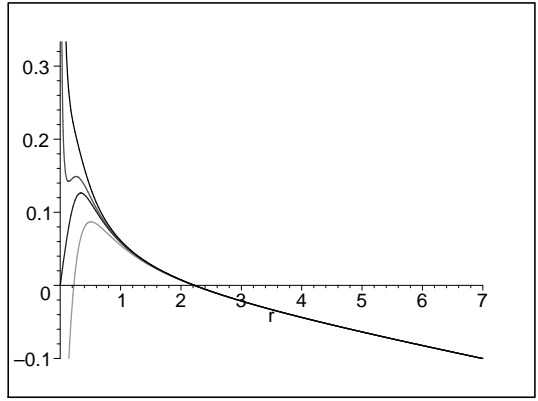}
\caption{The plots of the Hawking temperature $T_H$ vs. the (outer)
black hole horizon radius $r_+$
for
{
varying}
$\beta Q$ with a fixed 
$\Lambda>0$.
We consider $\beta Q={2/3,~1/2,~2/4.5,~1/3}$ (bottom to top curves) with $\beta=2$, $\Lambda=0.2$.}
\label{fig:T}
\end{figure}

\begin{figure}
\includegraphics[width=7cm,keepaspectratio]{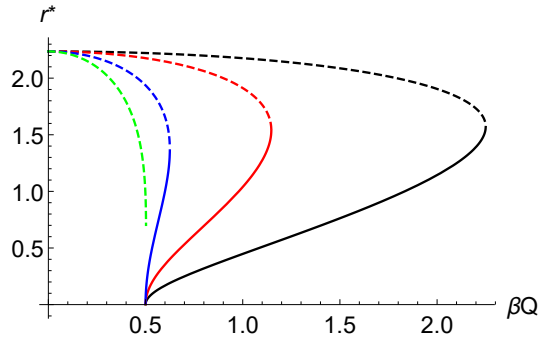}
\caption{The plots of the extremal horizons { $r^*_{H}$ and $r^*_{C}$ $(>r^*_{H})$} vs. $\beta Q$ for varying $\beta$.
We consider $\beta= 2,~1,~1/2,~1/3$ (right to left)
with $\Lambda=0.2$. The solid/dashed lines denote the `$+/-$' roots
in (\ref{r:ext}) and there is no `$+$' root for the last case.
}
\label{fig:r_star}
\end{figure}

The solution (\ref{f:sol2}) can have three horizons generally,
{\it i.e.}, two (inner, outer) black hole horizons, $r_-,r_+$,
and one cosmological horizon $r_{++}$. The Hawking temperature
for the outer black horizon $r_+$
is given by
\beq
T_H &\equiv& \f{\hbar}{4 \pi} \left. \f{df}{dr} \right|_{r=r_+} \no \\
&=&\f{\hbar}{4 \pi} \left[ \f{1}{r_+}-\La r_+ +2 \beta^2 r_+ \left(1-\sqrt{1+\f{Q^2}{\beta^2 r_+^4}} \right) \right]
\label{TH}
\eeq
and plotted in Fig. 1. There exist two extreme
limits of vanishing temperature, when the outer black hole horizon $r_+$ meets the inner horizon $r_-$ ({\it extremal black holes}) or the cosmological horizon $r_{++}$ ({\it Nariai solution}), at
\beq
r^*_{H/C}=\sqrt{
\f{\La-2 \beta^2
\pm  \sqrt{(\La -2\beta^2)^2+\La(\La-4 \beta^2)(4\beta^2 Q^2-1)}
}
{\La(\La-4 \beta^2)}
}
\label{r:ext}
\eeq
for a positive cosmological constant $\La$ ($r^*_H <r^*_C$)
(Figs. 1 and 2).
At the extreme points, the ADM mass $M$,
\beq
M=\f{r_+}{2} \left[ 1 -  \frac{\Lambda}{3} r_+^2
 + \frac{2 }{3} \beta^2 r_+^2 \left ( 1 - \sqrt{ 1+ \frac{Q^2} { \beta^2 r_+^{4} } } \right )
  + \frac{4}{3} \frac{Q^2} {r_+^{2}}~ {_2 F}_1 \left ( \frac{1}{2} , \frac{1}{4}; \frac{5}{4};
  \frac{-Q^2}{\beta^2 r_+^{4}} \right ) \right],
\eeq
becomes \cite{Kim:2016pky,Kim:2018ang}
\beq
M^*=\f{r^*}{3} \left[ 1
  + \frac{2 Q^2} {{r^*}^{2}}~ {_2 F}_1 \left ( \frac{1}{2} , \frac{1}{4}; \frac{5}{4};
  \frac{-Q^2}{\beta^2 {r^*}^{4}} \right ) \right].
  \label{M_star}
\eeq

The first law of black hole thermodynamics is found as
\beq
dM=T_H dS_{BH} +A_0(r_+) dQ,
\eeq
with the usual Bekenstein-Hawking formula for the black hole entropy
\beq
S_{BH}=\f{\pi r_+^2}{\hbar G},
\eeq
and the scalar potential \ci{Garc:1984,Cai,Li:2016}
\beq
A_0(r)=\frac{Q}{r}~ {_2 F}_1 \left ( \frac{1}{2} , \frac{1}{4}; \frac{5}{4};
  \frac{-Q^2}{\beta^2 {r}^{4}} \right ).
\eeq
The integrated first-law of black hole thermodynamics, called the {\it generalized} Smarr relation, is given by \cite{Cai,Li:2016,Smar:1972,Liu:2015} \footnote{The generalized Smarr relation without the last term has been obtained for the planar black holes in \cite{Li:2016,Liu:2015}. If we consider {\it topological} EBI black holes with the topological parameter $k=+1,0,-1$ for spherical, plane, and hyperbolic geometry generally, with the solid angle $\Omega_k$, one finds \cite{Cai} the last term as $(k/3) \sqrt{ 4\hbar  S_{BH}/{\Omega_k G}}$.}
\beq
M=\frac{2}{3} \left(S_{BH} T_{H} +Q A_0 (r_+) \right)+\frac{1}{3} \sqrt{\frac{ \hbar S_{BH}}{\pi G}}.
\eeq
Now, from the relation (\ref{r:ext}), one can classify the black holes
by the values of $\beta Q$ \cite{Kim:2016pky}.
\\

\begin{figure}
\includegraphics[width=4.8cm,keepaspectratio]{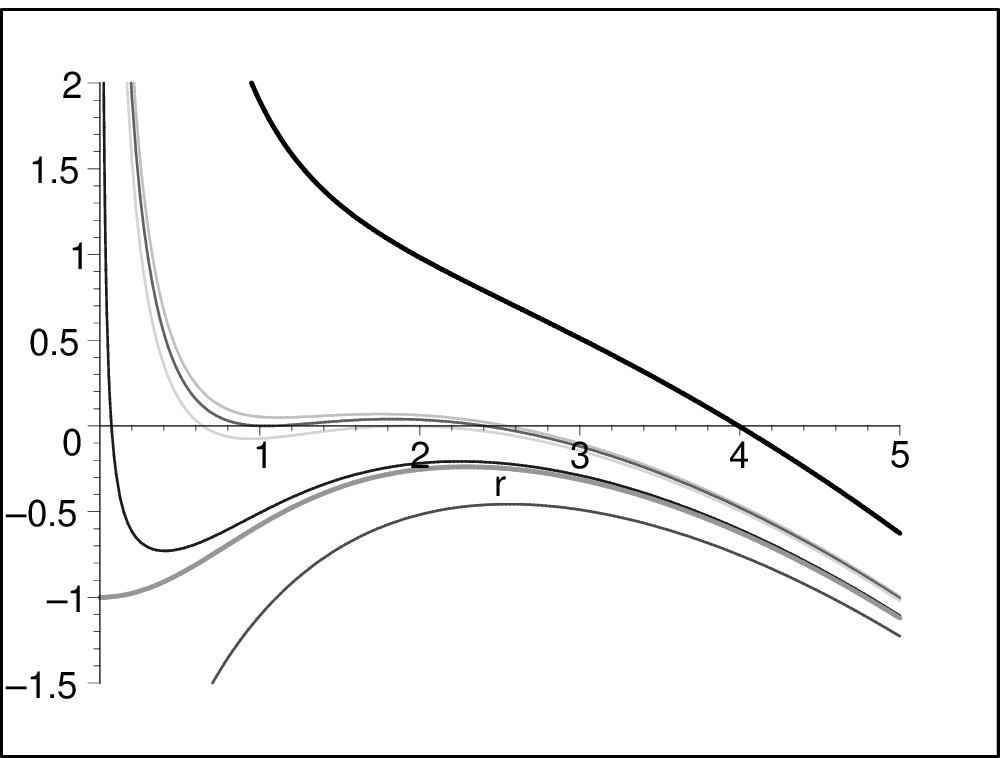}
\qquad
\includegraphics[width=4.8cm,keepaspectratio]{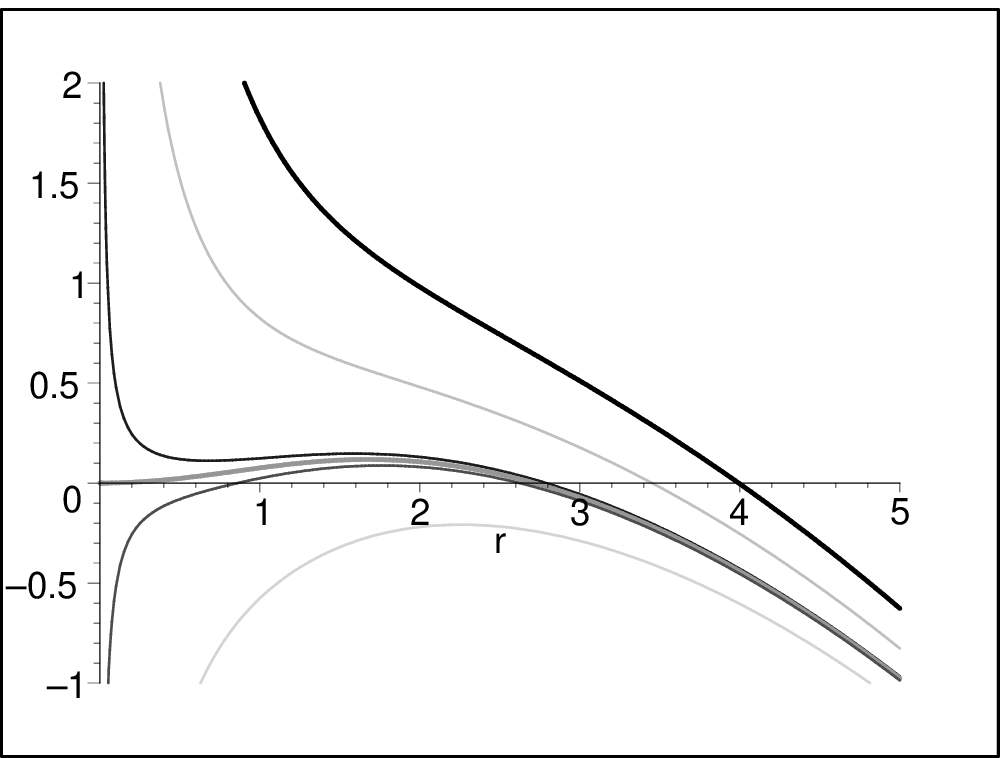}
\qquad
\includegraphics[width=4.8cm,keepaspectratio]{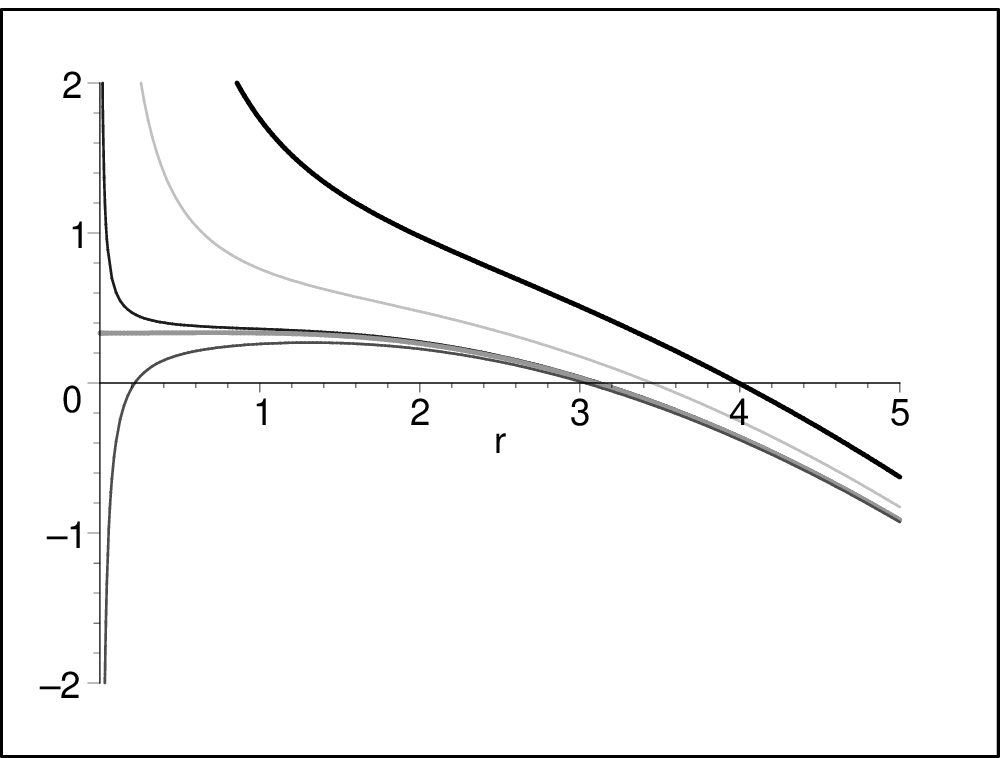}
\caption{The plots of $f(r)$ for varying
$M$ with a fixed $\beta Q$ ($\beta Q >1/2$ (left),
$\beta Q=1/2$ (center), $\beta Q<1/2$ (right)) and
$\Lambda>0$. We consider
{(top to bottom)} $M=0,~0.95,~1.2,~M_0,~1.5$ with
{ $\beta=1$, $M_0\approx 1.236$ (left);  $M=0,~0.85,~M_0,~0.9,~1.2$
with $\beta=1/2$, $M_0 \approx 0.874$ (center); and $M=0,~0.7,~M_0,~0.75,~1.2$
with $\beta=1/3$, $M_0 \approx 0.714$ (right)}, respectively for $Q=1, \Lambda=0.2$.}
\label{fig:f}
\end{figure}

\begin{figure}
\includegraphics[width=7cm,keepaspectratio]{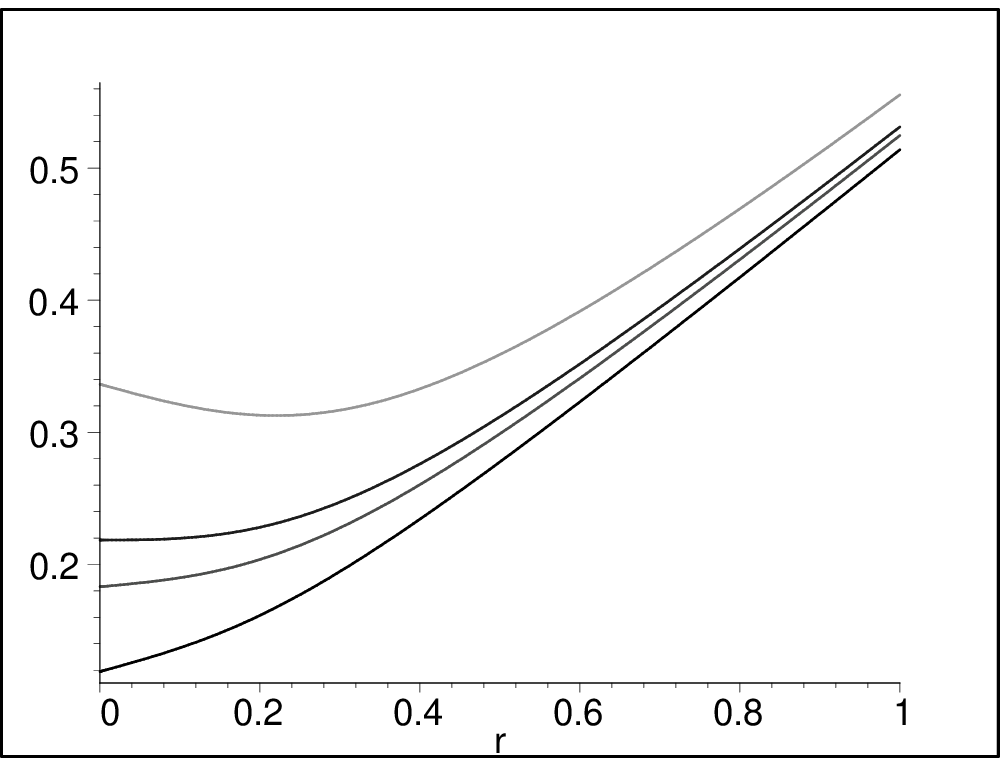}
\qquad
\includegraphics[width=7cm,keepaspectratio]{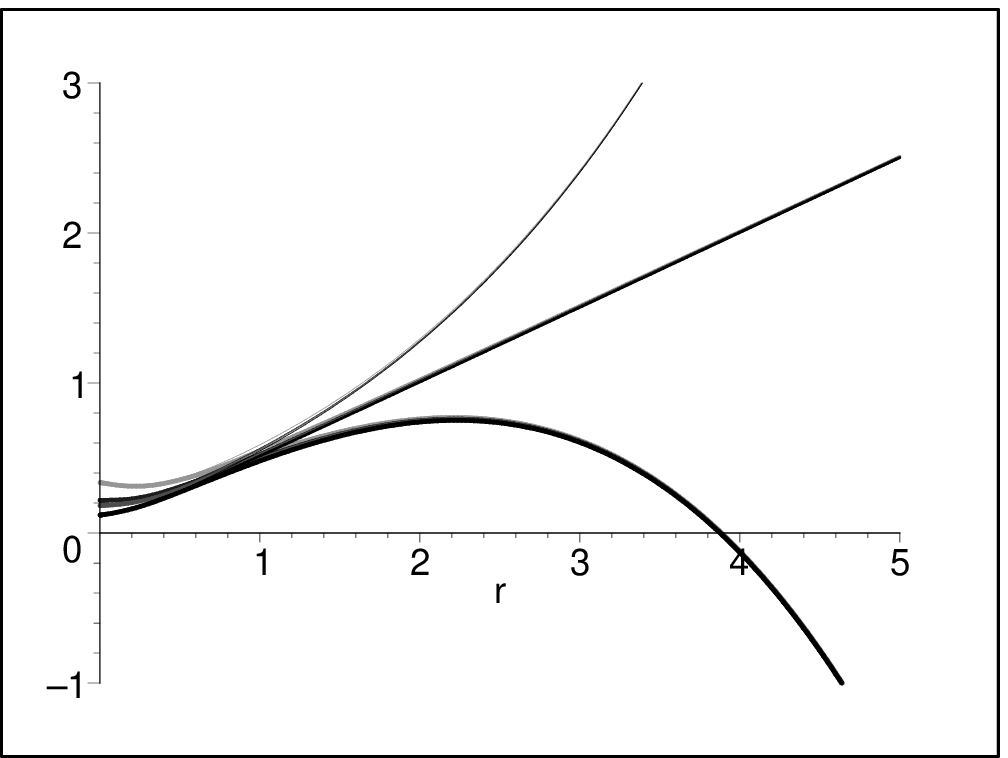}
\caption{The plots of the ADM mass $M$ vs. the black hole horizon radius $r_+$
for varying $\beta Q$ with a fixed
$\Lambda>0$. The marginal mass $M_0$ is given by the mass value at $r_+=0$. The top two curves in the left represent $M_0>M^*,~M_0=M^*$ for $\beta Q>1/2,~\beta Q=1/2$, respectively, with the extremal mass $M^*$, whereas the bottom two curves represent the cases where $M^*$ is absent for $\beta Q<1/2$. We consider $\beta Q={2/3,~1/2,~2/4.5,~1/3}$ (top to bottom) with $\beta=2$ and $\Lambda=0.2$. The effect of cosmological constant is not significant for small $r_+$ (left) but important for large $r_+$ (right). In the latter, we compare the dS case of $\Lambda=0.2$ (thick curve) with the flat $\Lambda=0$ (medium curve) and the AdS case of $\Lambda=-0.2$ (thin curve), respectively.
}
\label{fig:M}
\end{figure}

(i) $\beta Q >1/2$:
In this case, there are generally
three horizons (two smaller ones for black holes and the largest one for
the cosmological horizon) for $M_H^* \leq M \leq M_C^* ~(<M_0)$, depending on
$M$ and
$\La$
(Fig. \ref{fig:f}, left), where $M_H^*$ and $M_C^*$ denote the values of the extremal mass $M^*$ in (\ref{M_star}) at $r_H^*$ and $r_C^*$ for the extremal black holes and the (charged) Nariai solutions, respectively, with the vanishing Hawking temperature (Fig.1). When the mass is outside this range, {\it i.e.,} $M< M_H^*$ or $M > M_C^*$, the singularity at $r=0$
becomes naked as
in the RN black hole.\\

(ii) $\beta Q =1/2$: In this case, only the \Sch- de Sitter (Sch-dS) -like black holes with a non-degenerate black hole horizon $r_+$ are
possible for $M_0 < M \leq M_C^*$ (Fig. \ref{fig:f}, center). It is peculiar
that the black hole horizon $r_+$ shrinks to zero size, {\it i.e.},
{\it the point-like} horizon, for the marginal case $M=M_0$
(Fig. \ref{fig:M}). This corresponds to the extremal black hole, with the vanishing Hawking temperature (Fig.1), where the horizon {\it degenerates} at the origin. When the mass $M$ is smaller than the marginal mass $M_0$ or larger than $M_C^*$, the singularity at $r=0$ becomes naked. Note that, in this case, the GR limit $\beta \ra \infty$ does not exist.\\

(iii) $\beta Q <1/2$: This case is similar to the case (ii), except that the
marginal case $M=M_0$ has no (even a point) horizon so that its curvature singularity at $r=0$ is naked always (Fig. \ref{fig:f}, right), even though the (non-degenerate) black hole horizon can arbitrarily approach close to the point-like horizon $r_+=0$ in the limit $M\ra M_0$, with the {\it divergent} Hawking temperature (Fig. 1) \footnote{This behavior of Hawking temperature is similar to that of Sch. black hole. But the difference is that its mass $M$ has still a non-vanishing {\it remnant} $M_0$. If this tiny black hole with a unit charge $Q=\sqrt{4 \pi \epsilon_0 \alpha \hbar c}$~ ($\alpha$ is the fine-structure constant) is created at the Planck energy scale in the early universe, {\it i.e.}, $M_0 \approx \sqrt{\hbar c/G}$, we obtain $\beta \approx (9/2) \Gamma(3/4)^4 \pi^{-7/2}~ \alpha^{-3/2}~ (\epsilon_0 \hbar c)^{-1/2} G^{-1}$. If we consider $\alpha_{ Pl}\sim 1.5$ at the early universe, contrast to $\alpha \sim 10^{-2}$ at the current epoch, we have $\beta_{Pl}\sim 0.1~
(\epsilon_0 \hbar c)^{-1/2} G^{-1}$
and $\beta_{Pl} Q \approx 9 \Gamma(3/4)^4 \pi^{-3}
~\alpha_{Pl}^{-1} G^{-1} \sim 0.4~ G^{-1}$.
 }.

\section{Perturbation Equations}

In this section, we consider the gravitational perturbations of
an electrically charged black hole in EBI gravity with a positive cosmological constant $\La>0$, whose solutions for the metric and fields are given by (\ref{f:sol2}) and (\ref{Esol}), respectively.
We
follow the procedure of Chandrasekhar \cite{Chandrasekhar} for
the perturbations of RN solution but in the different conventions which
agree with those of Wald \cite{Wald:1984}
\footnote{We use a metric with signature $(-+++)$ and the Wald's conventions
for the Riemann and Ricci tensors which
differ from \cite{Chandrasekhar,Fernando:2004pc,Mell:1989}.}.
We generalize the earlier computations of \cite{Fernando:2004pc} to a positive cosmological constant but with some important corrections of errors.
To accommodate the procedure of \cite{Chandrasekhar}, it is useful to write the
metric (\ref{ansatz}) as
 \be
 ds^2 = -e^{2 \nu } dt^2 + e^{2 \psi} d\phi^2 +e^{2 \mu_2} dr^2 + e^{2 \mu_3 } d \theta^2
  \label{backgroundmetric}
 \ee
with
 \be
  e^{2 \nu }  \equiv
  N^2(r),~  e^{2 \mu_2 }  \equiv
  \frac{1}{f(r)} , ~e^{2 \mu_3} \equiv  r^2 , ~ e^{2 \psi} \equiv r^2 \sin^2 \theta .
  \label{backgroundpolars}
 \ee
We will consider (\ref{backgroundmetric}) as a background metric and consider
the {\it first-order}
metric perturbations. In our linear perturbations of the spherically symmetric system, we may restrict ourselves to axisymmetric modes of perturbations, without loss of generality \cite{Chandrasekhar}, whose metric may be generally written as
 \be
 ds^2 = -e^{2 \nu } dt^2 + e^{2 \psi} (d\phi - q_0 dt - q_2 dr - q_3 d \theta )^2 +e^{2 \mu_2} dr^2 + e^{2 \mu_3 } d \theta^2,
  \label{perturbedmetric}
 \ee
where the metric functions $\nu, \psi, \mu_2, \mu_3, q_0, q_2$, and $q_3$ are functions of only $t, r$, and $\theta$, due to axisymmetry.
There are two kinds of perturbations: The perturbations of $ q_0, q_2$, and $q_3$,
which are called ``axial" perturbations, induce a dragging effect and impart a
rotation to the black hole, whereas the increments of
$\delta \nu, \delta \psi, \delta \mu_2  $, and $\delta \mu_3$, which are called
``polar" perturbations, do not impart such rotation. These two perturbations
are decoupled
and can be treated independently \cite{Chandrasekhar}. In this paper, we will
focus only on the axial perturbations for simplicity.


To obtain the solutions for the perturbed metric (\ref{perturbedmetric}), we first need to consider the perturbations of BI and Einstein equations, (\ref{eomforBI}) and (\ref{eomforgrav}).
For this purpose, we adopt the tetrad formalism as in \cite{Chandrasekhar}
and use the Roman indices
$a, b = 0, 1, 2, 3$ for the tetrad indices and the Greek indices
$\mu, \nu =t, \phi, r, \theta$
for the curved coordinate indices.
The tetrad basis $ e^{a}_\mu $ satisfies $e^{a}_{\mu} e^{b}_{\nu} \eta_{ab}= g_{\mu\nu}$ with $\eta_{ab} = {\rm diag} (-1,+1,+1,+1) $.
For the metric given by
(\ref{perturbedmetric}), $ e^{a}_\mu $ can be obtained as
\beq
e^{0}_\mu  &= & ( e^{\nu} , 0,0,0) ,\no\\
e^{1}_\mu  &= & ( - q_0 e^{\psi} , e^{\psi} , - q_2 e^{\psi} , - q_3 e^{\psi} ) , \no \\
e^{2}_\mu  &= & ( 0,0, e^{\mu_2} , 0) , \no\\
e^{3}_\mu  &= & ( 0,0,0, e^{\mu_3} ) \label{tetrad}.
\eeq
Tensors in the two bases are related by the tetrads, for example,
\be
F_{\mu\nu } = e_{\mu}^{a} e_{\nu}^{b}F_{ab}, ~~
R_{\mu\nu } = e_{\mu}^{a} e_{\nu}^{b}R_{ab},  \label{F_tetrad}
\ee
for the field strength and the Ricci tensor. We will now consider equations of motion (\ref{eomforBI}) and (\ref{eomforgrav}) in the tetrad basis.

\subsection{The perturbed BI equations}

BI fields satisfy the Bianchi identities,
\beq
\nabla_{[\mu} F_{\nu\lambda]} =0, \label{F_Bianchi}
\eeq
in addition to the 
field equations (\ref{eomforBI}). The corresponding two
sets of equations for tetrad components of BI fields can be found from
the tetrads (\ref{tetrad}) and the relation (\ref{F_tetrad}).

First, from the Bianchi identities (\ref{F_Bianchi}), one can find four equations as
\bear
&& (e^{\psi + \mu_2} F_{12}  )_{,\theta} + (e^{\psi + \mu_3} F_{31}  )_{,r} =  0 , \label{no_monopole}
 \\
&&(e^{\psi + \nu} F_{01}  )_{,r} + (e^{\psi + \mu_2} F_{12}  )_{,t} =  0,  \label{Faraday law_3}  \\
&&(e^{\psi + \nu} F_{01}  )_{,\theta} + (e^{\psi + \mu_3} F_{13}  )_{,t} =  0,
\label{Faraday law_2}   \\
&&(e^{\psi + \mu_2} F_{02}  )_{,\theta} - (e^{\nu + \mu_3} F_{03}  )_{,r } +  (e^{\mu_2 + \mu_3} F_{23}  )_{,t }    \nonumber  \\
&&=e^{\psi + \nu} F_{01}  Q_{23} + e^{\psi + \mu_2} F_{12}  Q_{03} - e^{\psi + \mu_3} F_{13}  Q_{02},  \label{Faraday law_1}
\eear
from the assumed axisymmetry in the tetrads and BI fields.
Here,
$Q_{ab}$'s are defined as
\be
Q_{02} \equiv q_{0 , r} - q_{2, t} ,~~Q_{03} \equiv q_{0 , \theta} - q_{3, t},
~~ Q_{23} \equiv q_{2 , \theta} - q_{3, r}.
\ee
Note that (\ref{no_monopole}), (\ref{Faraday law_3}), and (\ref{Faraday law_2})
represent the absence of a $U(1)$ magnetic monopole and the Faraday's law of
induction in the curved space. However, (\ref{Faraday law_1}), which corresponds
the $\phi$-component of the Faraday's law, shows that there are ``induced" source
terms on the right hand side which correspond the {\it magnetic} currents
${\cal J}_{(m)} \sim \epsilon^{abcd} F_{ab} Q_{cd}$, due to {\it non-linear}
interactions between the electromagnetic
and
gravitational fields.

Second, from BI field equations (\ref{eomforBI}), which may be conveniently written as
\beq
\nabla_{\mu} D^{\mu\nu} =0, \label{eomforBI_2}
\eeq
where
\be
D^{\mu\nu} \equiv \frac{F^{\mu\nu}}{ \sqrt {1 + F^2 / \beta^2} },
\ee
one can find another set of four equations as
\bear
&& (e^{\psi + \mu_3} D
_{02}  )_{,r} + (e^{\psi + \mu_2} D
_{03}  )_{,\theta} =  0 , \label{Gauss_law} \\
&& - (e^{\psi + \nu} D_{23}  )_{,r} + (e^{\psi + \mu_2} D_{03}  )_{,t} =  0 ,
\label{Ampere_law_3}   \\
&&(e^{\psi + \nu} D_{23}  )_{,\theta} + (e^{\psi + \mu_3} D_{02}  )_{,t} =  0,
\label{Ampere_law_2}   \\
&& (e^{\mu_2+ \mu_3} D_{01}  )_{,t} + (e^{\nu + \mu_3} D_{12}  )_{,r} + (e^{\nu + \mu_2} D_{13}  )_{,\theta}   \nonumber  \\
&&=e^{\psi + \mu_3} D_{02}  Q_{02} + e^{\psi + \mu_2} D_{03}  Q_{03} - e^{\psi + \nu} D_{23}  Q_{23} \label{Ampere_law_1} .
\eear
Here, (\ref{Gauss_law}), (\ref{Ampere_law_3}), and (\ref{Ampere_law_2}) represent the Gauss's law and the Ampere's law (with the Maxwell's correction term) in the curved space. On the other hand, (\ref{Ampere_law_1}), which corresponds the $\phi$-component of the Ampere's law, shows the induced source terms which correspond the {\it electric} currents ${\cal J}_{(e)} \sim F_{ab} Q_{ab}$, due to non-linear interactions between the electromagnetic and gravitational fields, similar to (\ref{Faraday law_1}). These induced source terms in (\ref{Ampere_law_1}) and (\ref{Faraday law_1}) represent the {\it back-reactions} from the dynamical metric $q_0,q_2$ and $q_3$, which are also driven by $T_{\mu \nu}$ in the Einstein equations (\ref{eomforgrav}).

Up to now, we have not assumed any specific solution of the background BI fields.  Now, using that the only non-zero components of $F$ and $D$ in the background solution (\ref{Esol}) are
\be
F_{02} = - \frac{Q} {\sqrt{ r^4 + Q^2/\beta^2}} ,~~~
 D_{02} = - \frac{Q} {r^2},
\ee
and substituting $\psi , \nu , \mu_2 , \mu_3$ with $\psi + \delta \psi, \nu + \delta \nu , \mu_2 + \delta \mu_2 , \mu_3 + \delta \mu_3$ in the above equations,
we find the linearized versions of perturbed BI equations as
\bear
(r e^\nu F_{01}  )_{,r} + r e^{-\nu} F_{12,t} &=&  0,        \label{aximax1}  \\
 r e^\nu (F_{01} \sin \theta )_{,\theta} + r^2  F_{13,t} \sin \theta  &=&  0,    \label{aximax2}    \\
 r e^{-\nu}  D_{01,t} + (r e^\nu D_{12})_{,r} +  D_{13,\theta}   &=&
 - Q ( q_{0,2} - q_{2,0} ) \sin \theta , \label{aximax3}
\eear
\bear
 - (r e^\nu D_{23}  )_{,r} + r e^{-\nu} D_{03,t}  &=&  0 ,  \\
\delta D_{02,t} - \frac{Q} {r^2} ( \delta \psi + \delta \mu_3 )_{,t}+ \frac{e^\nu}{r \sin \theta} ( D_{23} \sin \theta )_{, \theta}  &=&  0,    \\
\left [ \delta F_{02}  - \frac{Q}{ \sqrt{r^4 + Q^2/\beta^2 }} (\delta \nu + \delta \mu_2 ) \right ]_{, \theta} + ( r e^\nu F_{30} )_{,r} + r e^{-\nu} F_{23, t}   &=& 0.
\eear

\subsection{ The perturbed Ricci tensor equations }

To find the metric perturbations of the Einstein equation (\ref{eomforgrav}),
we may conveniently write
the Einstein equations (\ref{eomforgrav}) as \footnote{The energy-momentum part
in the right-hand side of (\ref{Ricci_Eq1}) coincides with $``T_{\mu \nu}"$ for
Maxwell's electro magnetism. But, in BI theory, it is not valid generally.
This is one of the
errors in \cite{Fernando:2004pc}
which may affect the polar perturbations in EBI gravity.}
\beq
R_{\mu \nu}&=&\f{1}{2} \left(T_{\mu \nu}-\f{1}{2} g_{\mu \nu} {T^\ga}_{\ga} \right)+\La g_{\mu \nu}
\label{Ricci_Eq1} \\
&=&\frac{2}{ \sqrt{ 1+\f{F^2}{2 \beta^2} } } \left[ g^{\ga \de} F_{\mu \ga} F_{\nu \de}+  g_{\mu \nu} ~ \beta^2\left( 1- \sqrt{ 1+\f{F^2}{2 \beta^2} } \right) \right]+\La g_{\mu \nu}
\label{Ricci_Eq2}
\eeq
or
\beq
R_{ab}=\frac{2}{ \sqrt{ 1+\f{F^2}{2 \beta^2} } } \left[ \eta^{cd} F_{ac} F_{bd}+  \eta_{ab} ~ \beta^2\left( 1- \sqrt{ 1+\f{F^2}{2 \beta^2} } \right) \right]+\La \eta_{ab}
\label{Ricci_Eq_tetrad}
\eeq
in the tetrad basis, where $F^2 \equiv F_{\mu \nu}F^{\mu \nu} =F_{ab}F^{ab}$.

With a simple computation,
one can easily find that the cosmological constant term
does not appear explicitly in the perturbation of Einstein equations
(\ref{Ricci_Eq1}).
The lineralized form of the perturbed Ricci tensor equations are then given by
\beq
 \delta R_{ab} = \f{2}{ \sqrt{ 1 +\frac{F^2}{2\beta^2} } }
 \left[  -\f{1}{2} F_{cd} \delta F_{cd}
 \left( \f{ \eta_{ab} + \eta^{cd} F_{ac} F_{bd}/\beta^2 }
 {  1 +\frac{F^2}{2\beta^2}  } \right)
 + \eta^{cd} \left( \delta F_{ac}  F_{bd} + F_{ac} \delta  F_{bd}  \right) \right].
 \label{Ricci_pert}
\eeq
Since the only non-zero componsnt of the  background $F_{ab}$ is $F_{02}$, we replace $\delta F_{ab} $ by $ F_{ab}$ except for $\delta F_{02}$ for simplification, following \cite{Chandrasekhar}.
Explicit forms of the right-hand side of the perturbed Ricci tensors (\ref{Ricci_pert}) are given by
\bear
\delta R_{00} &&= - \delta R_{22} =  - \frac{2Q}{r^2} \delta F_{02} ,     \\
\delta R_{11} &&=  \delta R_{33} =  - \frac{2Q}{r^2}
\left( 1+\frac{Q^2}{\beta^2 r^4} \right)\delta F_{02}
   ,      \\
\delta R_{01} &&= - \frac{2 Q}{r^2}  F_{12} ,      \\
\delta R_{02} &&= 0 ,      \\
\delta R_{03} &&=  \frac{2 Q}{r^2}  F_{23} ,      \\
\delta R_{12} &&=  \frac{2 Q}{r^2}  F_{01} ,   \label{dr12=dt12}   \\
\delta R_{13} &&= 0 ,     \label{dr13=dt13}    \\
\delta R_{23} &&=  \frac{2 Q}{r^2}  F_{03} .
\eear

\subsection{The wave equations for the axial perturbations}
The linearized Einstein equations can be obtained by equating
$\delta R_{ab}$ of the left-hand side in the above equations with those which
can be computed from the Ricci tensors for the metric (\ref{perturbedmetric})
in the tetrad basis, whose explicit forms are given in the Appendix. As
mentioned before, the perturbation equations can be categorized into two kinds. One is the axial perturbations which are characterized by the non-vanishing of $q_0 , q_2 , q_3 , F_{01}, F_{12} $, and $F_{13}$.
The other is the polar perturbations which are characterized by the non-vanishing of $ \delta F_{02}, F_{03} , F_{23}, \delta \nu ,\delta \psi, \delta \mu_2$, and $\delta \mu_3$.
In this paper, we only consider the axial perturbations, which are relatively
easier and can be treated independently of the polar perturbations.
For this purpose, we first consider (\ref{dr12=dt12}) and (\ref{dr13=dt13}),
which are now given by
\bear
(r^2 e^{2\nu} Q_{23} \sin^3 \theta )_{,\theta} + r^4 Q_{02,t}   \sin^3 \theta  &=& 4 Q r e^\nu  F_{01} \sin^2 \theta ,   \label{daughterofdr12=dt12}  \\
 (r^2 e^{2\nu} Q_{23} \sin^3 \theta )_{,r} - r^2 e^{-2 \nu} Q_{03,t}   \sin^3 \theta  &=&0~.       \label{daughterofdr13=dt13}
\eear
Eliminating $D_{12}$ and $D_{13}$ from
(\ref{aximax3}) using
(\ref{aximax1}) and (\ref{aximax2}), we obtain
\be
 \left [ \frac{e^{2 \nu} } { \gamma} ( r e^\nu \gamma \widetilde{\cal D} )_{,r} \right]_{,r}  +  \frac{e^{ \nu} } { r} \left( \frac{  \widetilde{\cal D}_{,\theta}    } {  \sin \theta }  \right )_{,\theta} \sin \theta
-r e^{-\nu} \widetilde{\cal D}_{,t,t} = Q ( q_{0,2} - q_{2,0} )_{,t} \sin^2 \theta ,                         \label{daximaxi3}
\ee
where
\be
{\cal F}
\equiv F_{01} \sin \theta , ~~
\widetilde{\cal D}
\equiv D_{01} \sin \theta = \frac{{\cal F}}{\gamma},
\ee
with
\be
\gamma \equiv \sqrt {1 -\frac{F_{02}^2 }{\beta^2} } = \frac{1 } { \sqrt{ 1 + \frac{Q^2}{\beta^2 r^4} }  }~ .
\ee

With the substitution
\be
{\widetilde Q} (r, \theta ,t) \equiv r^2 e^{2\nu} Q_{23} \sin^3 \theta = \Delta  ( q_{2,3} - q_{3,2} ) \sin^3 \theta ,
\ee
(\ref{daughterofdr12=dt12}) and (\ref{daughterofdr13=dt13}) take the forms
\bear
\frac {1} {r^4 \sin^3 \theta}  \frac{ \partial {\tilde Q} } {\partial \theta }  &=& -( q_{0,2} - q_{2,0} )_{,t} + \frac { 4Q }{r^3 \sin^2 \theta}~ e^\nu \gamma \widetilde{\cal D}  ,   \label{gdaughterofdr12=dt12}  \\
\frac {\Delta} {r^4 \sin^3 \theta}  \frac{ \partial {\tilde Q} } {\partial r }  &=& ( q_{0,3} - q_{3,0} )_{,t} ~,    \label{gdaughterofdr13=dt13}
\eear
where $\Delta \equiv r^2 e^{2\nu}$.
Assuming that the perturbed fields $q_0 ,  q_2 , q_3 $, and ${\widetilde Q}$ have a time dependence $e^{-i \omega t}$
and eliminating $q_0$ from
(\ref{gdaughterofdr12=dt12}) and (\ref{gdaughterofdr13=dt13}), we obtain
\be
r^4 \left (  \frac {\Delta} {r^4}  \frac{ \partial {\tilde Q} } {\partial r } \right )_{,r} + \sin^3 \theta \left (  \frac {1} {\sin^3 \theta}  \frac{ \partial {\tilde Q} } {\partial \theta } \right )_{,\theta}
+ \omega^2 \frac{r^4}{\Delta} \tilde Q
 =4 Q e^\nu r  \left ( \frac{\gamma \widetilde{\cal D}   } { \sin^2 \theta } \right )_{,\theta} \sin^3 \theta  ~. \label{combined1}
\ee
Similarly, eliminating $(q_{0,2} - q_{2,0} )_{,t}$ from
(\ref{daximaxi3}) using
(\ref{gdaughterofdr12=dt12}), we obtain
\be
\left [ \frac{e^{2 \nu}}{\gamma} ( r e^\nu \gamma \widetilde{\cal D}   )_{,r} \right ]_{,r} + \frac{e^\nu}{r} \left ( \frac{\widetilde{\cal D} _{,\theta}}{\sin \theta} \right)_{,\theta} \sin \theta
+ \left  (\omega^2 r e^{-\nu}  - \frac{  4 Q^2 e^\nu \gamma}{ r^3} \right ) \widetilde{\cal D}
= - Q \frac{ {\widetilde Q}_{,\theta} } { r^4 \sin \theta } ~.       \label{combined2}
\ee

We can further separate the variables $r$ and $\theta$ in
(\ref{combined1}) and (\ref{combined2}) with
\be
{\tilde Q} (r ,\theta) =
{\cal Q}(r) C^{-3/2}_{l +2} ( \theta ) ,   \label{sepvartildeQ}
\ee
and
\be
\widetilde{\cal D}  (r ,\theta) = \frac {{\cal D}(r)} {\sin \theta} \frac{ d C^{-3/2}_{l +2} }{d\theta } = 3 {\cal D} (r) C^{-1/2}_{l +1} (\theta ),   \label{sepvartildeB}
\ee
where $C_{n}^m$ is the Gegenbauer function and we have used the recurrence relation,
\be
\frac{1}{\sin \theta} \frac{ dC_n^m} {d \theta} = - 2 m C_{n-1}^{m +1 } ~.
\ee
Substituting
(\ref{sepvartildeQ}) and (\ref{sepvartildeB}) in
(\ref{combined1}) and (\ref{combined2}), we obtain
two radial wave equations,
\be
\Delta \frac{d}{dr} \left ( \frac {\Delta}{r^4} \frac{d{\cal Q} }{dr} \right ) - \mu^2 \frac{ \Delta}{r^4} {\cal Q} + \omega^2 {\cal Q}
= - \frac{4 Q}{r^3} \mu^2 \Delta e^\nu \gamma {\cal D} , \label{radialQ}
\ee
\be
\left [ \frac{e^{2 \nu}}{\gamma} ( r e^\nu \gamma {\cal D}  )_{,r} \right ]_{,r} -( \mu^2 +2 ) \frac{e^\nu}{r} {\cal D}
+ \left  (\omega^2 r e^{-\nu}  - \frac{  4 Q^2 e^\nu \gamma}{ r^3} \right ) {\cal D}
= - Q \frac{{\cal Q} } { r^4 } ,       \label{radialB}
\ee
where
\be
\mu^2 \equiv
(l-1)(l+2)
\ee
and $l=0,1,2, \cdots$ denotes the orbital number.

Introducing the tortoise coordinate $r_*$,
\be
 \frac{d}{dr_*}=e^{2\nu} \frac{d}{dr}
\ee
and further defining ${\cal Q}$ and ${\cal D}$ as
\be
{\cal Q} \equiv r H_2 ,~~~ re^\nu \gamma {\cal D} \equiv - \frac { H_1 e^{- \frac{\varphi} {2 } } }{2 \mu}
\ee
with
\be
e^\varphi \equiv \frac{1} {\gamma} = \sqrt{ 1 + \frac{Q^2}{ \beta^2 r^4} } ,
\label{phi}
\ee
we find a pair of coupled equations of $H_1$ and $H_2$ \cite{Fernando:2004pc} as
\bear
\widetilde{\bf V} H_2 && = \frac{\Delta}{r^5} \left \{ \left[ \mu^2 r -( e^{2\nu} r )^\prime r + 3 e^{2 \nu} r \right] H_2
+ 2 \mu Q e^{- \frac{\varphi}{2} } H_1 \right \}     ,          \label{H2eqn}    \\
\widetilde{\bf V} H_1 && = \frac{\Delta}{r^5} \left \{ \left[ (\mu^2 +2) r + \frac{ 4 Q^2 e^{-\varphi} }{r}
+ r^3 \left ( \frac {(\varphi^\prime e^{2\nu} )^\prime }{2} + \frac {(\varphi^\prime)^2 e^{2\nu} }{4} \right ) \right ] H_1 + 2  \mu Q e^{-\frac{\varphi}{2} } H_2 \right \} ,
\label{H1eqn}
\eear
where
\be
\widetilde{\bf V} \equiv \frac{d^2}{ d r_*^2} + \omega^2~.
\ee

To find the one-dimensional Schr\"{o}dinger-type wave equations, we rewrite the coupled equations (\ref{H2eqn}) and (\ref{H1eqn}) in the matrix form,
\be
\widetilde{\bf V}\pmatrix { H_1 \cr  H_2  } =
\frac {\Delta}{r^5} \pmatrix {V_{11} & V_{12} \cr V_{21} & V_{22}} \pmatrix { H_1\cr  H_2  }  ,
\ee
where
\bear
V_{11} && \equiv V_1 = (\mu^2 +2) r + \frac{ 4 Q^2 e^{-\varphi} }{r}
+ r^3 \left ( \frac {(\varphi^\prime e^{2\nu} )^\prime }{2} + \frac {(\varphi^\prime)^2 e^{2\nu} }{4} \right ) ,     \\
V_{12} &&= V_{21} = 2 \mu Q e^{-\frac{\varphi}{2} },   \\
V_{22}  && \equiv V_2 = \mu^2 r -( e^{2\nu} r )^\prime r + 3 e^{2 \nu} r .
\eear
The matrix $V_{ij}$ can be diagonalized by the similarity transformation as in the RN case
\cite{Chandrasekhar}.
Then, the two decoupled, one-dimensional Schr\"{o}dinger-type equations are given by,
\be
\widetilde{\bf V} Z_i = U_i Z_i, ~~~(i=1,2),
\label{Regge-Wheeler}
\ee
where the
effective potentials, $U_1$ and $U_2$, which are {\it real-valued} functions, are given by
\bear
U_1 &&= \frac{\Delta}{r^5} \left[ \frac{ V_1 + V_2}{2} + \sqrt { \left ( \frac{V_1 - V_2}{2}  \right)^2 +V_{12}^2 }  \right ] ,
\label{V1}\\
U_2 &&= \frac{\Delta}{r^5} \left[ \frac{ V_1 + V_2}{2} - \sqrt { \left ( \frac{V_1 - V_2}{2}  \right)^2 + V_{12}^2 }  \right ].
\label{V2}
\eear
Here, the decoupled solutions $Z_1$ and $Z_2$ are related with the
coupled solutions $H_1$ and $H_2$, which are basically the perturbations
of the metric variable $Q_{23}$ and the fields strength $D_{01}$
\cite{Chandrasekhar},
\bear
\pmatrix { Z_1 \cr  Z_2  }
=\frac{1}{\sqrt{q_1 (q_1-q_2)}}\pmatrix {
 q_1 H_1+\sqrt{-q_1 q_2} H_2  \cr
-\sqrt{-q_1 q_2} H_1 +q_1 H_2 } ,
\eear
where
\bear
q_1 &&= \frac{ V_1 - V_2}{2} + \sqrt { \left ( \frac{V_1 - V_2}{2}  \right)^2 + V_{12}^2 }  ,  \\
q_2 &&= \frac{ V_1 - V_2}{2} - \sqrt { \left ( \frac{V_1 - V_2}{2}  \right)^2 + V_{12}^2 } .
\eear
The explicit form of $U_1 , U_2 ,q_1,$ and $q_2$ will be available upon substituting $e^{2 \nu}$ and $e^{\varphi}$ with the background solutions (\ref{backgroundpolars}) and (\ref{phi}), respectively, when we compute the QNMs in the next section.

\section{Quasi-Normal Modes for the Axial Perturbations: The WKB approach }

\subsection{The WKB approximation}

{Quasi-Normal Modes} (QNMs) are defined as the solutions which are purely
ingoing {as} $e^{-i (\omega t +k r_*)}$ near the (outer) black hole horizon $r_+$,
where $r_* = -\infty$. In this paper, in order to compute QNMs,
we will consider the WKB approximation
method \cite{Schutz:1985a,iyerwill,iyer}, which is simple to apply in our case
though there are several inherent limitations, as will be explained below. The
applicability depends also on several conditions, like the (asymptotic) boundary
conditions, and our dS space-time satisfies this condition which is a technical
reason of using the WKB approach in this paper. In this section, we briefly
summarize the basic results of the WKB approach.


The master equation for the WKB approach may be written as the one-dimensional Sch\"{o}dinger-type equation \footnote{In contrast to the usual Sch\"{o}dinger equation in quantum mechanics, the potential ${\bf Q}$ can be complex-valued generally.},
\beq
\f{d^2}{dx^2} \Phi (x)+{\bf Q}(x) \Phi(x)=0,
\label{Mater}
\eeq
where the {\it potential} function $-{\bf Q}(x)$
is assumed to be constant at the asymptotic boundaries $x=\pm \infty$,
although not necessary the same at both ends, and has {\it one} maximum
at a certain finite $x_0$ (Fig. 5). In the black hole case, $\Phi$
represents the radial part of the {perturbation}
with the usual time dependence of a positive-frequency mode
$e^{-i \omega t}$, as well as the appropriate angular dependence.
The coordinate $x$ is 
related to the {\it tortoise} coordinate
$r_*$ which ranges from $-\infty$, at the (outer) black hole horizon $r_+$,
to $+\infty$, at the spatial infinity $r=\infty$ for the asymptotically
flat \cite{iyerwill} or at the cosmological horizon $r_{++}$ for the
asymptotically dS case \cite{Mell:1989}{, as}
adopted in this paper.

\begin{figure}
\includegraphics[width=9cm,keepaspectratio]{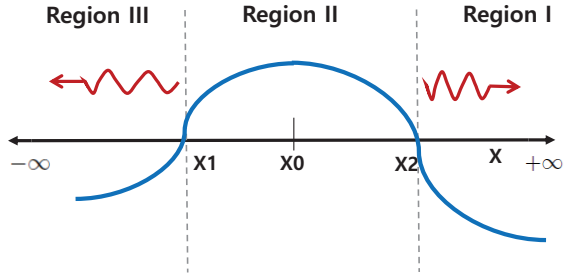}
\caption{A typical plot of the {\it real} part of the potential function
$-{\bf Q} (x)\equiv U-\omega^2$ with the effective potential $U$, which is independent of the frequency $\om \equiv \om_R-i \om_I$, and
the ({\it complex-valued}) ``energy"
$E \equiv \om^2=(\om_R^2-\om_I^2)-i (2 \om_I \om_R)$. Here, one (local)
maximum at $x_0$ (
$-{\bf Q}_0''\equiv U_0''<0$ ) and two turning points at $x_1$ and $x_2$ are assumed.
 }
\label{fig:Q}
\end{figure}

In order to solve the one-dimensional potential barrier problem of (\ref{Mater}), we adopt a {\it modified} WKB approach by matching the two exterior WKB solutions in regions I and III across the two turning points $x_1$ and $x_2$ simultaneously \cite{Bend}. In the interior region II, we expand the potential function $-{\bf Q}(x)$ near its maximum at $x=x_0$ and use it to connect the two exterior WKB solutions. For the black hole case with the quasi-normal mode boundary conditions, {\it i.e.}, purely ``outgoing" {(from the potential barrier)}, the result
may be written as the simple semi-analytic formula,
\beq
\frac{i {\bf Q}_0}{\sqrt{2 {\bf Q}_0''}}-i \sum^N_{k=1} \widetilde{\Lambda}_{2k}-\sum^{\widetilde{N}}_{k=1}\Omega_{2k+1}=n+\f{1}{2},
\label{WKB_Iyer}
\eeq
where the primes denote the derivatives in terms of $x$ and ${\bf Q}_0''>0$
due to the extremum of the potential $-{\bf Q}$
at $x_0$.
 Here, the subscripts $2k, 2k+1$ denote the order of the WKB approximations, $\widetilde{N}=N$ or $N \pm 1$, and $\widetilde{\Lambda}_{2k}$, $\Omega_{2k+1}$ are polynomials of the derivatives ${\bf Q}_0^{(m)}/{\bf Q}_0''$ and $\sqrt{{\bf Q}_0''}$. It is important to note that even and
 odd-order terms become pure-imaginary and real-valued, respectively, so that,
 when considering ${\bf Q}_0=\om^2- U_0$ in the black hole case, they play the
 distinct roles in obtaining the quasi-normal frequencies,
 $\om=\om_R-i \om_I$ \footnote{This fact does not seem to be well emphasized
 in the literature.
 This causes some misleading results, for example, in \cite{Fernando:2004pc} due to the related typos in the original work of the third-order terms in \cite{iyerwill} where $``i"$ factor in the even part of (\ref{WKB_Iyer}) is missing. Similar typos also appear in the 4th-order terms in \cite{Konoplya:2003ii} (Appendix A), in contrast to the correct $i$ factors in the corresponding 4th-order terms in \cite{iyerwill} (Appendix) and \cite{Matyjasek:2017psv} (private communications: We thank the authors for kindly sharing their results).}.

Then, the result up to the third WKB orders \cite{iyerwill,iyer} \footnote{The convergence in each order may not be guaranteed generally
since the expansions are just asymptotic ones. However, the popular
third-order result of \cite{iyerwill} can also be obtained as the lower-order
results of the phase integral approach which allows the {\it convergent}
expansions with controlled accuracy \cite{Galt:1991} and works well in many
cases. For extensions to higher orders, see \cite{iyerwill} (Appendix) (4th order),
\cite{Konoplya:2003ii} (6th order), and \cite{Matyjasek:2017psv}
(13th order). See also \cite{Kono:2019} for a recent review.}, with
${\bf Q}\equiv \om^2- U$ from (\ref{Regge-Wheeler}), is given by

\be
\label{WKBQ}
\omega^2 =\left[U_0+\sqrt{-2U_0''}~\widetilde{\Lambda}_2\right]
-i\left(n+\frac{1}{2}\right)\sqrt{-2U_0''}~\left(1+\widetilde{\Omega}_3\right)
\ee
with the {\it real}-valued WKB correction terms,
\beq
\widetilde{\Lambda}_2 &=&\frac{1}{\sqrt{-2U_0''}}
\left[\frac{1}{32}
\left(\frac{U_0^{(4)}}{U_0''}\right)\left(1+4\alpha^2\right)
-\frac{1}{288}\left(\frac{U_0^{(3)}}{U_0''}\right)^2 (7+60\alpha^2)\right],
\label{Lambda2}
\\
\widetilde{\Omega}_3&=&\frac{1}{-2U_0''}
\left[\frac{5}{6912}\left(\frac{U_0^{(3)}}{U_0''}\right)^4 (77+188\alpha^2)
-\frac{1}{384}\left(\frac{{U_0^{(3)}}^{2}U_0^{(4)}}{{U_0''}^{3}}\right)
(51+100\alpha^2)\right. \label{Omega3}\\ \nonumber
&&+\left.\frac{1}{2304}\left(\frac{U_0^{(4)}}{U_0''}\right)^2(67+68\alpha^2)
+\frac{1}{288}\left(\frac{U_0^{(3)}U_0^{(5)}}{{U_0''}^{2}}\right)(19+28\alpha^2)
-\frac{1}{288}\left(\frac{U_0^{(6)}}{U_0''}\right)(5+4\alpha^2)\right],
\eeq
where
$\alpha \equiv n+\frac{1}{2}~(n=0,1,2, \cdots),
$
$\widetilde{\Omega}_3 \equiv {\Omega}_3/(n+1/2)$, and the primes and the
superscript $(m)$ of $U^{(m)}_0$ denote
the derivatives of the effective potential $U(r_{\ast})$
with respect to the tortoise coordinate $r_{\ast}$, evaluated at its
maximum point $r_0$. The maximum point $r_0$ is determined by
$
{d U }/{d r_*}=e^{2 \nu}
{d U }/{d r}=0,
$
which does not have analytic solutions in our case, where the effective potentials $U_i$ in (\ref{V1}) and (\ref{V2}) are quite
complicated so that $r_0$ can be found only numerically. Then, QNMs can be
found numerically by solving (\ref{WKBQ}) from the numerical
computations of (\ref{Lambda2}) and (\ref{Omega3}) as a function of
the parameters of the {theory.}


\begin{figure}[ht]
\begin{center}
{\includegraphics[width=5.0cm,keepaspectratio]{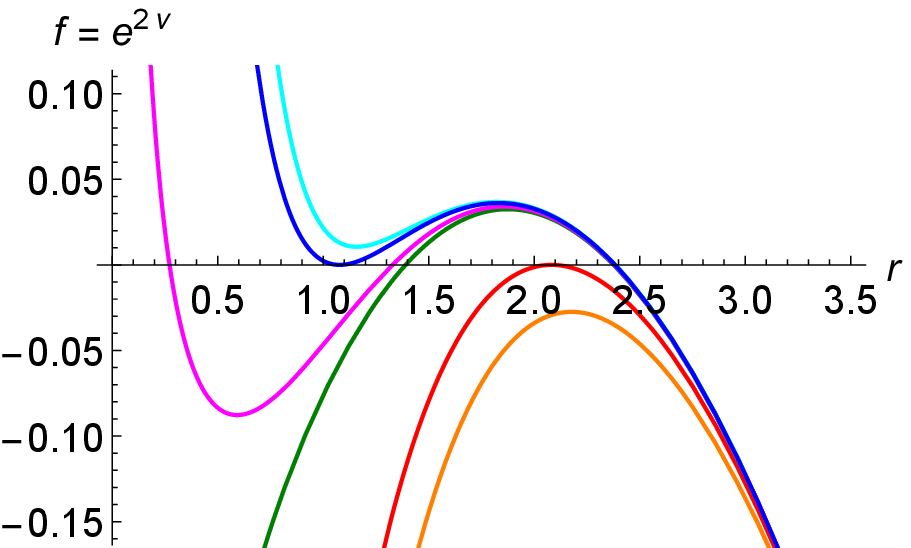}}
{\includegraphics[width=5.0cm,keepaspectratio]{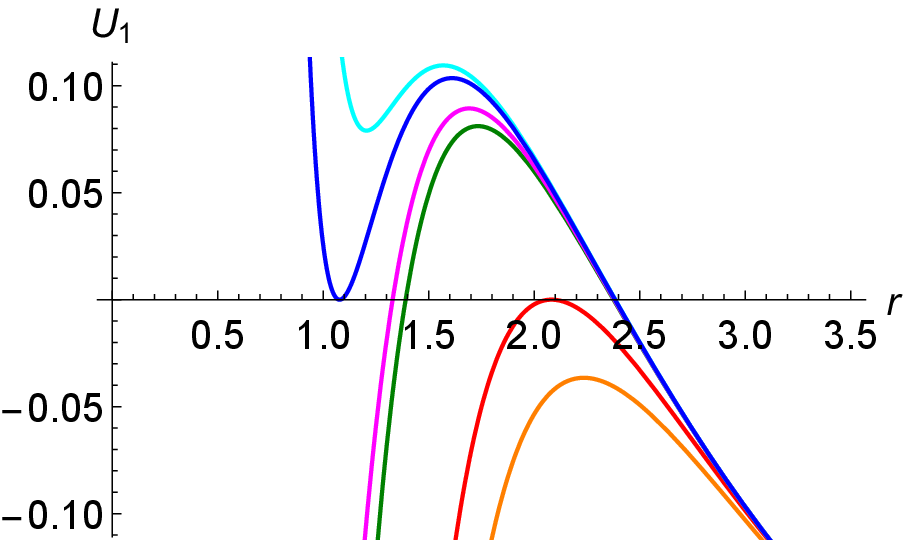}}
{\includegraphics[width=5.0cm,keepaspectratio]{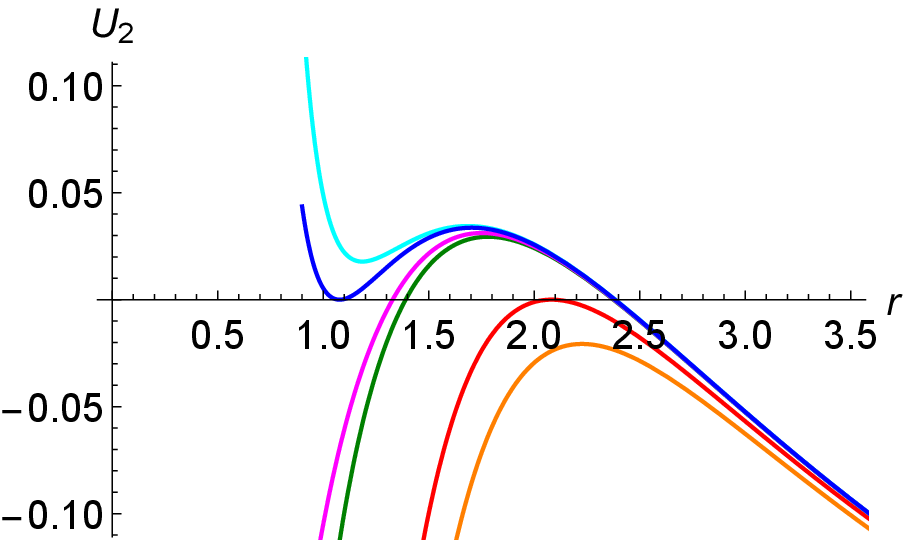}}
\end{center}
\vspace{-0.6cm}
\caption{The plots of $f=e^{2 \nu (r)}$ and the effective potentials $U_1(r)$ and $U_2(r)$ for varying the BI parameter $\beta$. Here, we consider $\beta=0.05,~ \beta_{\rm min},~ 0.5,~ 0.64,~ \beta_{\rm max},~ 2.0$ (bottom to top)
with $\beta_{\rm min} \approx 0.103,~ \beta_{\rm max} \approx 1.166$ when
$M=0.95$, $Q=1$, $l=2$, and $\Lambda =0.2$.
}
\label{figI}
\end{figure}

\begin{figure}[ht]
\begin{center}
{\includegraphics[width=7.5cm,keepaspectratio]{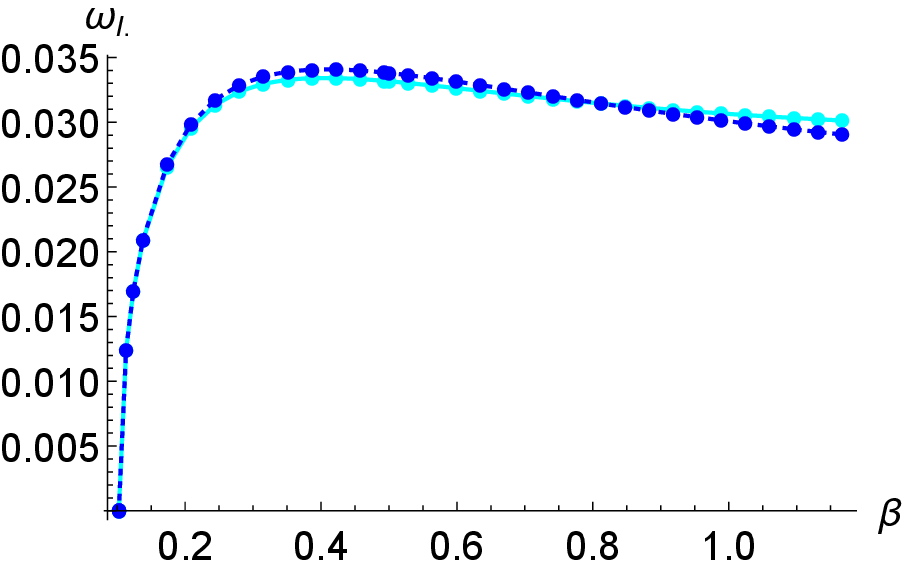}
}
$\;\;\;\;\;${\includegraphics[width=7.5cm,keepaspectratio]{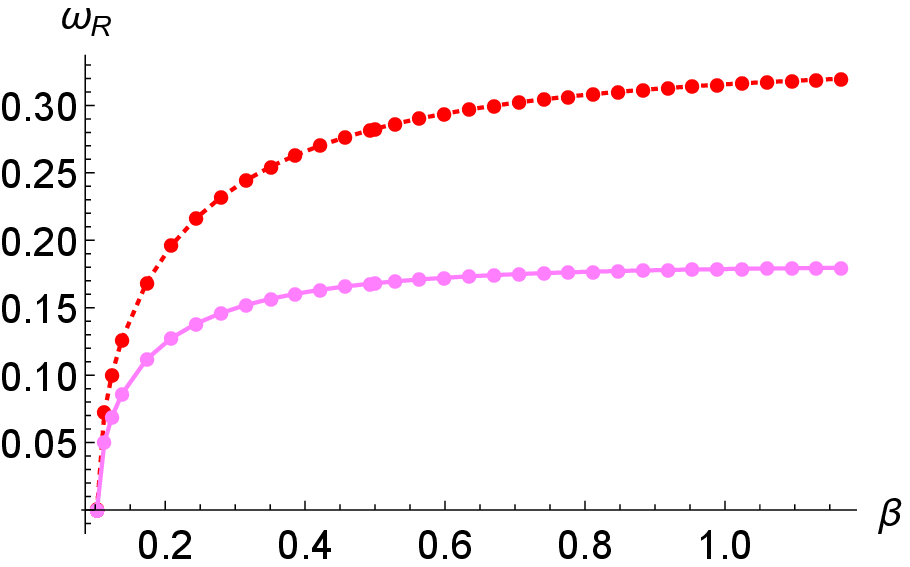}
}
\end{center}
\vspace{-0.6cm}
\caption{{\footnotesize The plots of {\it lowest} ($n=0$) $\omega=\omega_R -i \omega_I$ vs. $\beta$ for $U_1$ (dashed curves) and $U_2$ (solid curves)
when $M=0.95$, $Q=1$, $l=2$, and $\Lambda =0.2$.
}}
\label{figII}
\end{figure}

\subsection{QNMs
vs. $\beta$}

Figs. 6 and 7 show the effective potentials and corresponding {\it lowest} ({\it i.e.,} fundamental) QNMs with $n=0$ as functions of the BI parameter $\beta$. As can be seen in Fig.~\ref{figI} (left), there are one
degenerate {\it event} horizon
for $\beta=\beta_{\rm min}$,
and one degenerate black hole horizon
with one
cosmological horizon for $\beta=\beta_{\rm max}$. In between them,
$\beta_{\rm min}< \beta< \beta_{\rm max}$, there are generally three
horizons (two black hole horizons and one cosmological horizon), depending
on $\beta$. Only for this dS black hole regime,
the WKB formula (\ref{WKBQ}) for QNMs may apply.
Fig. \ref{figII} shows that
the black holes are stable for
the range of $\beta_{\rm min}<\beta\leq\beta_{\rm max}$ since their metric perturbations are decaying (which is called also as the``ring-down" phase) with the decay time $\tau \sim \om_I^{-1}$ for $\om_I>0$. Moreover, the results show
$\om \approx 0$ \footnote{Hereafter, when we denote
$\om \approx 0$, it means zero with the accuracy of $10^{-9} \sim 10^{-8}$.},
{\it i.e.,} the ``frozen" mode,
at $\beta=\beta_{\rm min}\approx 0.103$ and no QNMs
beyond $\beta_{\rm max} \approx 1.166$, where the WKB formula (\ref{WKBQ})
 does not apply, as expected from the behaviors of the metric functions and
 the effective potentials in Fig. 6. As
$\beta$ increases,
the {\it negative} imaginary parts of QNMs,
$\om_I$ increase first monotonically and reach
maximum points around $\beta Q=1/2$ (here, $\beta=0.5$ with $Q=1$), which
divides the Sch-type and the RN-type \cite{Kim:2016pky}, and then decrease
monotonically.
Beyond the above range, 
$\beta <\beta_{\rm min}$ or
$\beta >\beta_{\rm max}$, one can not say anything about its
stability
since the WKB formula for QNMs can not apply due to the
lack of the required boundary conditions and we need other
analysis to test its stability. For example, the WKB formula does not say
anything about the stability
of the {\it naked singularity}
for $\beta>\beta_{\rm max}$.
On the other hand, the real parts of QNMs
$\om_R$ increase monotonically and saturate to maximum values, corresponding to
those of the RN-dS case \ci{Kokk:1988,Mell:1989},
showing two decoupled oscillation modes with a rough relation,
$\om_{R(1)} \sim 2 \om_{R(2)}$. This is in contrast to the imaginary parts
with a rough relation, $\om_{I(1)}\sim \om_{I(2)}$, so that the real-to-imaginary frequency ratios are
$\om_{R(1)}/\om_{I(1)} \sim 2 \om_{R(2)}/\om_{I(2)} \sim 10$.


\begin{figure}[!htbp]
\begin{center}
{\includegraphics[width=5cm,keepaspectratio]{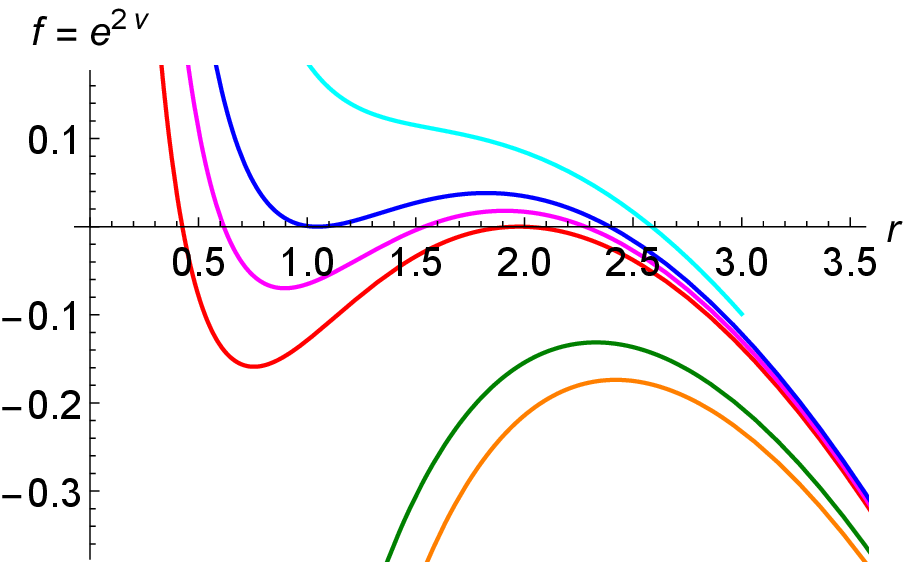}}
{\includegraphics[width=5cm,keepaspectratio]{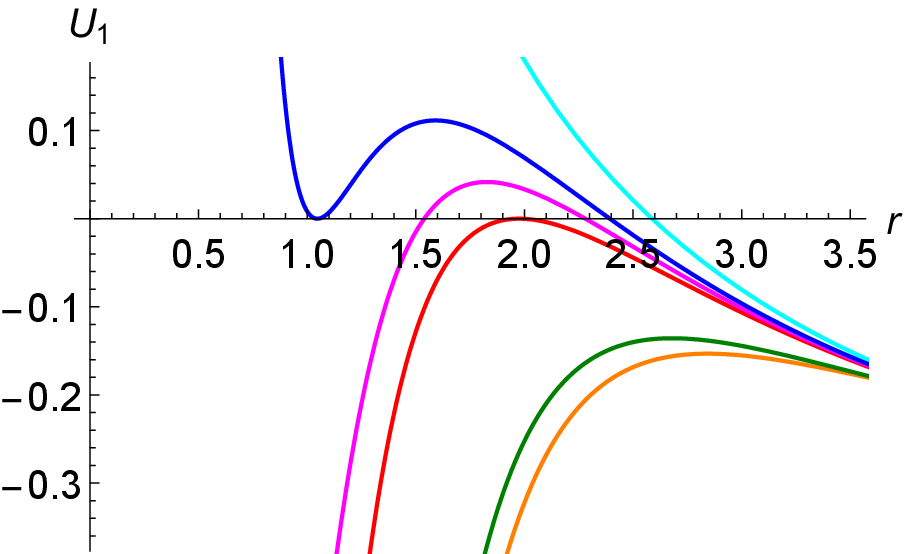}}
{\includegraphics[width=5cm,,keepaspectratio]{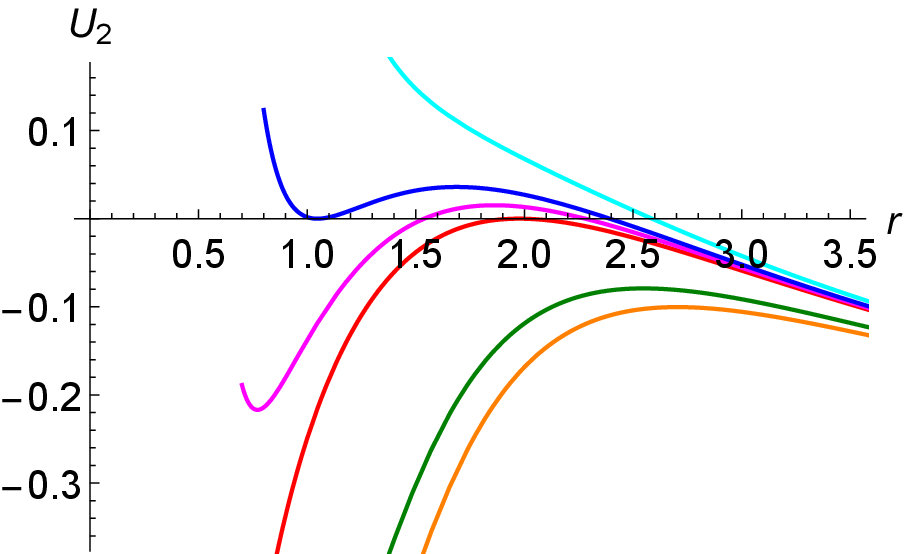}}
\end{center}
\vspace{-0.6cm}
\caption{{\footnotesize The plots of $f=e^{2 \nu (r)}$ and the effective potentials $U_1(r)$ and $U_2(r)$ for varying the electric charge $Q$. Here, we consider $Q=0.05,~ 0.5,~ Q_{\rm min},~ 0.9680,~ Q_{\rm max},~ 1.1$ (bottom to top) with $Q_{\rm min} \approx 0.9320,~ Q_{\rm max} \approx 1.0041$ when
$M=0.95$, $\beta=1$, $l=2$, and $\Lambda =0.2$.
}}
\label{figV}
\end{figure}

\begin{figure}[!htbp]
\begin{center}
{\includegraphics[width=7.5cm,keepaspectratio]{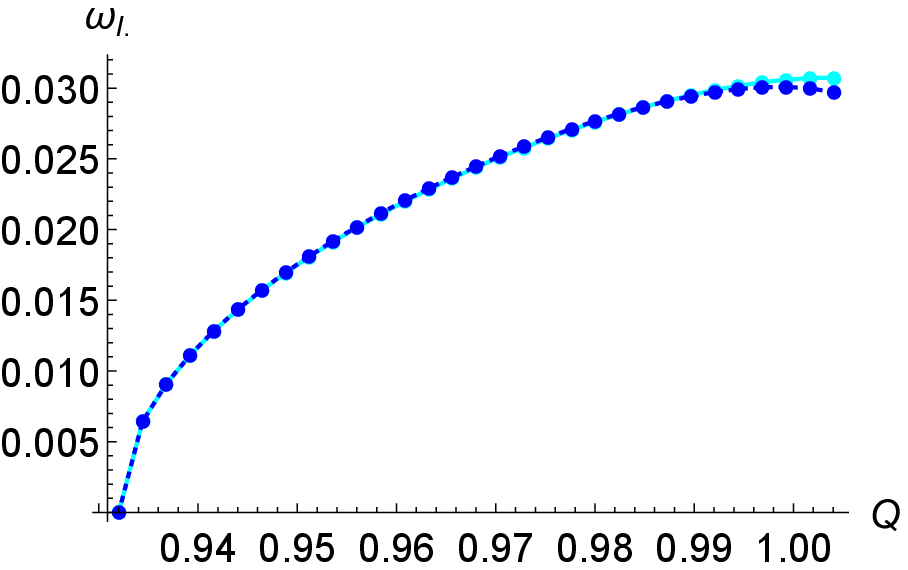}}
$\;\;\;\;\;${\includegraphics[width=7.5cm,keepaspectratio]{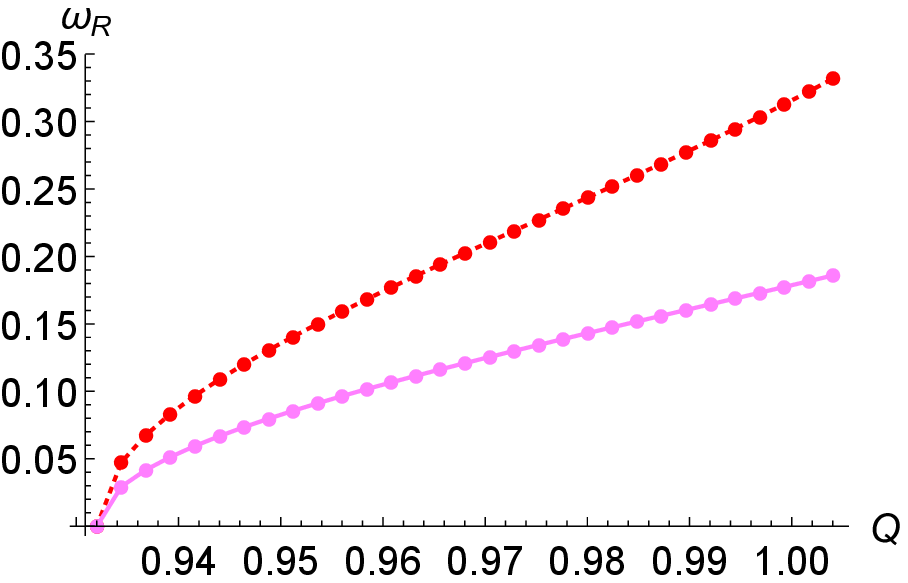}}
\end{center}
\vspace{-0.6cm}
\caption{{\footnotesize The plots of {\it lowest} ($n=0$) $\omega=\omega_R -i \omega_I$ vs. $Q$ for $U_1$ (dashed curves) and $U_2$ (solid curves)
when $M=0.95$, $\beta=1$, $l=2$, and $\Lambda =0.2$. The results show
$\om \approx 0$ at $Q=Q_{\rm min}\approx 0.9320$ and no QNMs
beyond $Q_{\rm max} \approx 1.0041$.
}}
\label{figVI}
\end{figure}

\subsection{QNMs
vs. $Q$}

Figs. 8 and 9 show the effective potentials and corresponding lowest QNMs as functions of the electric charge $Q$. As can be seen in Fig.~\ref{figV}, there are one
degenerate event horizon ({\it charged Nariai} solution) with one Cauchy horizon for $Q=Q_{\rm min}$,
and one degenerate black hole horizon with one
cosmological horizon for $Q=Q_{\rm max}$. In between them,
$Q_{\rm min}< Q< Q_{\rm max}$, there are generally three (two black hole and one cosmological) horizons, depending on $Q$. The results show
$\om \approx 0$
at $Q=Q_{\rm min}\approx 0.9320$ and no QNMs beyond $Q_{\rm max} \approx 1.0041$,
as expected from the behaviors of the metric functions and the effective
potentials in Fig. 8. The behaviors of $\om_I$ and $\om_R$
are almost the same as those of Fig. 7 for $\beta < 0.5$
with the similar relations, $\om_{R(1)} \sim 2 \om_{R(2)}$,
$\om_{I(1)}\sim \om_{I(2)}$ and
$\om_{R(1)}/\om_{I(1)} \sim 2 \om_{R(2)}/\om_{I(2)} \sim 10$, and
other discussions are similar.

\begin{figure}[!htbp]
\begin{center}
{\includegraphics[width=5cm,keepaspectratio]{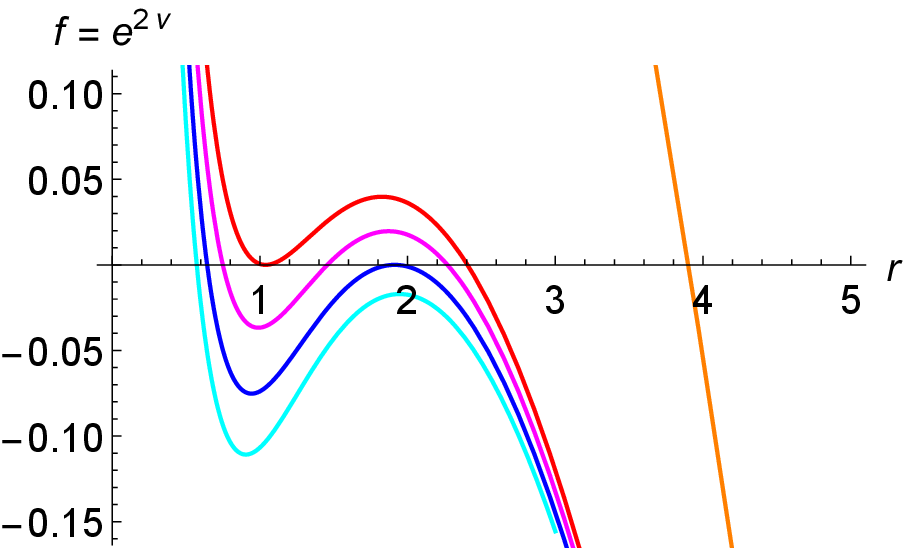}}
{\includegraphics[width=5cm,keepaspectratio]{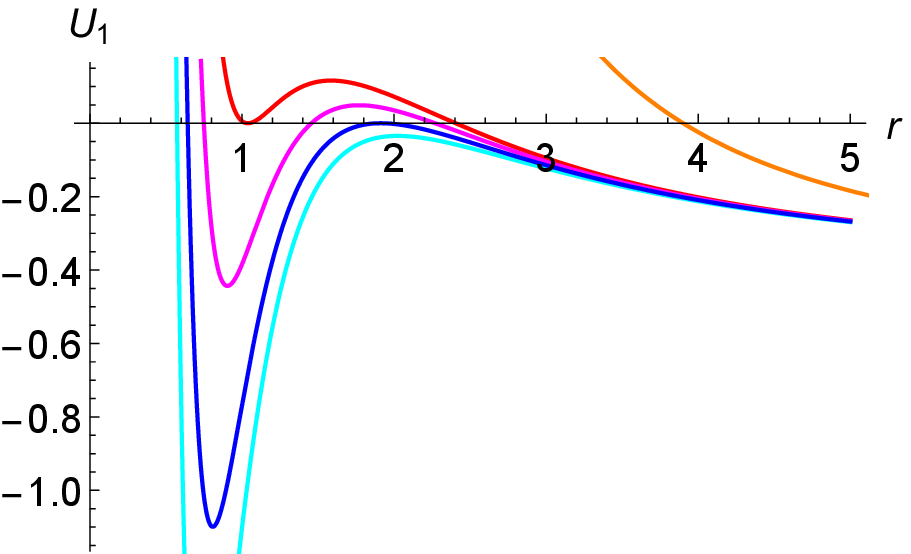}}
{\includegraphics[width=5cm,,keepaspectratio]{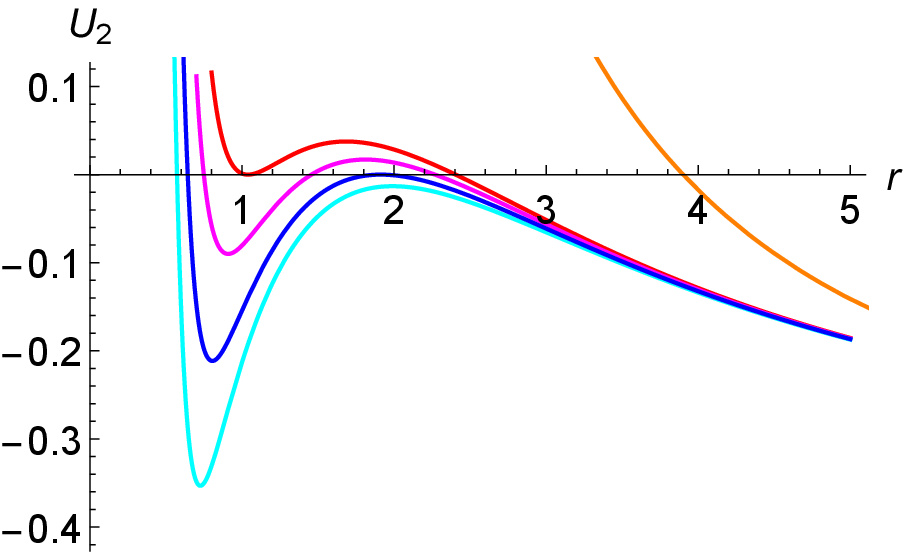}}
\end{center}
\vspace{-0.6cm}
\caption{{\footnotesize The plots of $f=e^{2 \nu (r)}$ and the effective potentials $U_1(r)$ and $U_2(r)$ for varying the mass $M$ with a fixed $\beta Q>1/2$. Here, we consider $M=0.1,~ M_{\rm min},~ 0.9650,~ M_{\rm max},~ 1.0$ (top to bottom) with $M_{\rm min} \approx 0.9464, ~M_{\rm max} \approx 0.9836$ when
$\beta=1$, $Q=1$, $l=2$, and $\Lambda =0.2$.
}}
\label{figIII}
\end{figure}

\begin{figure}[!htbp]
\begin{center}
{\includegraphics[width=7.5cm,keepaspectratio]{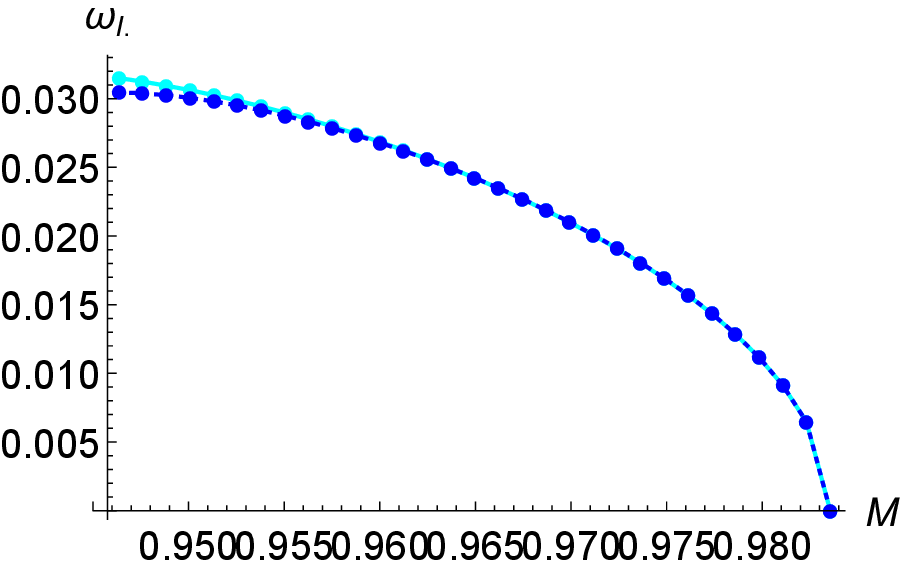}}
$\;\;\;\;\;${\includegraphics[width=7.5cm,keepaspectratio]{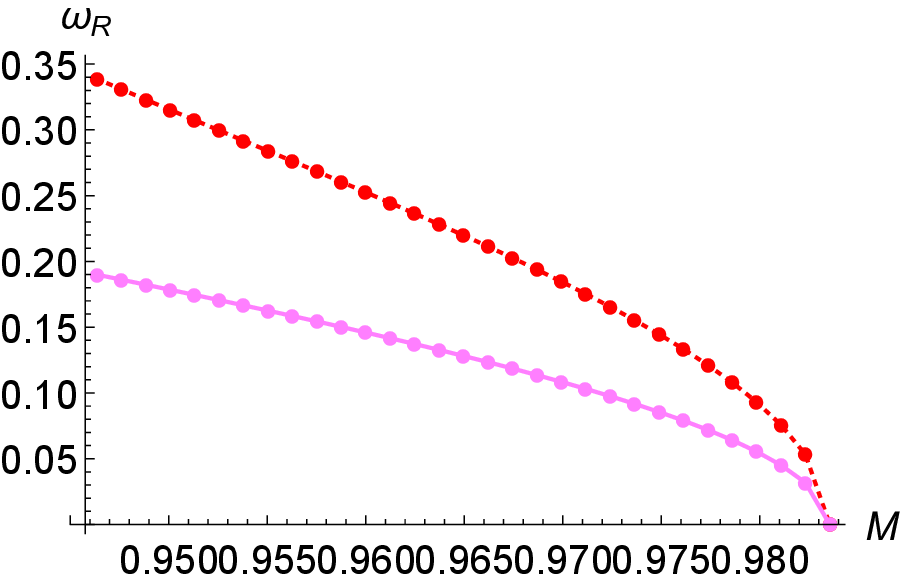}}
\end{center}
\vspace{-0.6cm}
\caption{{\footnotesize The plots of {\it lowest} ($n=0$) $\omega=\omega_R -i \omega_I$ vs. $M$ for $U_1$ (dashed curves) and $U_2$ (solid curves)
when $\beta=1$, $Q=1$, $l=2$, and $\Lambda =0.2$.
}}
\label{figIV}
\end{figure}

\begin{figure}[!htbp]
\begin{center}
{\includegraphics[width=5cm,keepaspectratio]{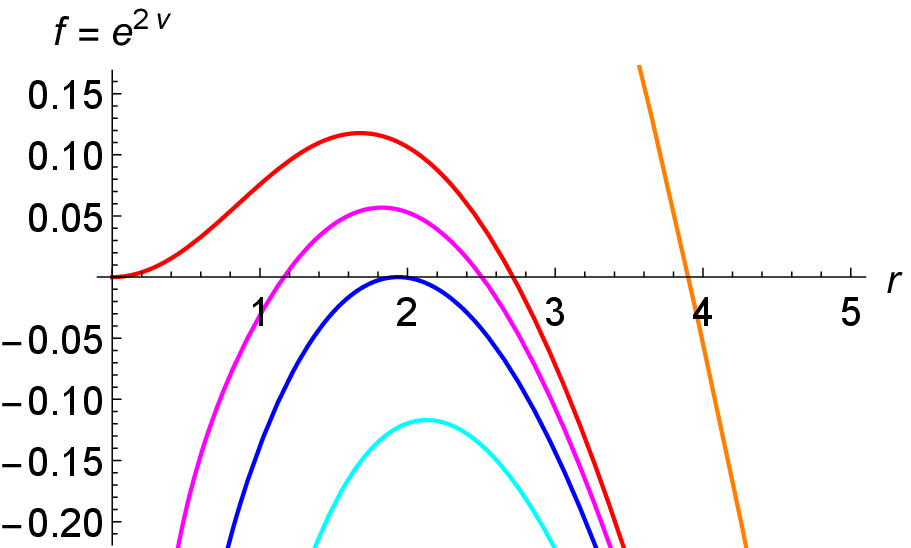}}
{\includegraphics[width=5cm,keepaspectratio]{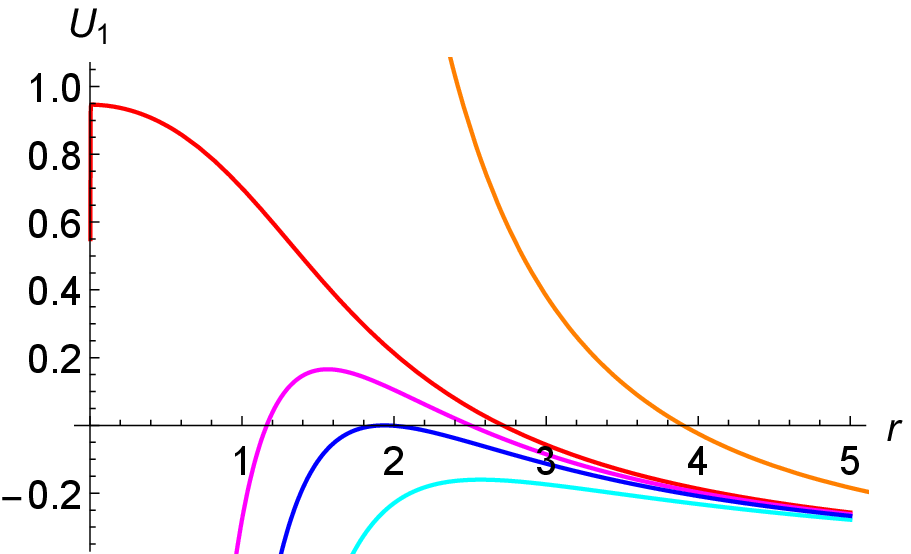}}
{\includegraphics[width=5cm,,keepaspectratio]{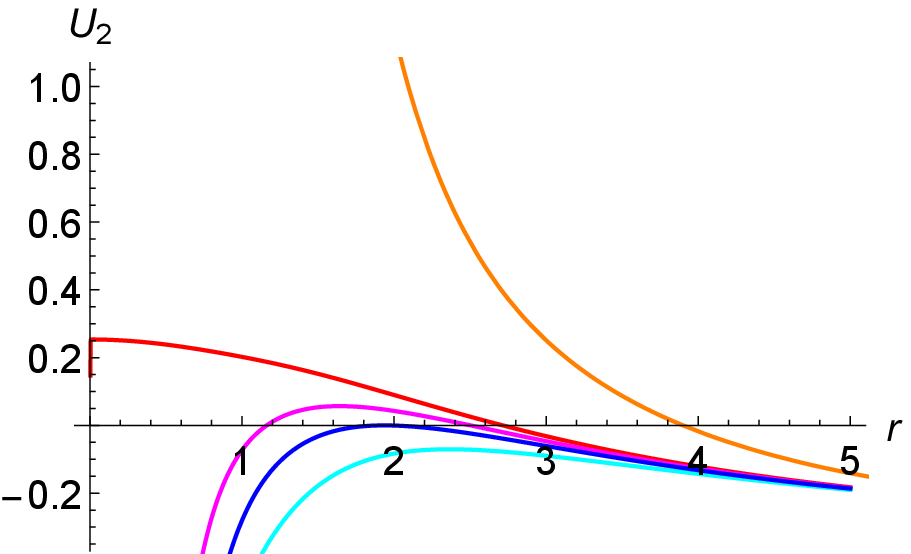}}
\end{center}
\vspace{-0.6cm}
\caption{{\footnotesize The plots of $f=e^{2 \nu (r)}$ and the
effective potentials $U_1(r)$ and $U_2(r)$ for varying the mass $M$ with a
fixed $\beta Q=1/2$. Here, we consider
$M=0.1,~ M_{\rm min},~ 0.9275,~ M_{\rm max},~1.0$ (top to bottom) with
$M_{\rm min}=M_0 \approx 0.8740,~ M_{\rm max} \approx 0.9810$ when
$\beta=1$, $Q=1$, $l=2$, and $\Lambda =0.2$.
}}
\label{figIII}
\end{figure}

\begin{figure}[!htbp]
\begin{center}
{\includegraphics[width=7.5cm,keepaspectratio]{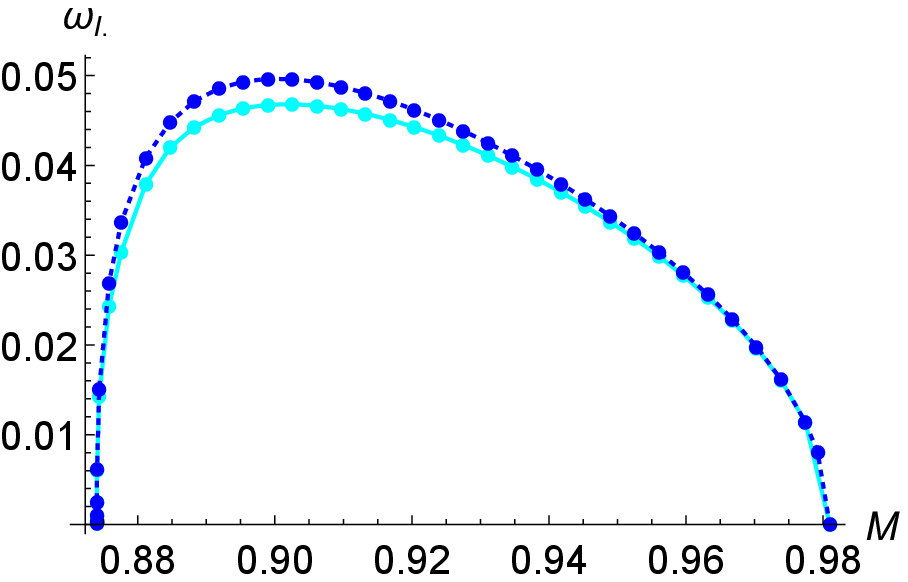}}
$\;\;\;\;\;${\includegraphics[width=7.5cm,keepaspectratio]{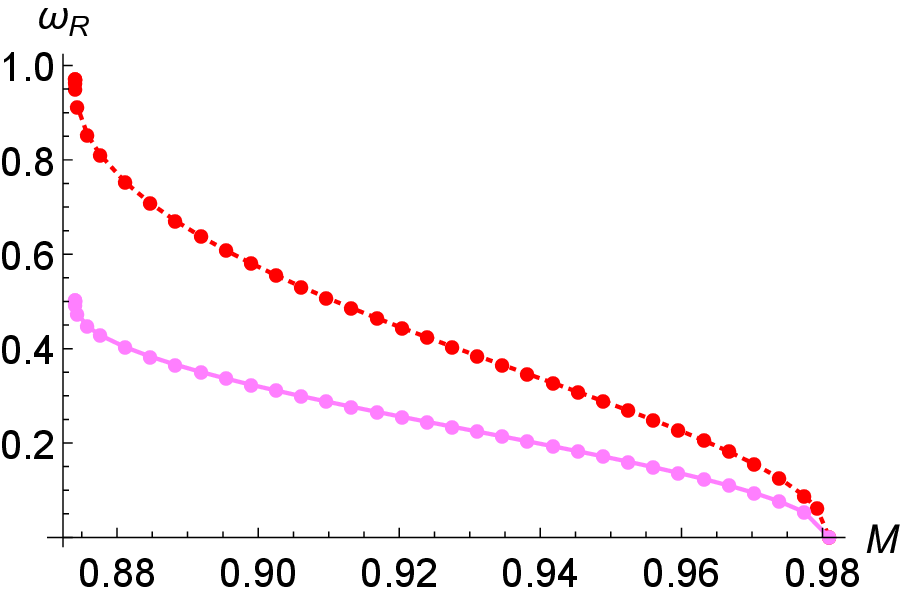}}
\end{center}
\vspace{-0.6cm}
\caption{{\footnotesize The plots of {\it lowest} ($n=0$) $\omega=\omega_R -i \omega_I$ vs. $M$ for $U_1$ (dashed curves) and $U_2$ (solid curves)
when $\beta=1/2$, $Q=1$, $l=2$, and $\Lambda =0.2$.
}}
\label{figIV}
\end{figure}

\begin{figure}[!htbp]
\begin{center}
{\includegraphics[width=5cm,keepaspectratio]{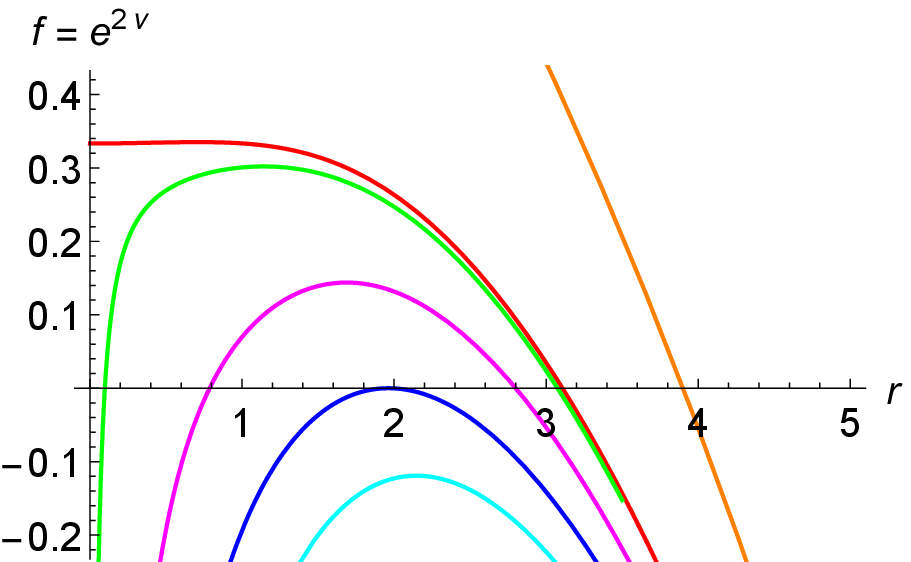}}
{\includegraphics[width=5cm,keepaspectratio]{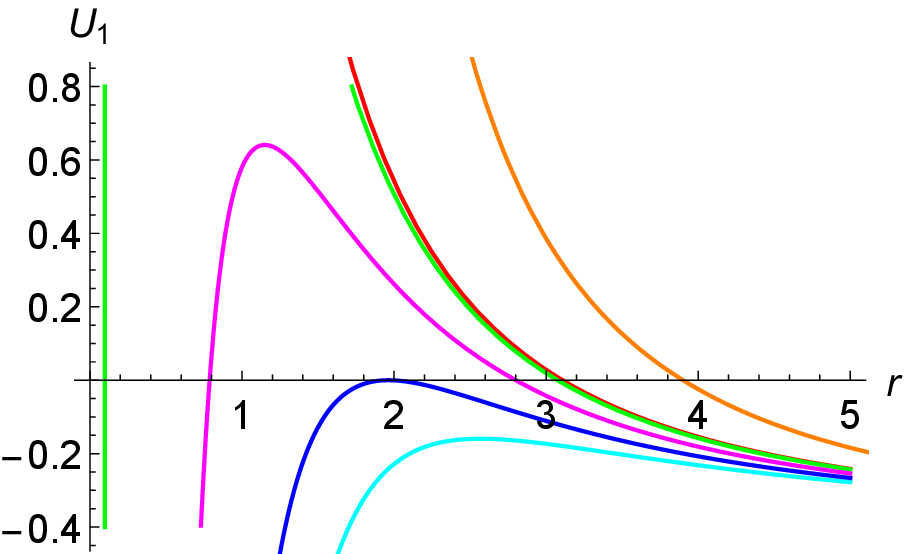}}
{\includegraphics[width=5cm,,keepaspectratio]{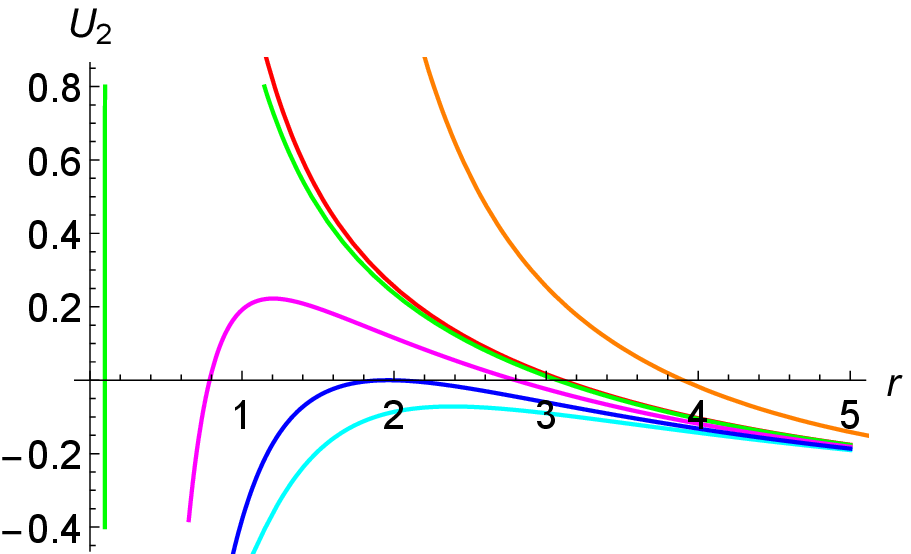}}
\end{center}
\vspace{-0.6cm}
\caption{{\footnotesize The plots of $f=e^{2 \nu (r)}$ and the
effective potentials $U_1(r)$ and $U_2(r)$ for varying the mass $M$ with a
fixed $\beta Q<1/2$. Here, we consider
$M=0.1,~ M_{\rm min},~ 0.73,~ 0.8455, ~M_{\rm max},1.0$ (top to bottom) with
$M_{\rm min} \approx 0.7136,~ M_{\rm max} \approx 0.9774$ when
$\beta=1/3$, $Q=1$, $l=2$, and $\Lambda =0.2$.
}}
\label{figIII}
\end{figure}

\begin{figure}[!htbp]
\begin{center}
{\includegraphics[width=7.5cm,keepaspectratio]{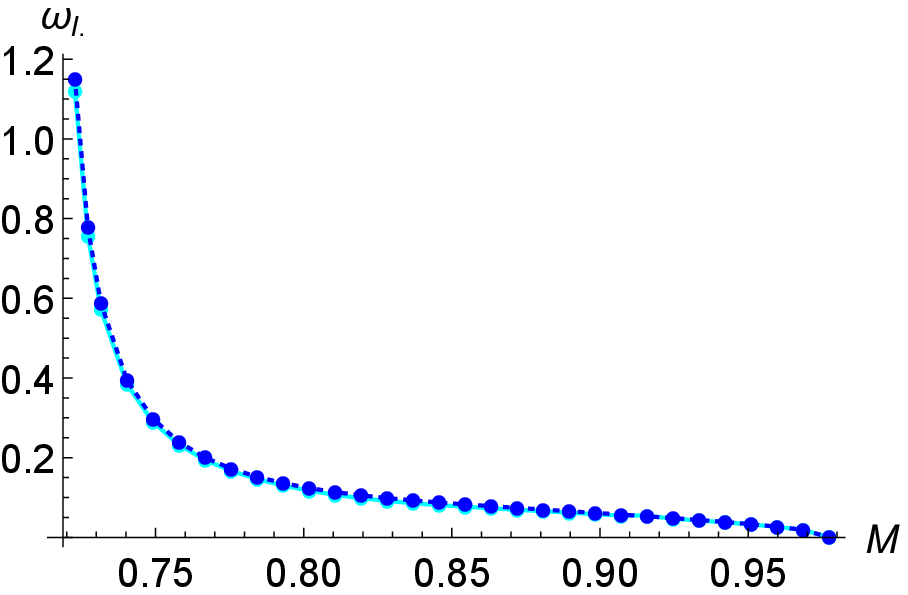}}
$\;\;\;\;\;${\includegraphics[width=7.5cm,keepaspectratio]{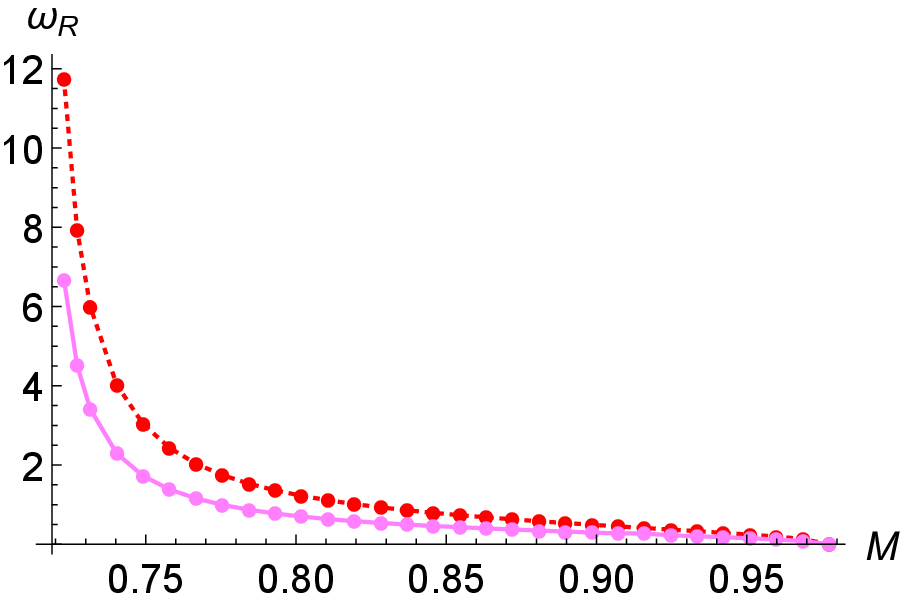}}
\end{center}
\vspace{-0.6cm}
\caption{{\footnotesize The plots of {\it lowest} ($n=0$) $\omega=\omega_R -i \omega_I$ vs. $M$ for $U_1$ (dashed curves) and $U_2$ (solid curves)
when $\beta=1/3$, $Q=1$, $l=2$, and $\Lambda =0.2$.
}}
\label{figIV}
\end{figure}

\begin{figure}[!htbp]
\begin{center}
{\includegraphics[width=7.5cm,keepaspectratio]{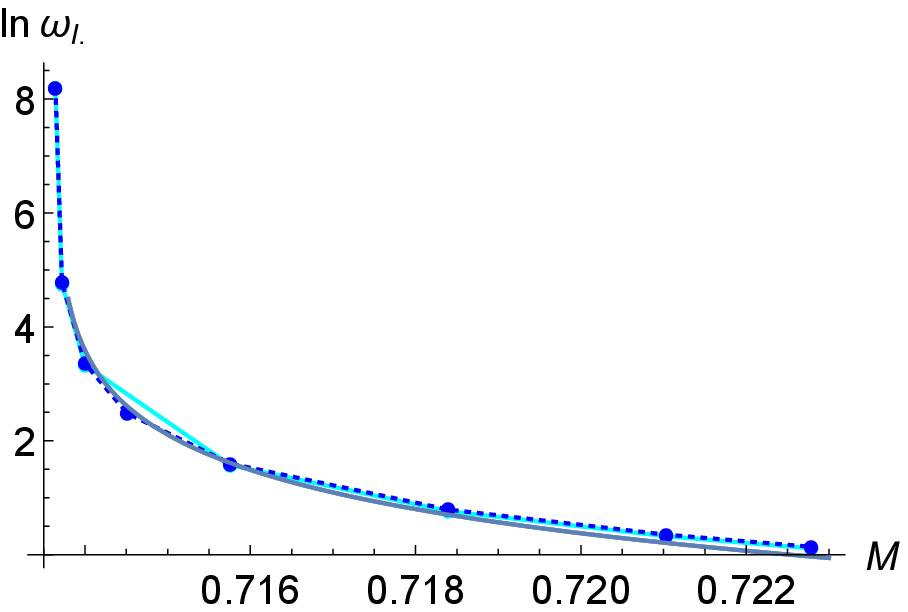}}
$\;\;\;\;\;${\includegraphics[width=7.5cm,keepaspectratio]{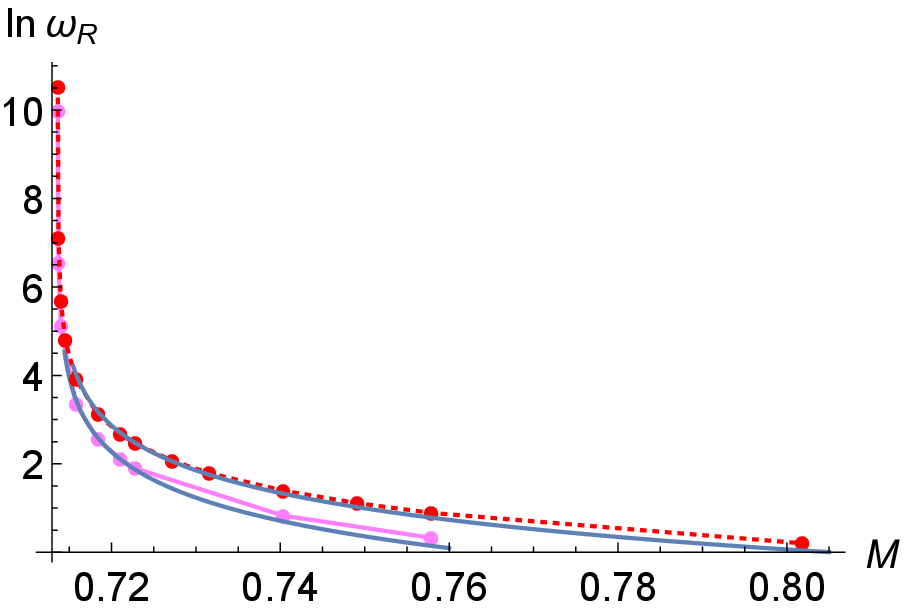}}
\end{center}
\vspace{-0.6cm}
\caption{{\footnotesize The logarithmic plots of {\it lowest} ($n=0$) $\omega_R$ and $\omega_I$ near $M=M_{\rm min} \approx 0.7136$,
corresponding to Fig. 15. The blue lines represent the best-fit curves
near the limit, given by (\ref{omega_singular_fit}).
}}
\label{figIV}
\end{figure}

\begin{figure}[!htbp]
\begin{center}
{\includegraphics[width=7.5cm,keepaspectratio]{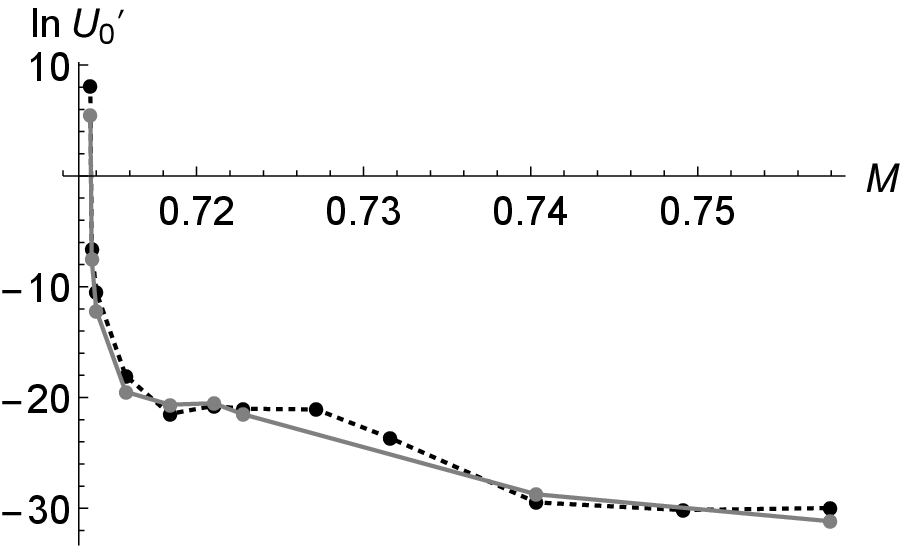}}
\end{center}
\vspace{-0.6cm}
\caption{{\footnotesize The plots of ${\rm ln} U_0'$
for $U_1$ (dashed curves) and $U_2$ (solid curves) in Figs. 15 and 16.
}}
\label{figIV}
\end{figure}
\subsection{QNMs
vs. $M$}

Here,
we divide into three cases depending on the values of $\beta Q$, which divides Sch-type and RN-type by $\beta Q=1/2$ \cite{Kim:2016pky}. First of all, Figs. 10 and 11 show the effective potentials and corresponding lowest QNMs as functions of the mass $M$ for a RN-type black hole with fixed $\beta Q>1/2$. As can
be seen in Fig. 10,
there are one degenerate event horizon
(chraged Nariai solution) with one Cauchy horizon
for $M=M_{\rm max}$, and one degenerate black hole horizon with one
cosmological horizon for $M=M_{\rm min}$. In between them,
$M_{\rm min}< M< M_{\rm max}$, there are generally three
horizons, depending on $M$. Fig. 11 shows that
$\om \approx 0$ at $M=M_{\rm max}\approx 0.9836$ and no QNMs
below $M_{\rm min} \approx 0.9464$, corresponding to replacing
$Q_{\rm min} \ra M_{\rm max}$ and $Q_{\rm max} \ra M_{\rm min}$ in Fig. 9.

Figs. 12 and 13 show the cases for $\beta Q=1/2$. The basic difference
from the case of $\beta Q>1/2$ in Figs. 10 and 11 is that there is only one
{\it degenerate} {(black hole) horizon} at the origin $r=0$, {\it i.e.}, the {\it point-like}
horizon, with the {\it vanishing} Hawking temperature $T_H=0$ and $\om_I \approx 1.844 \times 10^{-4}$ (for $U_1$), $1.840 \times 10^{-4}$ (for $U_2$) at $M=M_{\rm min}=M_0\approx 0.8740$. On the other hand, $\om_I \approx 0$ at $M=M_{\rm max}\approx 0.9810$
and $\omega_I$ has a maximum point about $M \sim 0.9$. However, the
behavior of $\om_R$ is similar to the case of $\beta Q>1/2$ in Fig. 10,
except the
rapid but {\it finite} oscillation near $M_{\rm min}$.

Figs. 14 and 15 show the cases for a Sch-type black hole with $\beta Q<1/2$. The basic difference
from the case of $\beta Q=1/2$ in Figs. 12 and 13 is that there is
no (even a point) black hole horizon
in addition to the
cosmological horizon for $M=M_{\rm min}$, even though the
(non-degenerate) black hole horizon can arbitrarily approach close to
$r_+=0$ as $M\ra M_0$, with the {\it divergent} Hawking
temperature (Fig. 1). One more remarkable difference is that, as
$M \ra M_{\rm min}$, $\omega$
is
{\it divergent} as
\beq
\omega_I \approx  (M-M_{\rm{min}})^{-a} e^{-b},~~\omega_R \approx  (M-M_{\rm{min}})^{-c} e^{-d}
\label{omega_singular_fit}
\eeq
with $(a,~ b,~ c,~ d) \approx (1.1223,~ 5.2870,~ 1.0752,~ 2.5693),~
 (1.1241,~ 5.3062,~ 1.0950,~ 3.2682)$
for $U_1$, $U_2$, respectively, within the limited
results due to numerical errors \footnote{In the
numerical computations of WKB formula (93), we find that the ``smallness"
of $U_0'$, which should be ``0" by definition, can be a
good barometer of the numerical errors. As $M \ra M_{\rm min}$,
the (non-vanishing) magnitude of $U_0'$ is increasing as shown in Fig. 17.
}. There is
one degenerate
horizon where the black hole horizon meets the cosmological horizon for $M=M_{\rm max}$ and generally two (one for black hole
and one for cosmological) horizons in between them,
$M_{\rm min}< M< M_{\rm max}$. However, Fig. 15 shows $\om \approx 0$
at $M_{\rm max}\approx 0.9774$, similarly to the case of $\beta Q>1/2$
in Fig. 11.

\begin{figure}[!htbp]
\begin{center}
{\includegraphics[width=5cm,keepaspectratio]{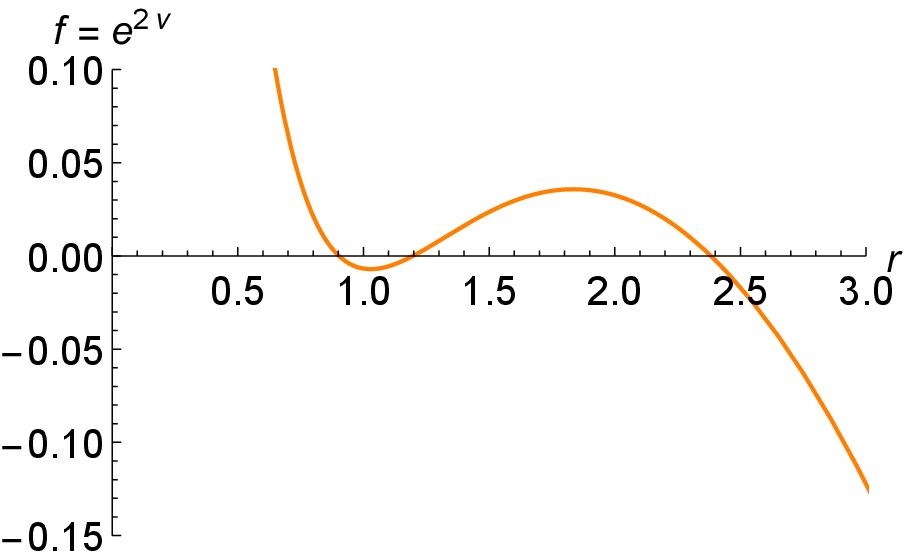}}
{\includegraphics[width=5cm,keepaspectratio]{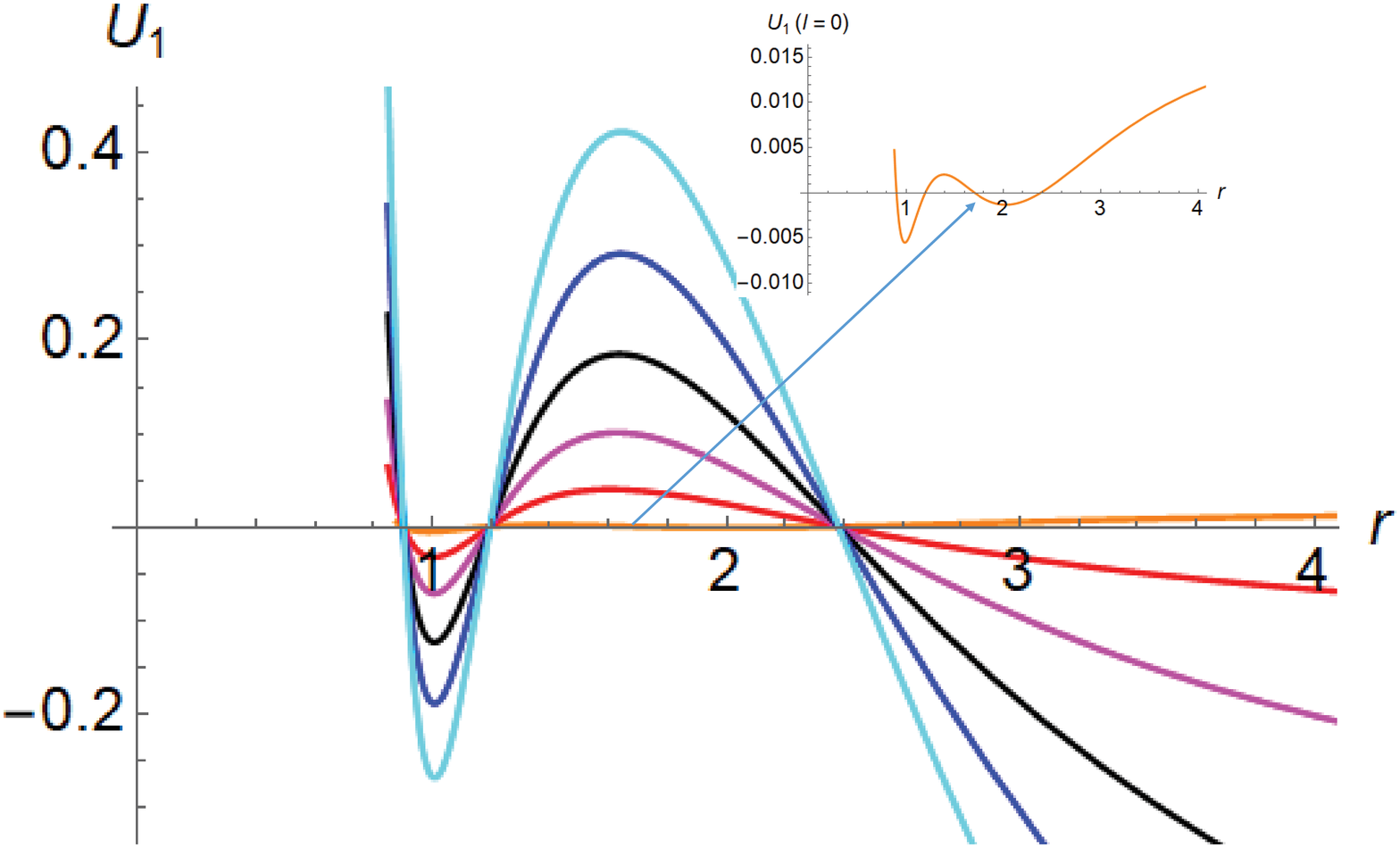}
}
{\includegraphics[width=5cm,,keepaspectratio]{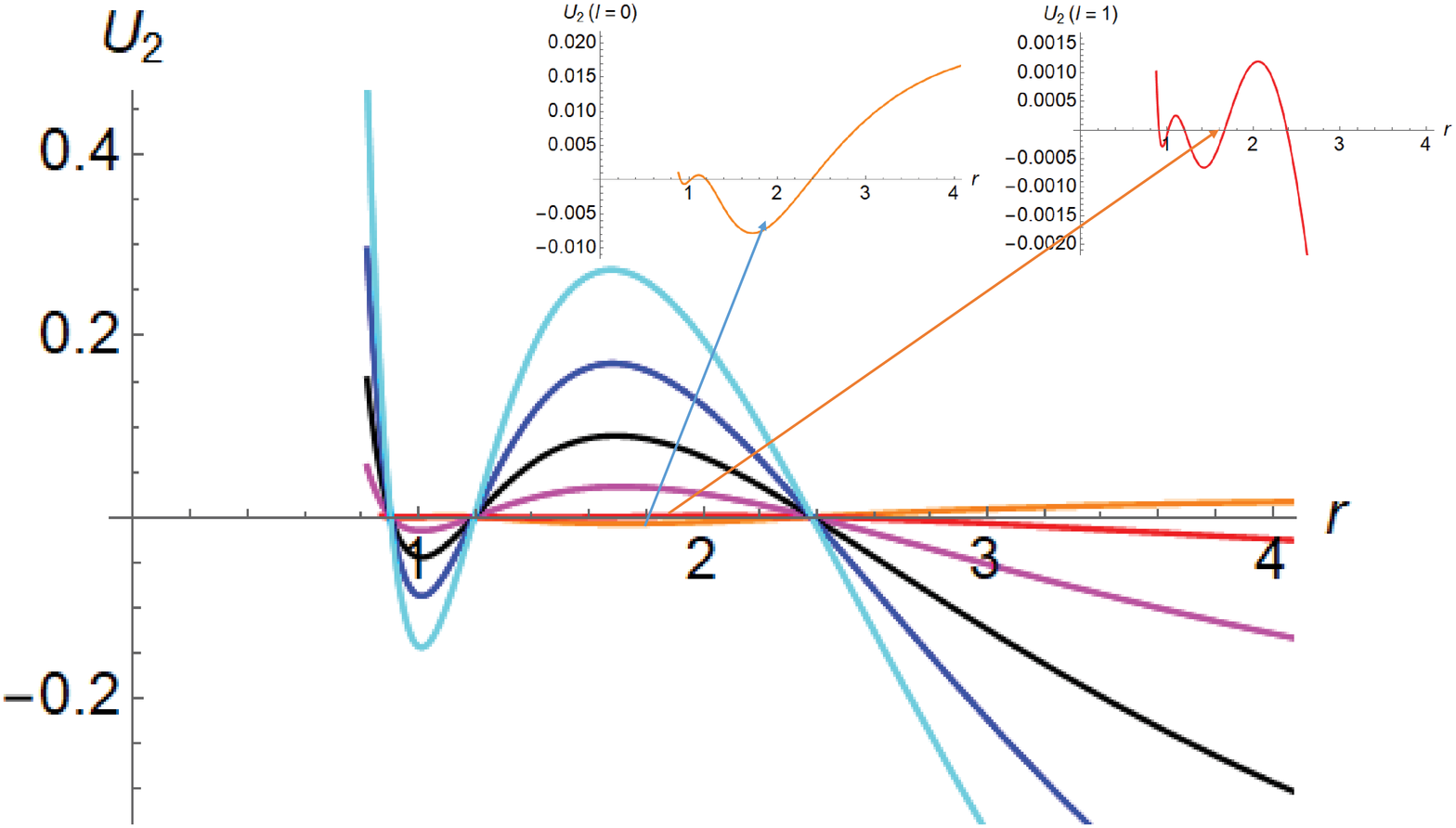}
}
\end{center}
\vspace{-0.6cm}
\caption{{\footnotesize The plots of $f=e^{2 \nu (r)}$ and the effective
potentials $U_1(r)$ and $U_2(r)$ for varying the orbital number $l$ with
a fixed $\beta Q>1/2$. Here, we consider from $l=0$ to $l=5$
(bottom to top, barrier region) when $M=0.95$,
$\beta=1$, $Q=1$, and $\Lambda =0.2$ with three (two for black hole and
one for cosmological) horizons.
}}
\label{figVII}
\end{figure}

\begin{figure}[!htbp]
\begin{center}
{\includegraphics[width=7.5cm,keepaspectratio]{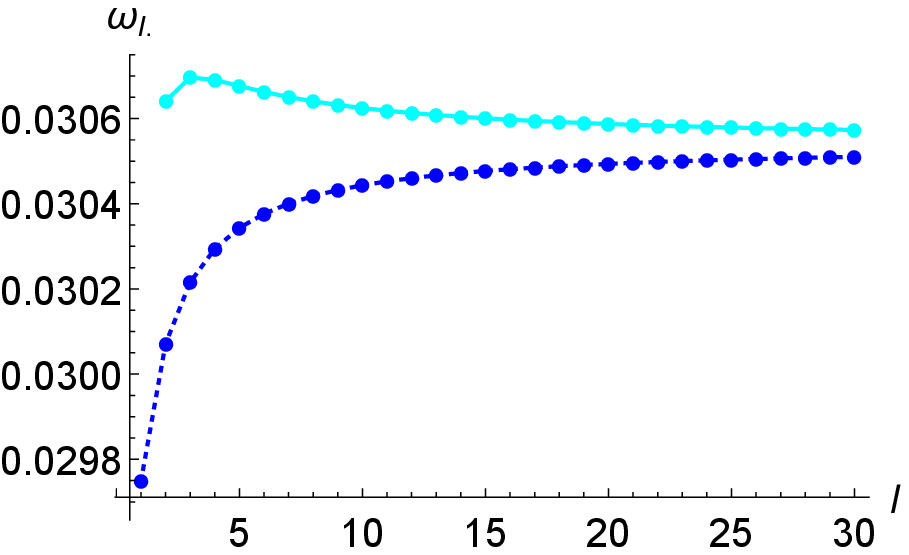}}
$\;\;\;\;\;${\includegraphics[width=7.5cm,keepaspectratio]{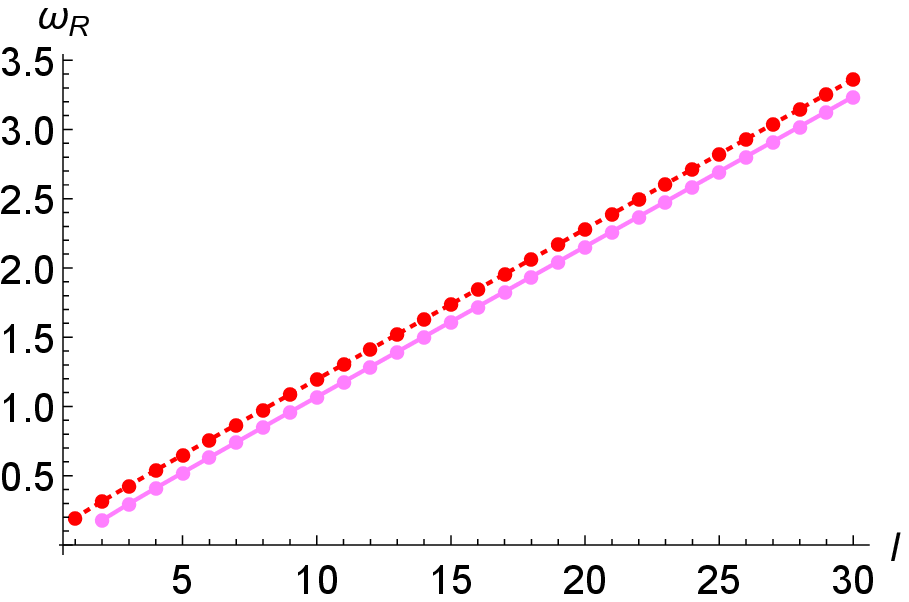}}
\end{center}
\vspace{-0.6cm}
\caption{{\footnotesize The plots of {\it lowest} ($n=0$) $\omega=\omega_R -i \omega_I$ vs. $l$ for $U_1$ (dashed curves, from $l=1$ to $l=30$) and $U_2$ (solid curves, from $l=2$ to $l=30$)
when $M=0.95$, $\beta=1$, $Q=1$, and $\Lambda =0.2$.
}}
\label{figVIII}
\end{figure}

\begin{figure}[!htbp]
\begin{center}
{\includegraphics[width=5cm,keepaspectratio]{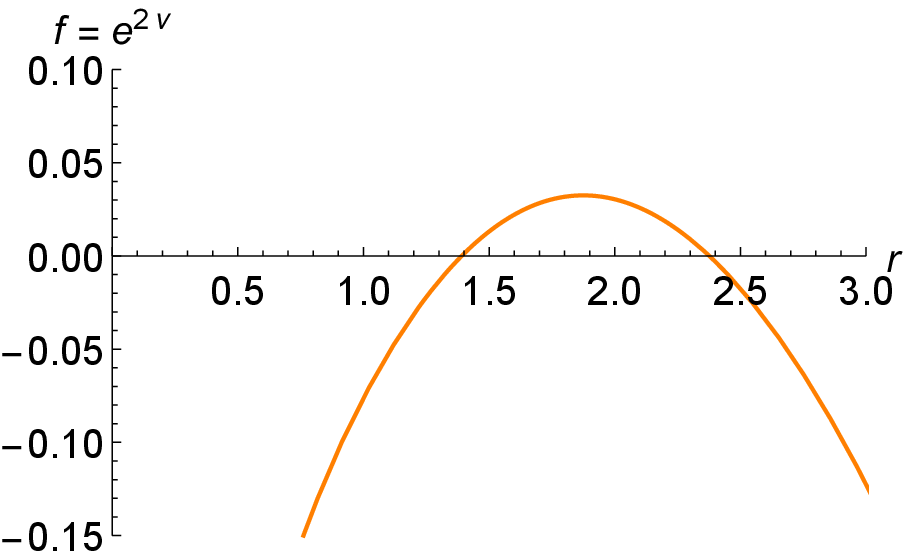}}
{\includegraphics[width=5cm,keepaspectratio]{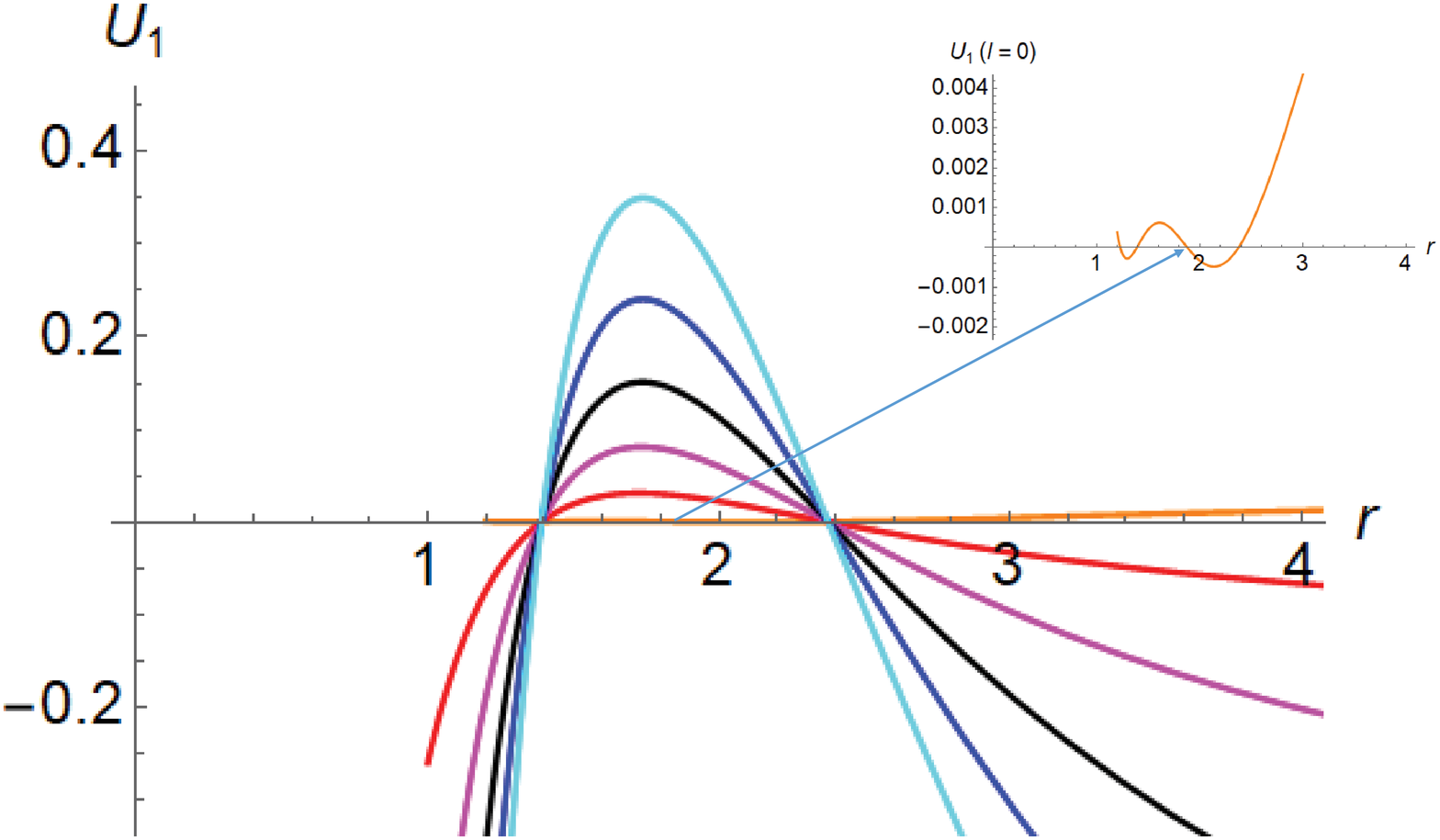}}
{\includegraphics[width=5cm,,keepaspectratio]{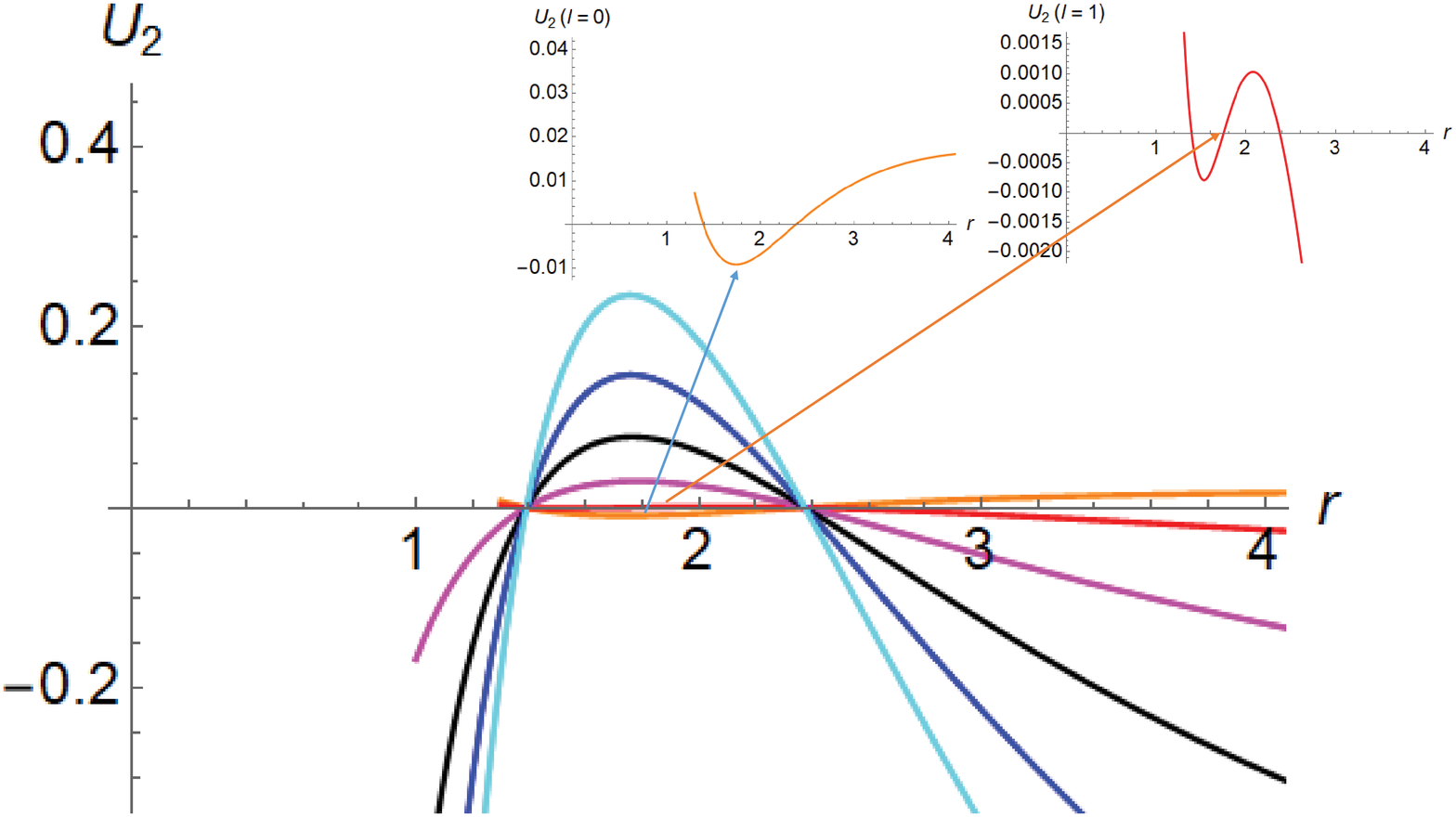}}
\end{center}
\vspace{-0.6cm}
\caption{{\footnotesize The plots of $f=e^{2 \nu (r)}$ and the effective
potentials $U_1(r)$ and $U_2(r)$ for varying the orbital number
$l$ with a fixed $\beta Q=1/2$. Here, we consider from $l=0$ to $l=5$
(bottom to top, barrier region) when $M=0.95$,
$\beta=1/2$, $Q=1$, and $\Lambda =0.2$ with two (one for black hole and one
for cosmological) horizons.
}}
\label{figVII}
\end{figure}

\begin{figure}[!htbp]
\begin{center}
{\includegraphics[width=7.5cm,keepaspectratio]{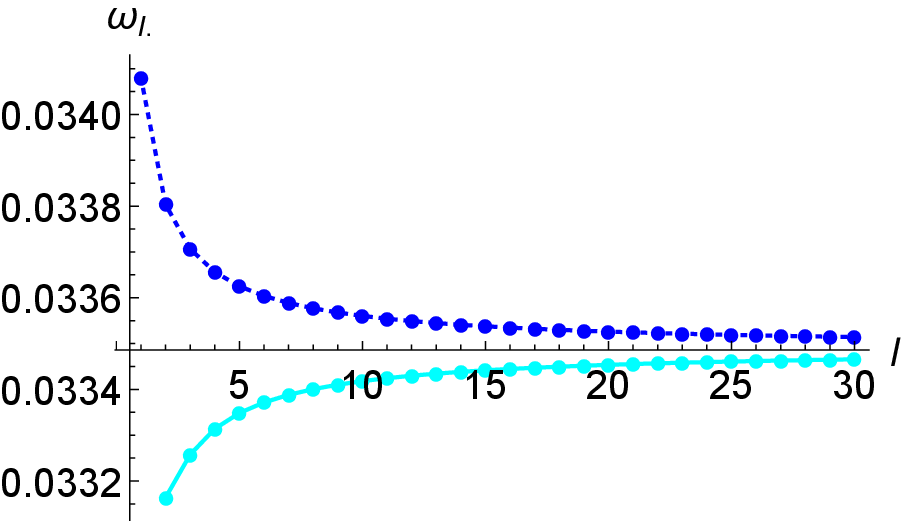}}
$\;\;\;\;\;${\includegraphics[width=7.5cm,keepaspectratio]{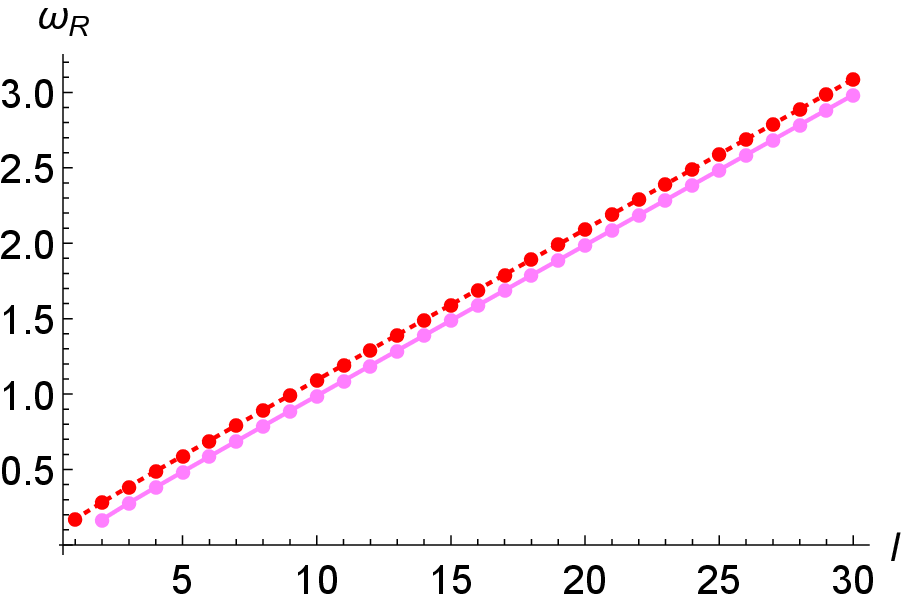}}
\end{center}
\vspace{-0.6cm}
\caption{{\footnotesize The plots of {\it lowest} ($n=0$) $\omega=\omega_R -i \omega_I$ vs. $l$ for $U_1$ (dashed curves, from $l=1$ to $l=30$) and $U_2$ (solid curves, from $l=2$ to $l=30$)
when $M=0.95$, $\beta=1/2$, $Q=1$, and $\Lambda =0.2$.
}}
\label{figVIII}
\end{figure}

\begin{figure}[!htbp]
\begin{center}
{\includegraphics[width=5cm,keepaspectratio]{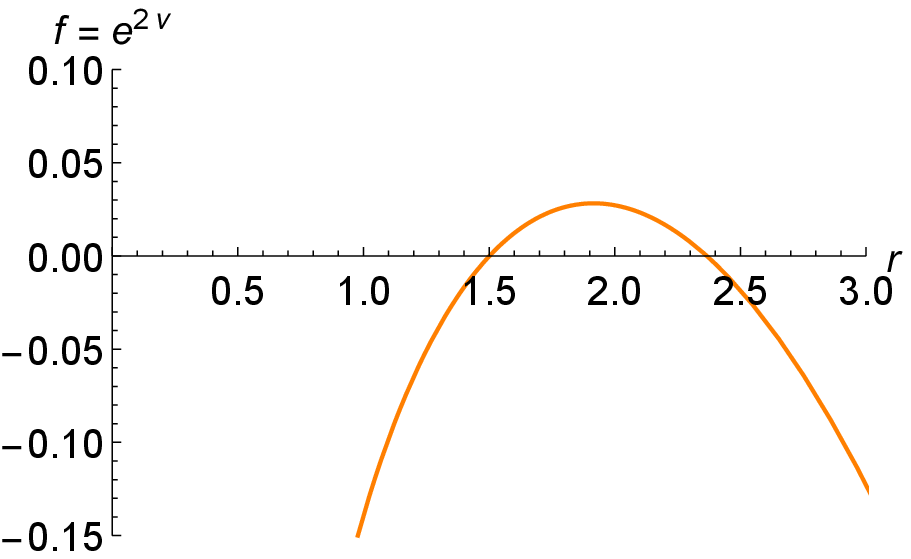}}
{\includegraphics[width=5cm,keepaspectratio]{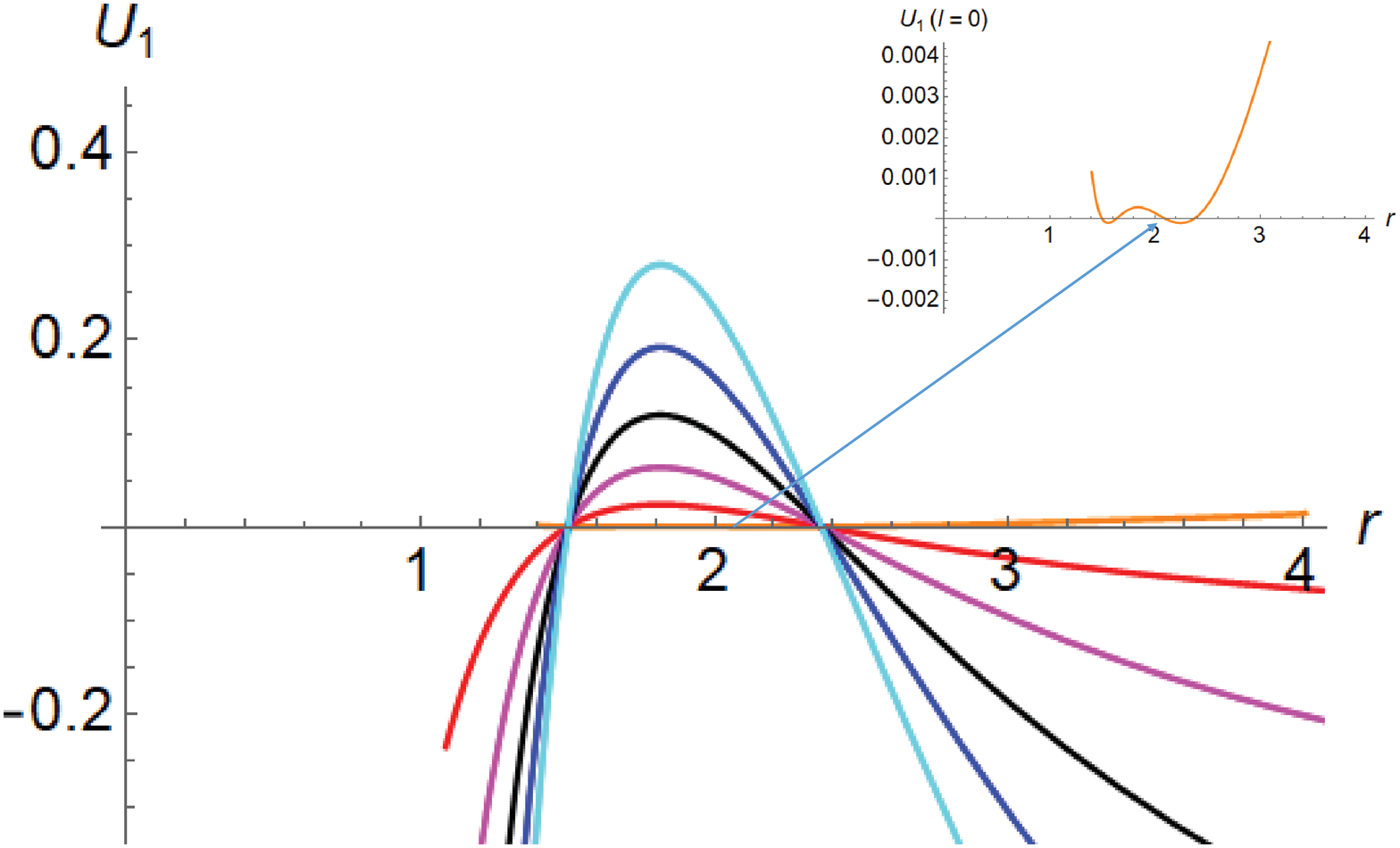}}
{\includegraphics[width=5cm,,keepaspectratio]{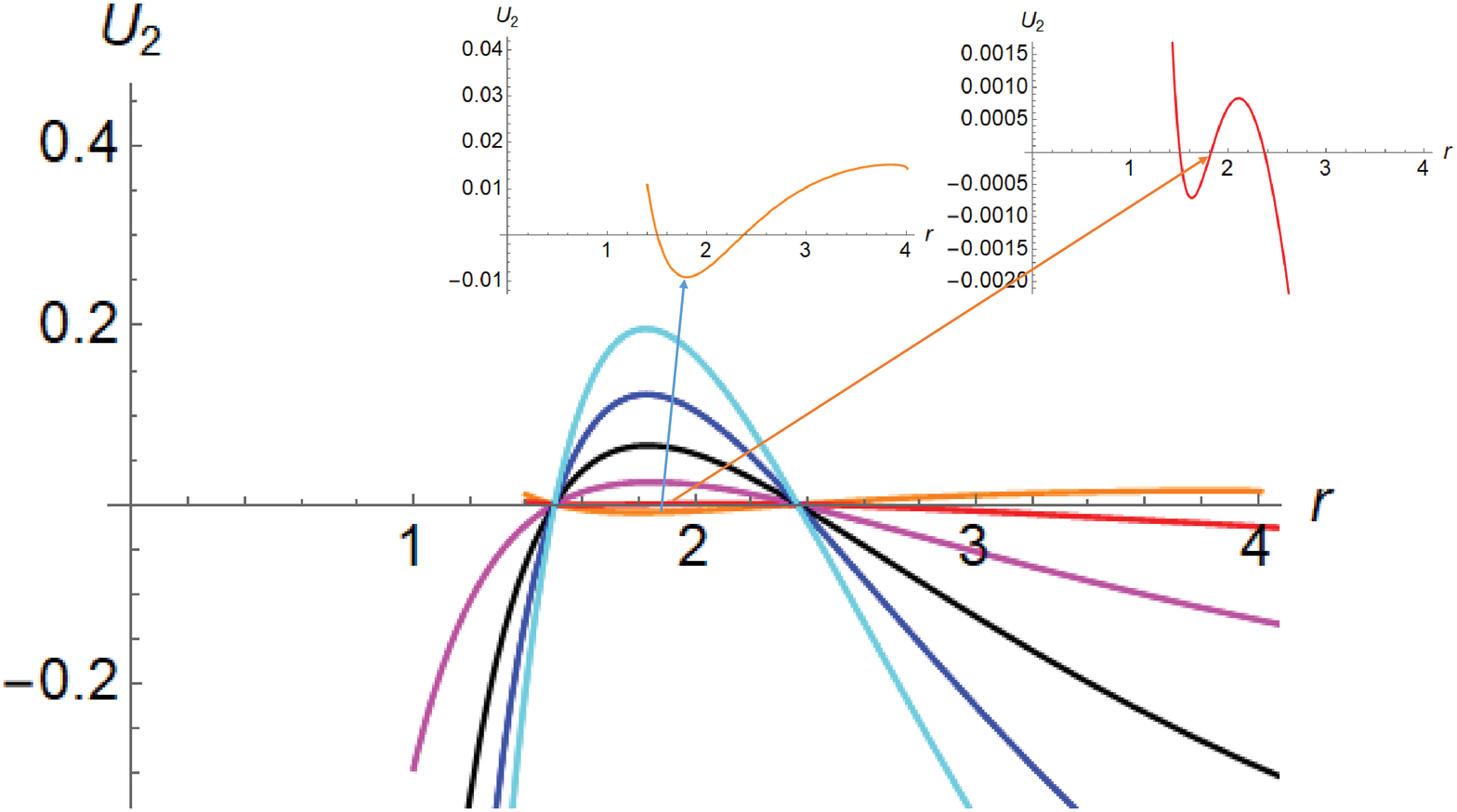}}
\end{center}
\vspace{-0.6cm}
\caption{{\footnotesize The plots of $f=e^{2 \nu (r)}$ and the effective
potentials $U_1(r)$ and $U_2(r)$ for varying the orbital number
$l$ with a fixed $\beta Q<1/2$. Here, we consider from $l=0$ to $l=5$
(bottom to top, barrier region) when $M=0.95$,
$\beta=1/3$, $Q=1$, and $\Lambda =0.2$ with two (one for black hole and
one for cosmological) horizons.
}}
\label{figVII}
\end{figure}

\begin{figure}[!htbp]
\begin{center}
{\includegraphics[width=7.5cm,keepaspectratio]{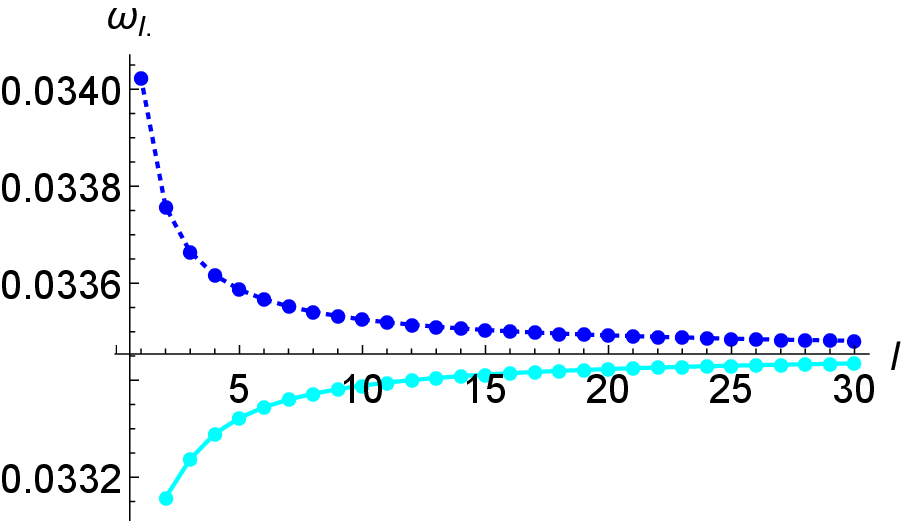}}
$\;\;\;\;\;${\includegraphics[width=7.5cm,keepaspectratio]{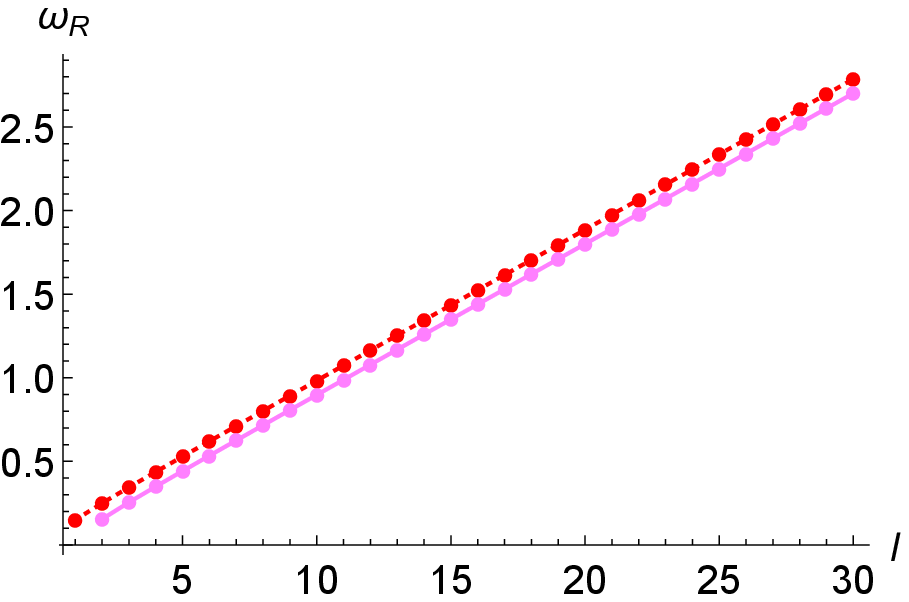}}
\end{center}
\vspace{-0.6cm}
\caption{{\footnotesize The plots of {\it lowest} ($n=0$) $\omega=\omega_R -i \omega_I$ vs. $l$ for $U_1$ (dashed curves, from $l=1$ to $l=30$) and $U_2$ (solid curves, from $l=2$ to $l=30$)
when $M=0.95$, $\beta=1/3$, $Q=1$, and $\Lambda =0.2$.
}}
\label{figVIII}
\end{figure}

\subsection{QNMs
vs. $l$}

We also divide into three cases depending on the values of
$\beta Q$.
Figs. 18 and 19 show the effective potentials and corresponding lowest QNMs as functions of the orbital number
$l$ for a fixed $\beta Q>1/2$. As
shown in Fig. 18,
there are three (two for black hole and one for cosmological) horizons independent of $l$ (Fig. 18 (left)). On the other hand, the effective potentials depend on $l$, representing the angular momentum barriers for higher $l$ with a single maximum. However, for {\it lower} $l$,
there is no local maximum (but a minimum) in the effective
potential ($l=0$ case for $U_2$) or there are ``three turning points" with
one additional turning point within the usual two turning points, {\it i.e.,}
the outer black hole horizon at $r_+ \approx 1.2$ and the cosmological
horizon at $r_{++} \approx 2.4$ ($l=0$ case for $U_1$ and $l=1$ case for $U_2$) so that the usual WKB formula does not apply \footnote{For  three-turning-point problems, a complex matrix WKB approach may be applied \cite{Galt:1991}. See also \cite{Kono:2019} for other restrictions on using the WKB formula.}.
One remarkable thing in the result of Fig. 19 is that, in contrast to other varying parameters, $\om_{I(1)}$ and $\om_{I(2)}$ respond differently to the varying $l$, while $\om_{R(1)}$ and $\om_{R(2)}$ respond almost identically.
In particular, in the large $l$ limit, $\om_I$ approaches asymptotically to a limiting value $\om_I \approx 0.03055$, while $\om_R$ shows a {\it linear} dependence $\om_R = \sigma l$ with $\sigma \approx 0.1092$, similar to earlier results \cite{iyer}.

Figs. 20 and 21 show the case for $\beta Q=1/2$. The main difference from $\beta Q>1/2$ case in Figs. 18 and 19 is the ``magnitude flip" between $\om_{I(1)}$ and $\om_{I(2)}$ in  Fig. 21. In the large $l$ limit, $\om_I$ approaches asymptotically to a limiting value $\om_I \approx 0.03349$, while $\om_R$ shows a {\it linear} dependence $\om_R = \sigma l$ with $\sigma \approx 0.1006$. On the other hand, for {\it lower} $l$, {\it i.e.}, $l=0$ for $U_1$ and $l=0,1$ for $U_2$, the usual WKB formula does not apply, due to either ``three turning points" ($l=0$ for $U_1$ and $l=1$ for $U_2$) or ``no local maximum" ($l=0$ for $U_2$) within the usual two turning points $r_+ \approx 1.4$ and $r_{++} \approx 2.4$.

Figs. 22 and 23 show the cases for $\beta Q<1/2$ but the results do not show any qualitative difference from $\beta Q=1/2$ in Figs. 20 and 21. In the large $l$ limit, $\om_I$ approaches asymptotically to a limiting value, $\om_I \approx 0.03345$, while $\om_R$ shows a {\it linear} dependence $\om_R = \sigma l$ with $\sigma \approx 0.0910$. For {\it lower} $l$, {\it i.e.}, $l=0$ for $U_1$ and $l=0,1$ for $U_2$, the usual WKB formula does not apply, due to either ``three turning points" ($l=0$ for $U_1$ and $l=1$ for $U_2$) or ``no local maximum" ($l=0$ for $U_2$) within the usual two turning points $r_+ \approx 1.5$ and $r_{++} \approx 2.4$.\\

\begin{figure}[!ht]
\begin{center}
{\includegraphics[width=5cm,keepaspectratio]{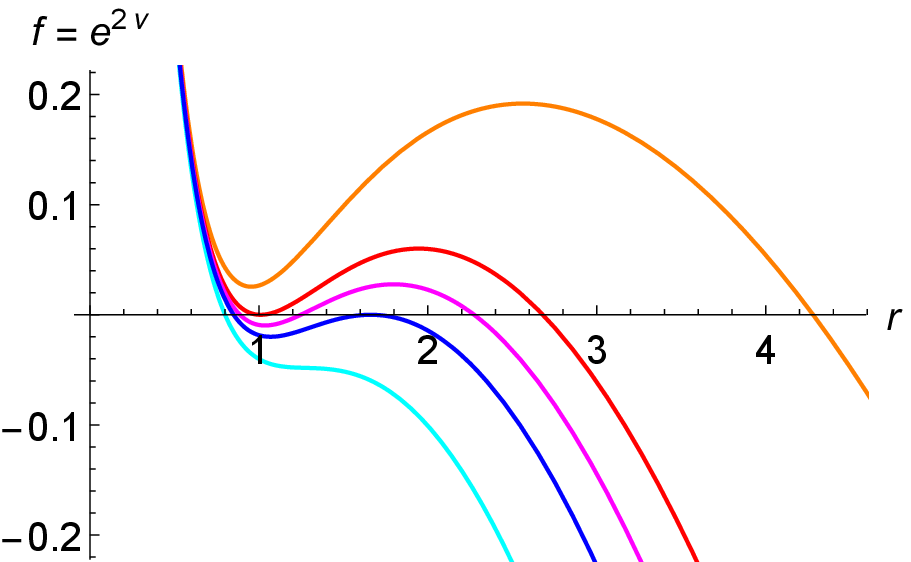}}
{\includegraphics[width=5cm,keepaspectratio]{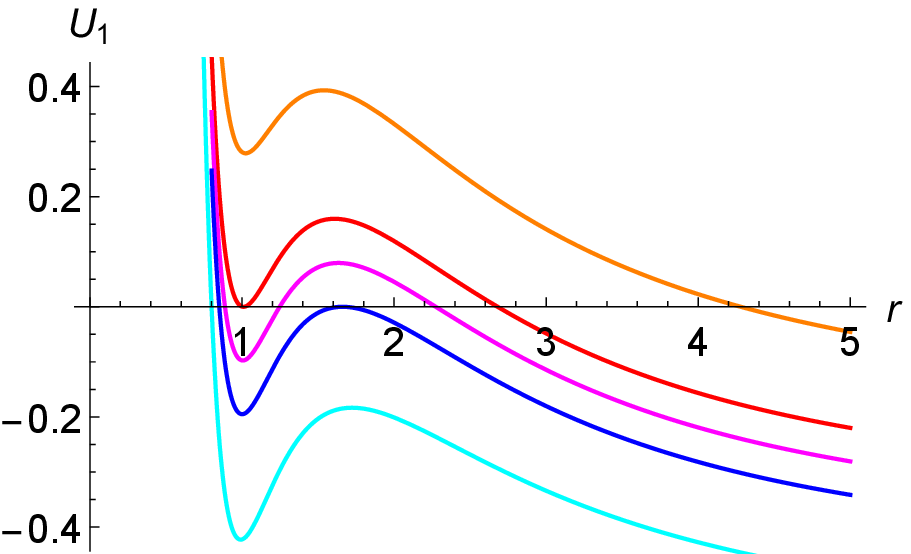}}
{\includegraphics[width=5cm,,keepaspectratio]{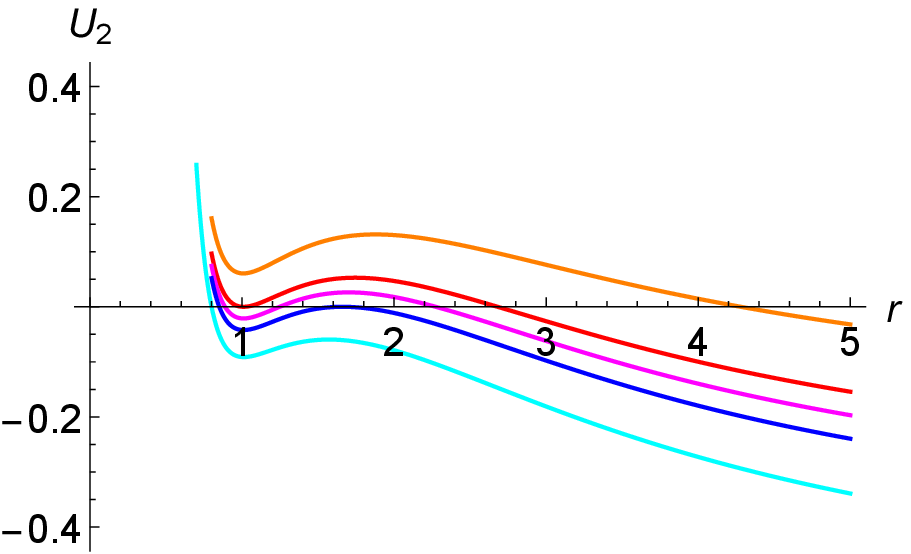}}
\end{center}
\vspace{-0.6cm}
\caption{{\footnotesize The plots of $f=e^{2 \nu (r)}$ and the effective potentials, $U_1(r)$ and $U_2(r)$ for varying
the cosmological constant
$\Lambda$ with a fixed $\beta Q>1/2$. Here, we consider $\Lambda=0.1,~ \Lambda_{\rm min},~ 0.2074,~ \Lambda_{\rm max},~ 0.3$ (top to bottom) with $\Lambda_{\rm min} \approx 0.1796$ and $\Lambda_{\rm max} \approx 0.2352$ when
$\beta=1$, $Q=1$, $l=2$, and $M =0.95$.
}}
\label{figIX}
\end{figure}

\begin{figure}[!ht]
\begin{center}
{\includegraphics[width=7.5cm,keepaspectratio]{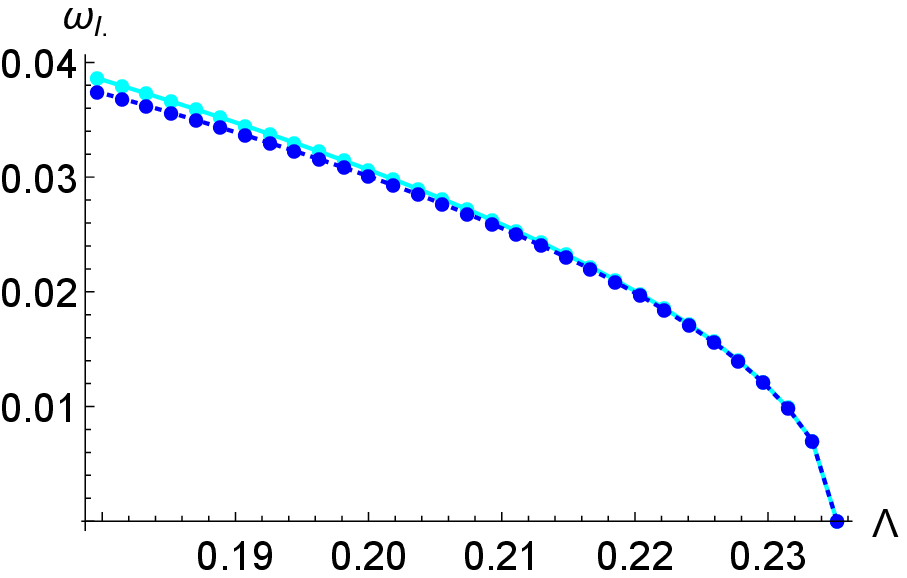}}
$\;\;\;\;\;${\includegraphics[width=7.5cm,keepaspectratio]{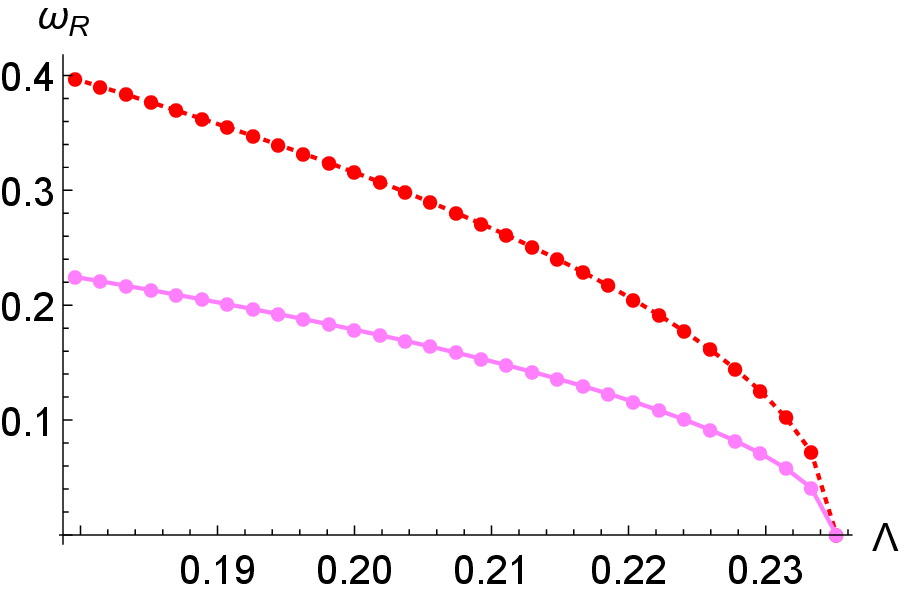}}
\end{center}
\vspace{-0.6cm}
\caption{{\footnotesize The plots of {\it lowest} ($n=0$) $\omega=\omega_R -i \omega_I$ vs. $\Lambda$ for $U_1$ (dashed curves) and $U_2$ (solid curves)
when $\beta=1$, $Q=1$, $l=2$, and $M =0.95$.
}}
\label{figX}
\end{figure}

\begin{figure}[!ht]
\begin{center}
{\includegraphics[width=5cm,keepaspectratio]{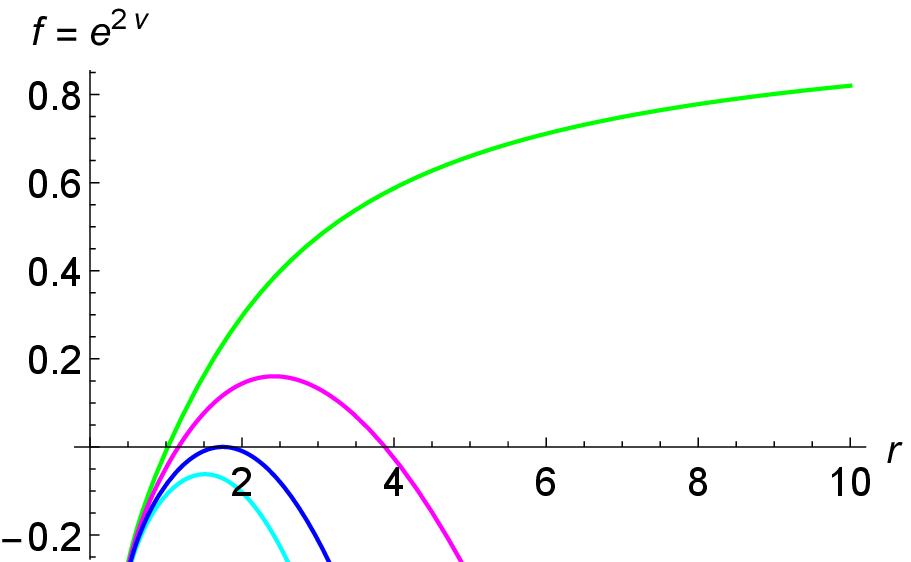}}
{\includegraphics[width=5cm,keepaspectratio]{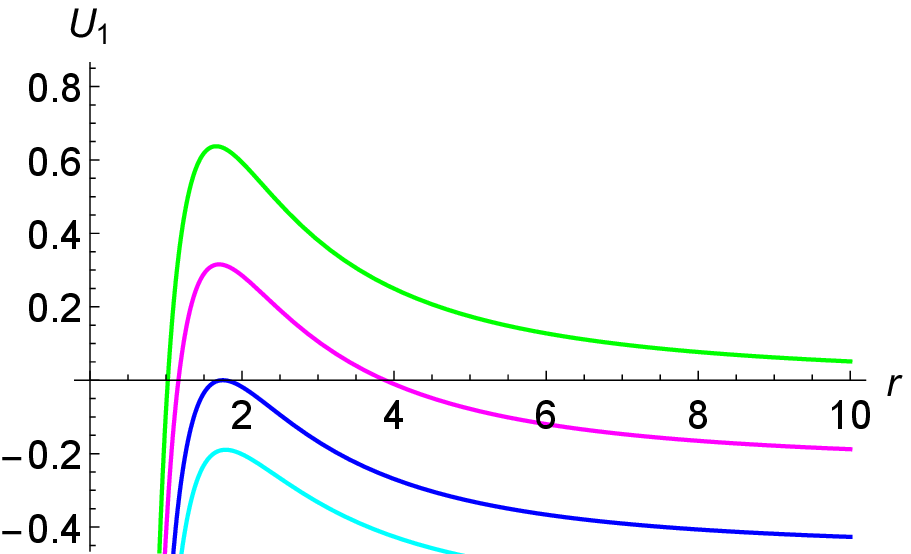}}
{\includegraphics[width=5cm,,keepaspectratio]{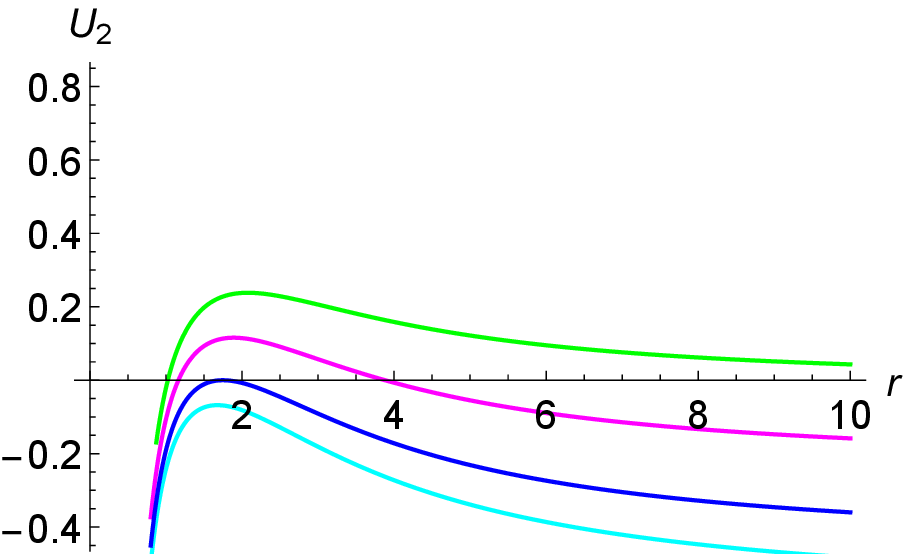}}
\end{center}
\vspace{-0.6cm}
\caption{{\footnotesize The plots of $f=e^{2 \nu (r)}$ and the
effective potentials $U_1(r)$ and $U_2(r)$ for varying the cosmological constant $\Lambda$ with a
fixed $\beta Q=1/2$. Here, we consider
$\Lambda=0,~ 0.1149,~ \Lambda_{\rm max},~ 0.3$ (top to bottom) with $\Lambda_{\rm max} \approx 0.2298$ when
$\beta=1/2$, $Q=1$, $l=2$, and $M =0.95$.
}}
\label{figIX}
\end{figure}

\begin{figure}[!ht]
\begin{center}
{\includegraphics[width=7.5cm,keepaspectratio]{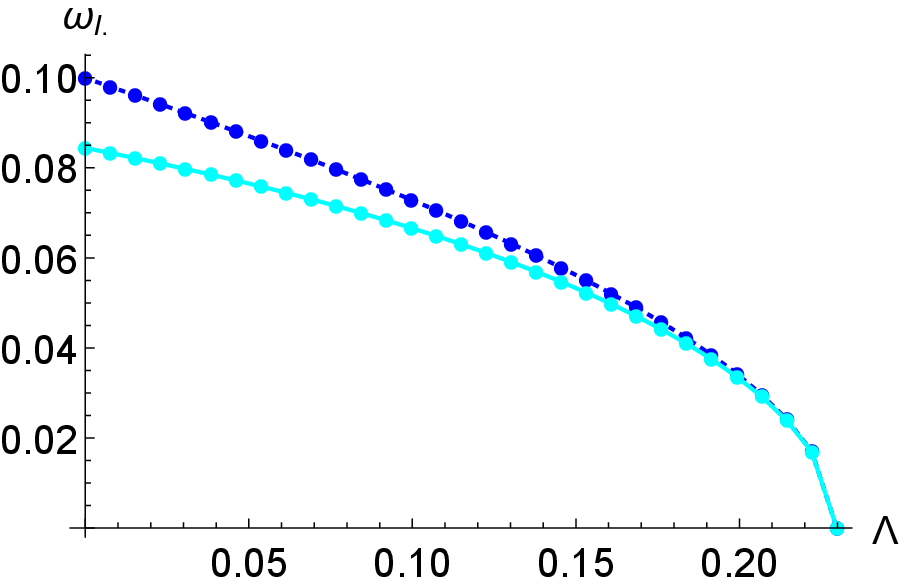}}
$\;\;\;\;\;${\includegraphics[width=7.5cm,keepaspectratio]{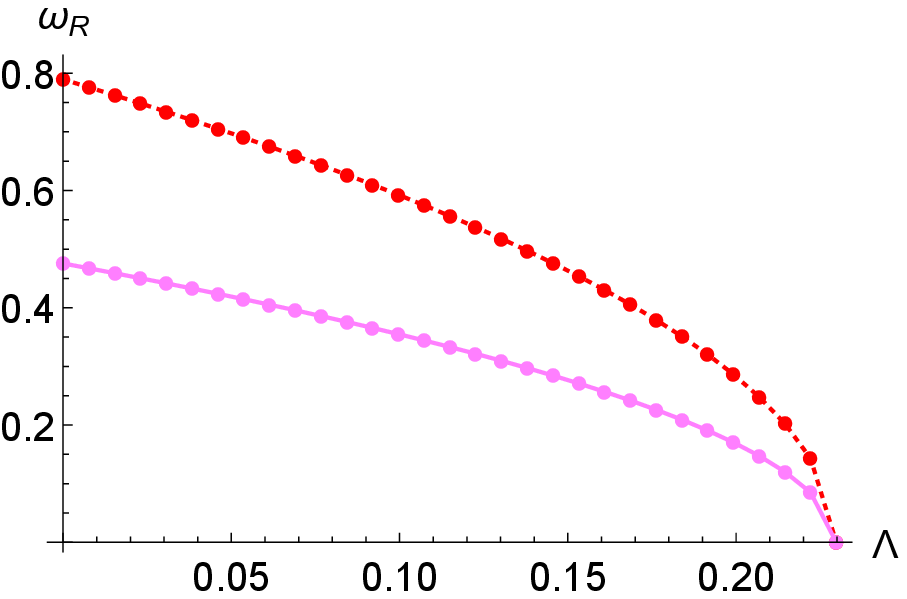}}
\end{center}
\vspace{-0.6cm}
\caption{{\footnotesize The plots of {\it lowest} ($n=0$) $\omega=\omega_R -i \omega_I$ vs. $\Lambda$ for $U_1$ (dashed curves) and $U_2$ (solid curves)
when $\beta=1/2$, $Q=1$, $l=2$, and $M =0.95$.
}}
\label{figX}
\end{figure}

\begin{figure}[!ht]
\begin{center}
{\includegraphics[width=5cm,keepaspectratio]{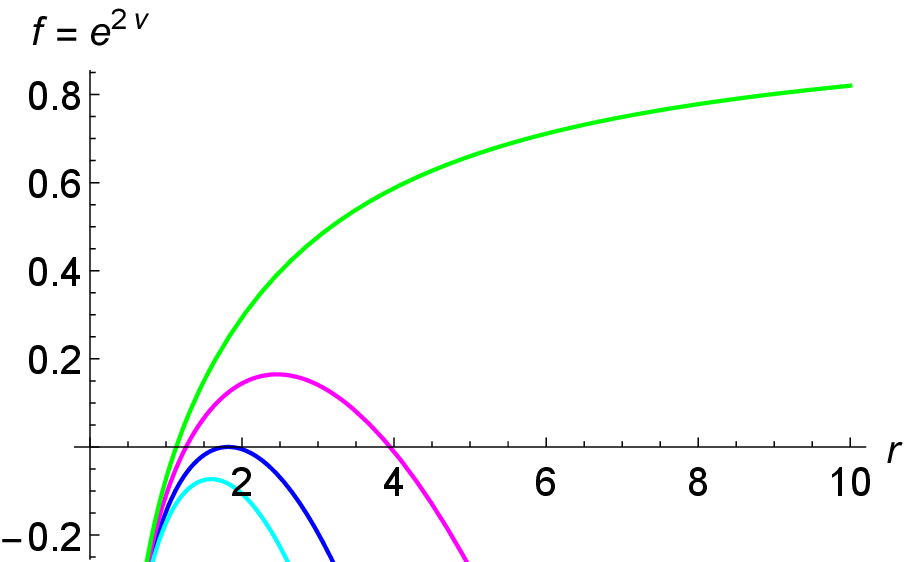}}
{\includegraphics[width=5cm,keepaspectratio]{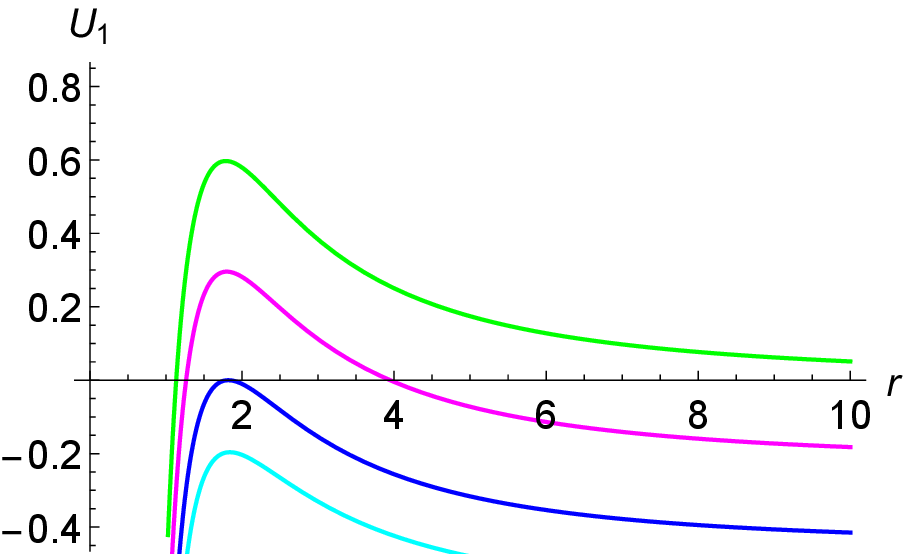}}
{\includegraphics[width=5cm,,keepaspectratio]{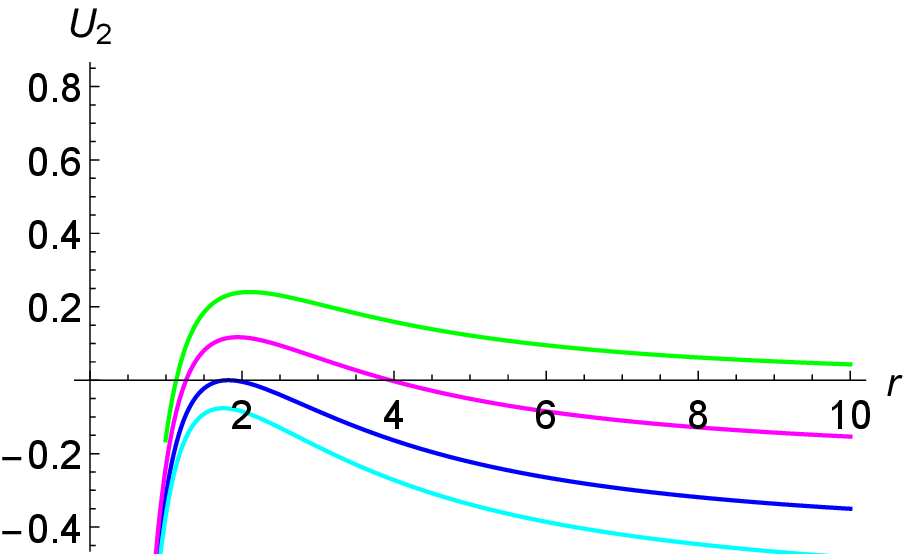}}
\end{center}
\vspace{-0.6cm}
\caption{{\footnotesize The plots of $f=e^{2 \nu (r)}$ and the
effective potentials $U_1(r)$ and $U_2(r)$ for varying the cosmological constant $\Lambda$ with a
fixed $\beta Q<1/2$. Here, we consider
$\Lambda=0,~ 0.1121,~ \Lambda_{\rm max},~ 0.3$ (top to bottom) with $\Lambda_{\rm max} \approx 0.2243$ when
$\beta=1/3$, $Q=1$, $l=2$, and $M =0.95$, with the similar horizon structures as in the case of $\beta Q=1/2$ in Fig. 26.
}}
\label{figIX}
\end{figure}

\begin{figure}[!ht]
\begin{center}
{\includegraphics[width=7.5cm,keepaspectratio]{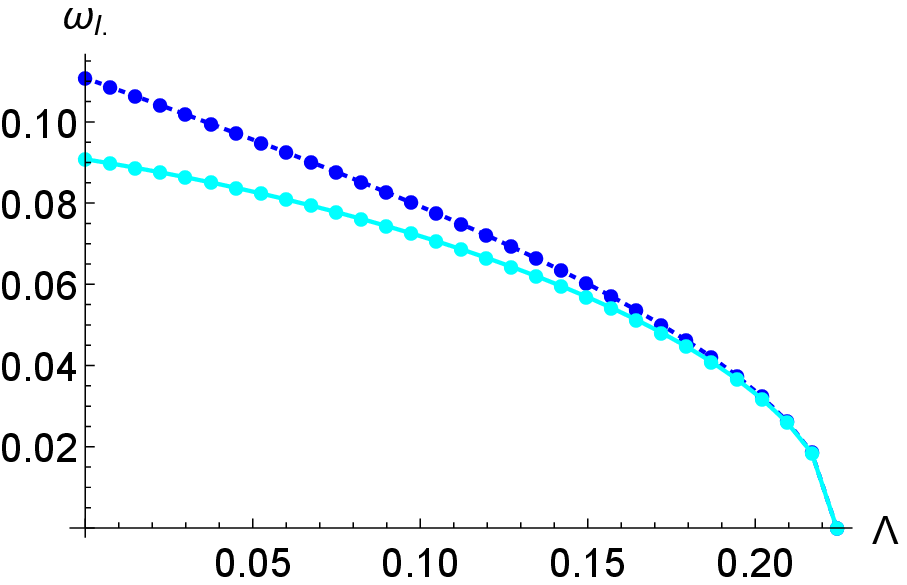}}
$\;\;\;\;\;${\includegraphics[width=7.5cm,keepaspectratio]{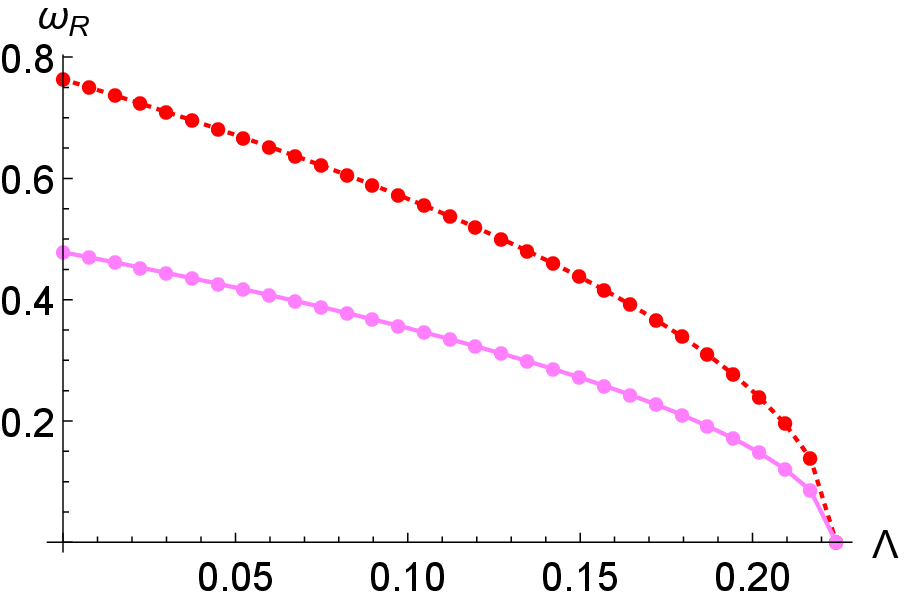}}
\end{center}
\vspace{-0.6cm}
\caption{{\footnotesize The plots of {\it lowest} ($n=0$) $\omega=\omega_R -i \omega_I$ vs. $\Lambda$ for $U_1$ (dashed curves) and $U_2$ (solid curves)
when $\beta=1/3$, $Q=1$, $l=2$, and $M =0.95$.
}}
\label{figX}
\end{figure}

\subsection{QNMs
vs. $\Lambda$}

As in the previous two cases, we also divide into three cases depending on the values of $\beta Q$.
Figs. 24 and 25 show the effective potentials and corresponding {lowest} QNMs as functions of the cosmological constant $\Lambda$ for a fixed $\beta Q>1/2$. As can be seen in Fig. 24,
there are one degenerate event horizon (charged Nariai solution) with one Cauchy horizon for $\Lambda=\Lambda_{\rm max}$, and one degenerate black
hole horizon with one cosmological horizon for $\Lambda_{\rm min}$.
In between them, $\Lambda_{\rm min}< \Lambda < \Lambda_{\rm max}$, there
are generally three
horizons depending on $\Lambda$, similar to Fig. 10.
Fig. 25 shows $\om \approx 0$ at $\Lambda_{\rm max} \approx 0.2352$ and no QNMs below $\Lambda_{\rm min} \approx 0.1796$, similar to the case in
Fig. 11.

Figs. 26 and 27 show the case for $\beta Q=1/2$.
As can be seen in Fig. 26,
there are one degenerate event horizon
for $\Lambda=\Lambda_{\rm max}$, and one non-degenerate black
hole horizon for $\Lambda=``0"$, {\it i.e.}, the {\it flat} case.
In between them, $0< \Lambda< \Lambda_{\rm max}$, there are generally
two (one for black hole and one for cosmological) horizons. The main difference in Fig. 27 from $\beta Q>1/2$ case in Fig. 25 is the ``magnitude flip" between $\om_{I(1)}$ and $\om_{I(2)}$, as in Figs. 13 and 21.

Figs. 28 and 29 show the case for $\beta Q<1/2$ but there is no qualitative difference in the result from $\beta Q=1/2$ case in Fig. 27.

\section{Exact Solution}

Generally, finding analytic expressions for QNMs is difficult and one needs to consider their numerical computations. However, there exist some special cases where exact solutions can be found. In this section, we consider the exact solution near the (charged) Nariai solution where the black hole horizon $r_+$ and the cosmological horizon $r_{++}$ merge. To this end, we first note that the metric function $f(r)=e^{2\nu (r)}$ in ({\ref{ansatz}) and (\ref{backgroundmetric}) can be written \cite{Card:2003,Moli:2003}, near the Nariai limit $r_+ \ra r_{++}$, as
\beq
f(r)=\f{2 \kappa_+}{r_{++}-r_+} (r_{++}-r) (r-r_+) +{\cal O} (\epsilon^3),
\eeq
where $\epsilon \equiv ({r_{++}-r_+})/{r_+} \ll 1$ and $\kappa_+ \equiv (1/2)(df/dr)|_{r =r_+}$ is the surface gravity at the horizon $r_+$, which is related to the Hawking temperature $T_H=\hbar \kappa /(2 \pi)$ in (\ref{TH}). In this limit, the tortoise coordinate $r_*$, for the physical region $r_+ \leq r \leq r_{++}$, can be obtained as
\beq
r_* &\equiv&\int^r f^{-1}(r) dr \no \\
&=&\f{1}{2 \kappa_+} \rm{ln} \left(\f{r-r_+}{r_{++}-r} \right)+{\cal O} (\epsilon^3),
\label{tortoise_Nariai}
\eeq
where we have chosen the integration constant such that $r_* = -\infty$ at $r=r_+$ and $r_* = +\infty$ at $r=r_{++}$. Then, one can invert the relation (\ref{tortoise_Nariai}) to get
\beq
r= \f{r_+ + r_{++} e^{2 \kappa_+ r_*}}{1+e^{2 \kappa_+ r_*}} +{\cal O} (\epsilon^3)
\label{r_Nariai}
\eeq
and
\beq
f(r_*)\equiv f(r(r_*))=\f{ \kappa_+ ( r_{++}-r_+)}{2 ~\rm{cosh^2} (\kappa_+ r_*) }
+{\cal O} (\epsilon^3).
\label{f_Nariai}
\eeq
Now, substituting (\ref{r_Nariai}) and (\ref{f_Nariai}) in the effective potentials $U_i$ of (\ref{V1}) and (\ref{V2}), one can find

\beq
U_i =\f{U_i^{(0)}}{\rm{cosh^2} (\kappa_+ r_*)} +{\cal O} (\epsilon^3),
\label{PT}
\eeq
where
\bear
U_1^{(0)} &\equiv & \frac{\kappa_+ ( r_{++}-r_+) }{2 r_+^3} \left[ \frac{ V_1^{(0)} + V_2^{(0)}}{2} + \sqrt{ \left ( \frac{V_1^{(0)} - V_2^{(0)}}{2}  \right)^2 +{V_{12}^{(0)}}^2 }  \right ] ,
\label{V1_0}\\
U_2^{(0)} &\equiv &\frac{\kappa_+ ( r_{++}-r_+) }{2 r_+^3} \left[ \frac{ V_1^{(0)} + V_2^{(0)}}{2} - \sqrt{ \left ( \frac{V_1^{(0)} - V_2^{(0)}}{2}  \right)^2 +{V_{12}^{(0)}}^2 }  \right ]
\label{V2_0}
\eear
and
\bear
V_1^{(0)} &\equiv&  (\mu^2 +2) r_+ +
\frac{ 4 Q^2}{r_+}\left(1+\f{Q^2}{\beta^2 r_+^4} \right)^{-1/2}
+~ {\cal O} (\epsilon) ,     \label{V1_0}
\\
V_{12}^{(0)} &\equiv & 2 \mu Q \left(1+\f{Q^2}{\beta^2 r_+^4}\right)^{-1/4} + ~ {\cal O} (\epsilon),  \label{V12_0}   \\
V_2^{(0)} &\equiv &  \mu^2 r_+ +~ {\cal O} (\epsilon).  \label{V2_0}
\eear
Here, we have used $f(r_*) \sim {\cal O} (\epsilon^2)$ and $df/dr=-2\kappa_+ \rm{tanh}(\kappa_+ r_*)+{\cal O} (\epsilon^3)\sim {\cal O} (\epsilon^2)$ from $\kappa_+ \sim {\cal O} (\epsilon)$ near the Nariai limit. The potential in (\ref{PT}) is known as P\"{o}shl-Teller potential \cite{Posc:1933} and its QNMs can be solved analytically \cite{Ferr:1984} as
\beq
\omega=\kappa_+ \left[ -i \left(n+\f{1}{2} \right)+ \sqrt{\f{U_i^{(0)}}{\kappa_+^2}-\f{1}{4}}\right],
\label{omega_exact}
\eeq
where $n=0, 1, 2, \cdots$ is the overtone mode number. In Figs. 30 - 33, considering the dependence on the Hawking temperature $T_H$ or black-hole horizon radius $r_+$, the analytic result of QNMs in (\ref{omega_exact}) are compared with the numerical results based on the WKB approximations in Sec. IV. The results show quite good agreements for the imaginary parts $\om_I$ even beyond the Nariai limit (Figs. 30 and 31). The best-fit curves for the numerical results in Fig. 30 near the Nariai limit  are $\omega_{I(1)}/T_H \approx 3.1559,~ 3.0757,~ 3.013 $ and $\omega_{I(2)}/T_H \approx 3.1555,~ 3.0887,~ 2.996$ for $U_1$ and $U_2$ cases when $\beta=1,~ 1/2,~ 1/3$, respectively, while the analytic result is $\omega_I/T_H=\pi$ from (\ref{omega_exact}) for the lowest mode, $n=0$. This result may indicate the accuracy of our WKB approach itself. This is contrary to the real parts $\om_R$ which depart from numerical results beyond the Nariai limit (Figs. 32 and 33). But this would be partly due to the non-constant or scale-dependent nature of the real part of $\omega/\kappa_+$ in (\ref{omega_exact}) for the EBI-dS case, which may be compared with the RN-dS case in \cite{Moli:2003} or the Sch-dS case in \cite{Moss:2001,Card:2003}.

\begin{figure}[!htbp]
\begin{center}
\includegraphics[width=4.8cm,keepaspectratio]{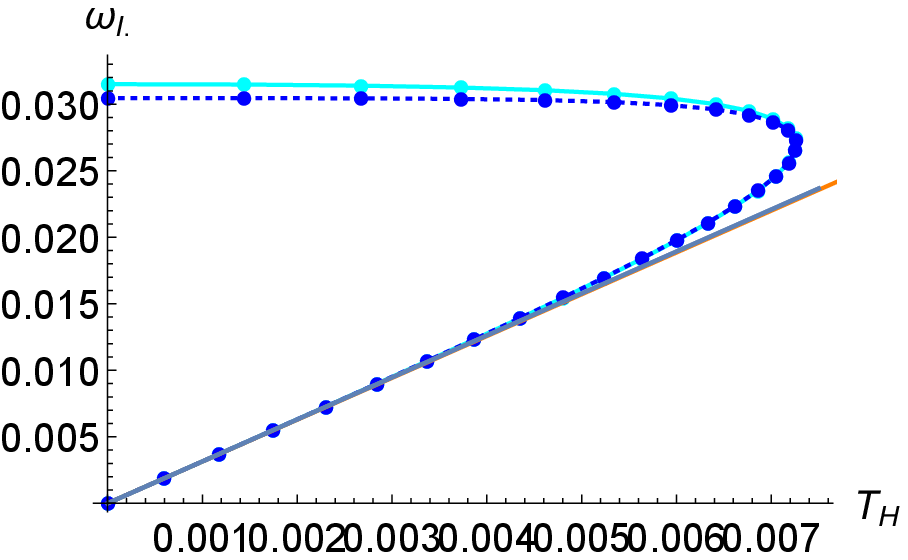}
\qquad
\includegraphics[width=4.8cm,keepaspectratio]{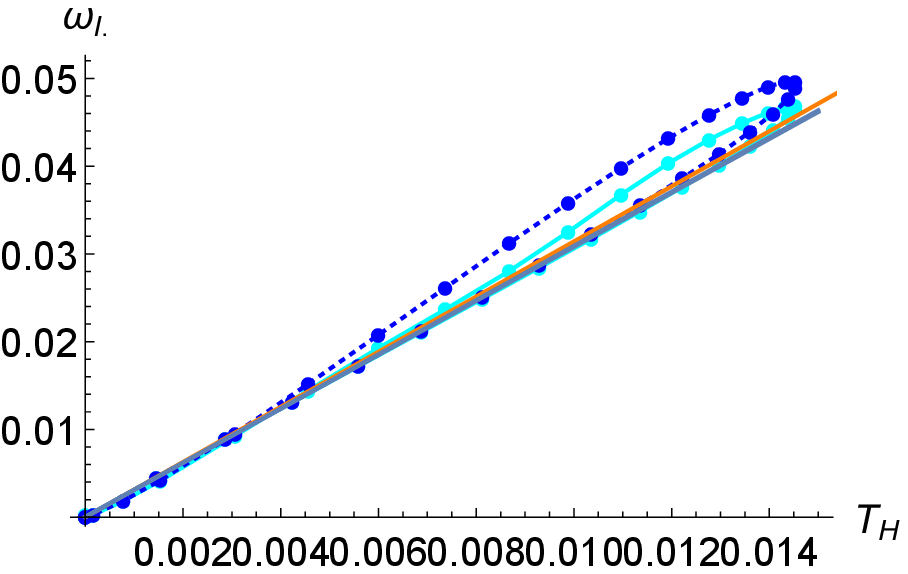}
\qquad
\includegraphics[width=4.8cm,keepaspectratio]{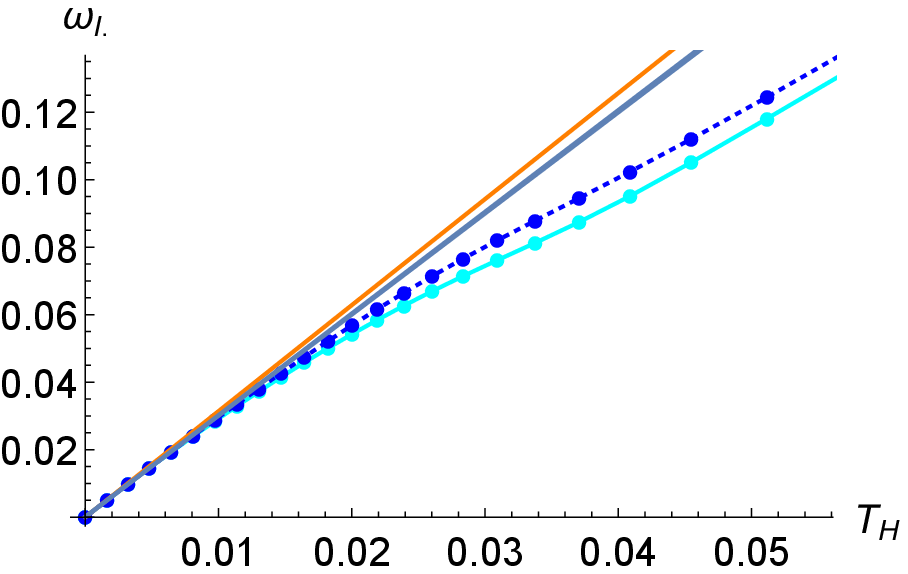}
\end{center}
\vspace{-0.6cm}
\caption{{\footnotesize The plots of {\it lowest} $\omega_I$ vs. Hawking temperature $T_H$ for $U_1$ (dashed curves) and $U_2$ (solid curves)
when $\beta=1$ (left), $1/2$ (center), $1/3$ (right), and $Q=1$, $l=2$, $\Lambda =0.2$, $\hbar \equiv 1$. The orange lines denote the exact solutions (\ref{omega_exact}) near the Nariai limit at the bottom and the agreements with the numerical results (the blue lines represent the best-fit curves near the limit) are quite good.
}}
\label{figIV}
\end{figure}

\begin{figure}[!htbp]
\begin{center}
\includegraphics[width=4.8cm,keepaspectratio]{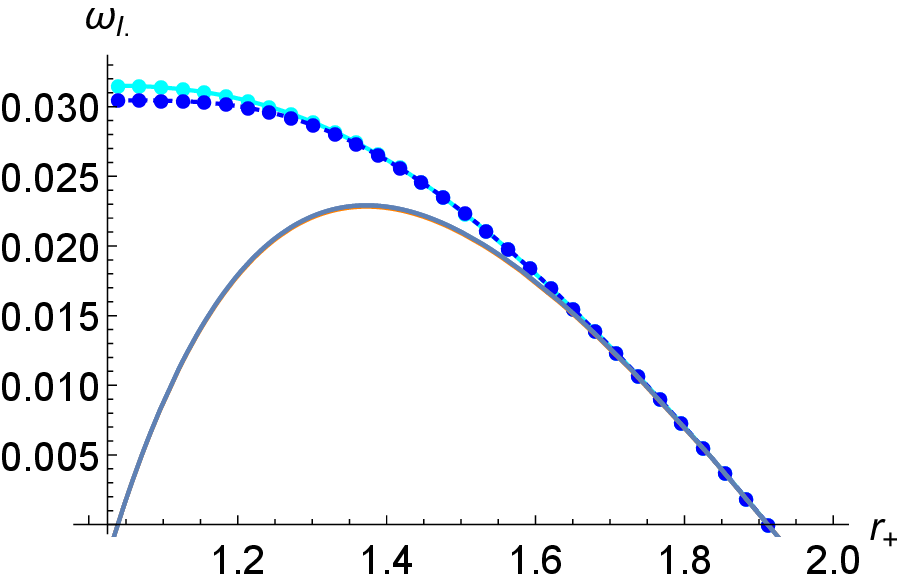}
\qquad
\includegraphics[width=4.8cm,keepaspectratio]{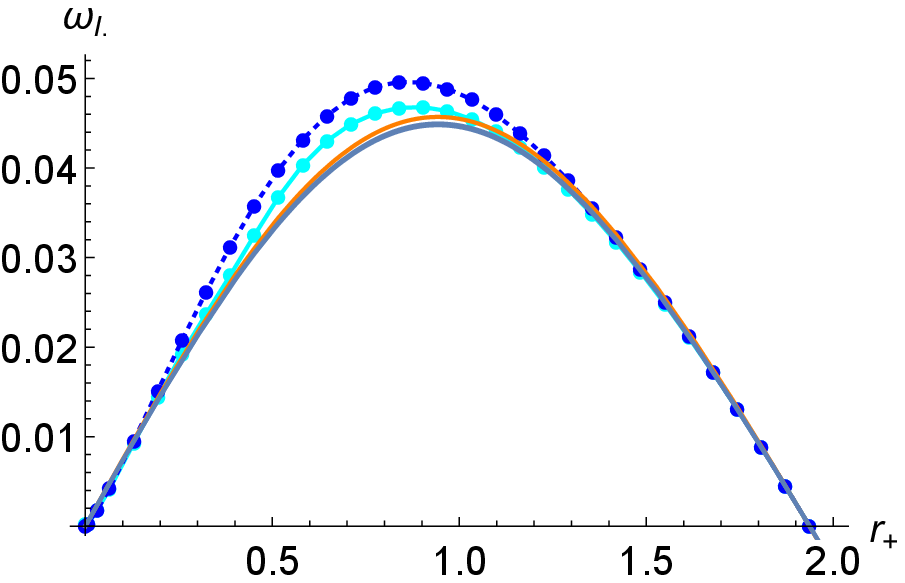}
\qquad
\includegraphics[width=4.8cm,keepaspectratio]{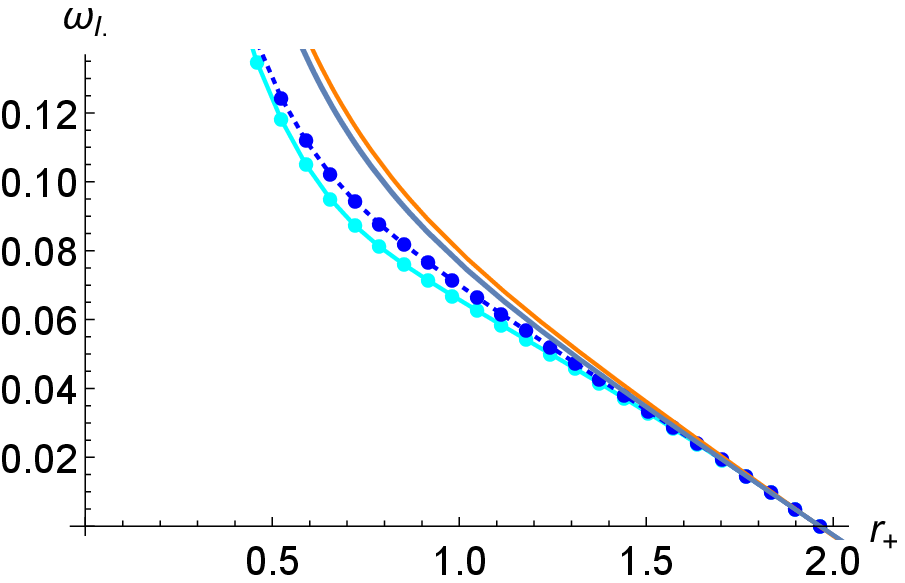}
\end{center}
\vspace{-0.6cm}
\caption{{\footnotesize The replots of Fig. 30
in terms of $r_+$, corresponding to $T_H$.
}}
\label{figIV}
\end{figure}

\begin{figure}[!htbp]
\begin{center}
\includegraphics[width=4.8cm,keepaspectratio]{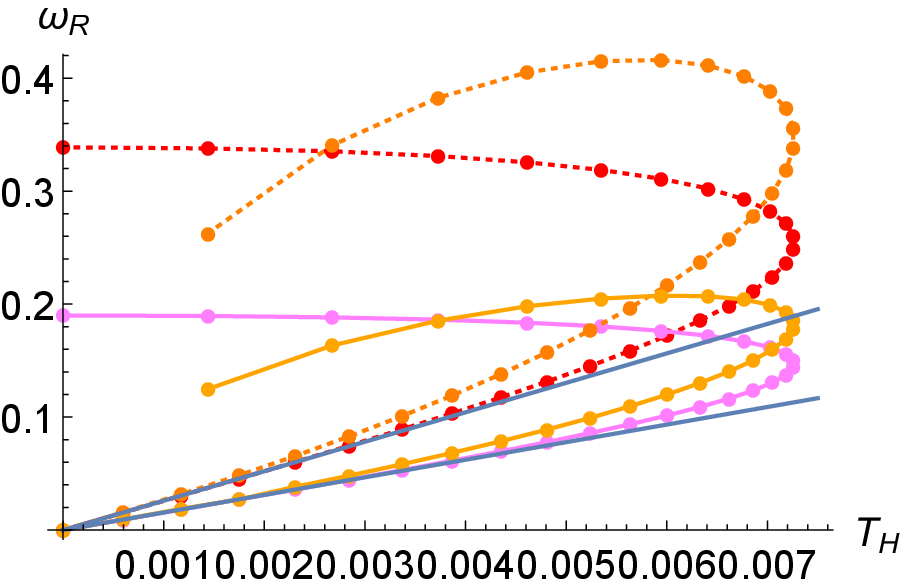}
\qquad
\includegraphics[width=4.8cm,keepaspectratio]{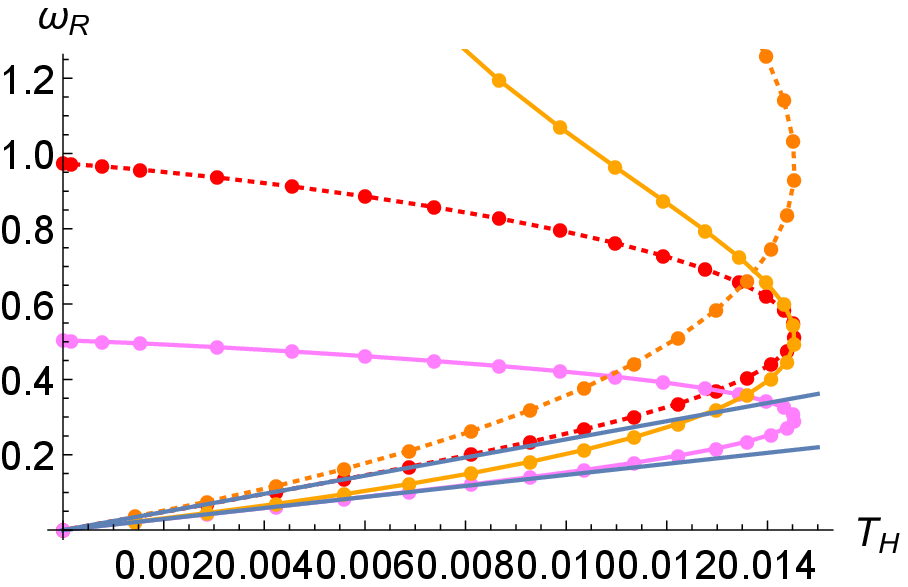}
\qquad
\includegraphics[width=4.8cm,keepaspectratio]{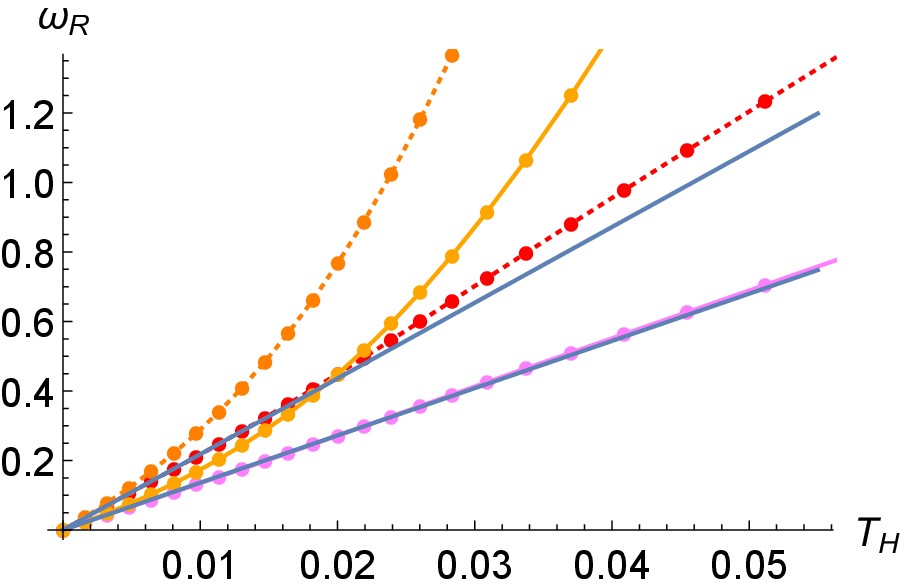}
\end{center}
\vspace{-0.6cm}
\caption{{\footnotesize The plots of {\it lowest} $\omega_R$,
associated with their imaginary correspondents in Fig. 30.
The orange lines denote the exact solutions near the Nariai limit
at the bottom and the blue lines represent the best-fit curves
near the limit.
}}
\label{figIV}
\end{figure}

\begin{figure}[!htbp]
\begin{center}
\includegraphics[width=4.8cm,keepaspectratio]{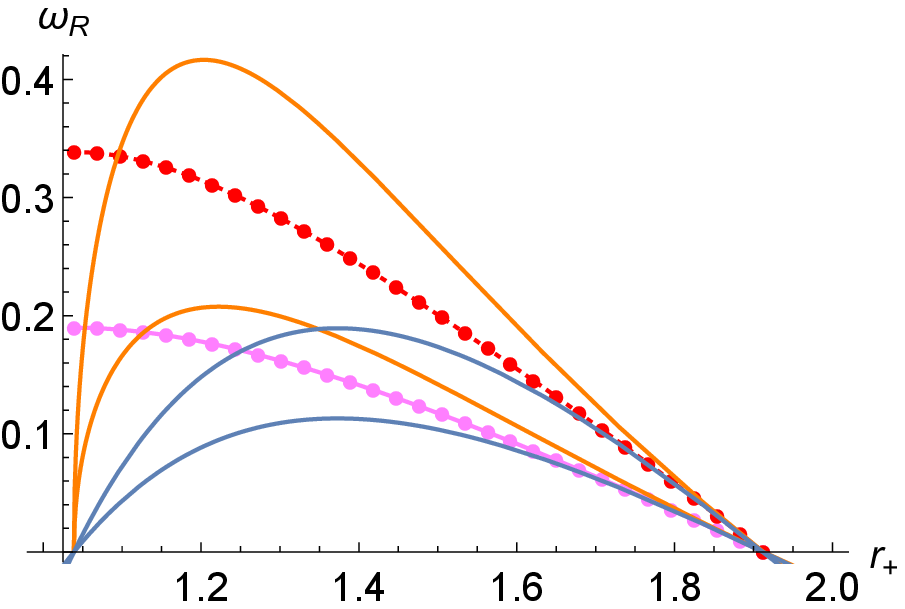}
\qquad
\includegraphics[width=4.8cm,keepaspectratio]{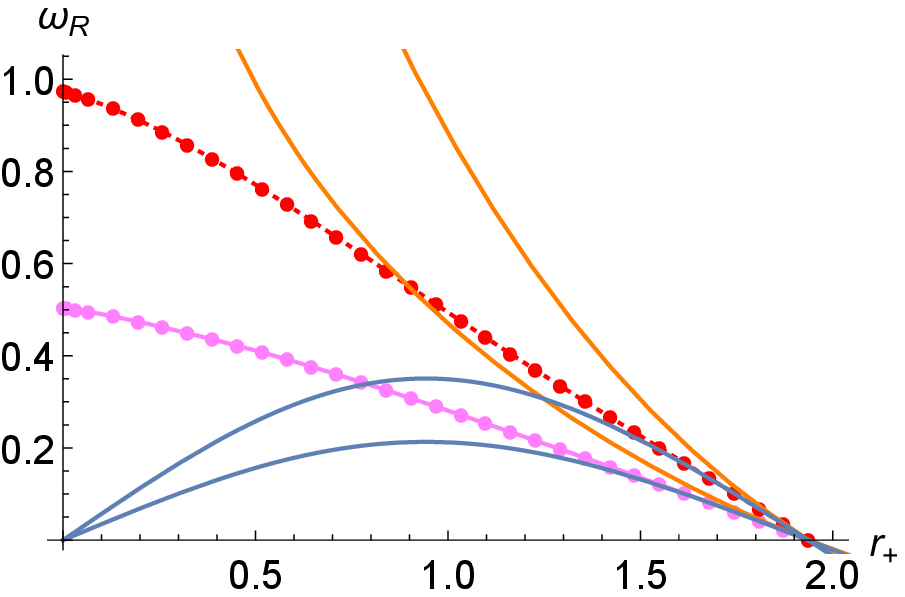}
\qquad
\includegraphics[width=4.8cm,keepaspectratio]{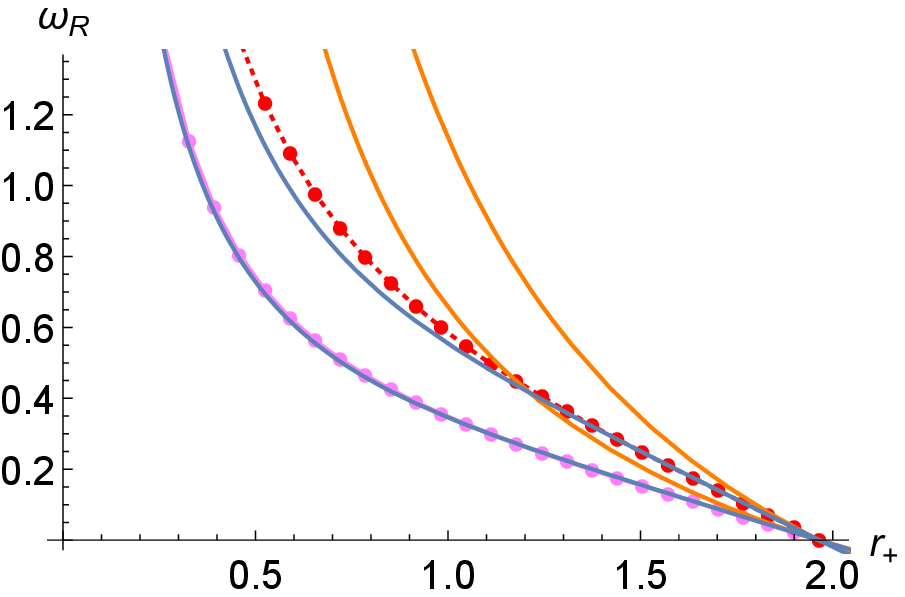}
\end{center}
\vspace{-0.6cm}
\caption{{\footnotesize The replots of Fig. 32
in terms of $r_+$, corresponding to $T_H$.
}}
\label{figIV}
\end{figure}

\begin{figure}[!htbp]
\begin{center}
{\includegraphics[width=7.5cm,keepaspectratio]{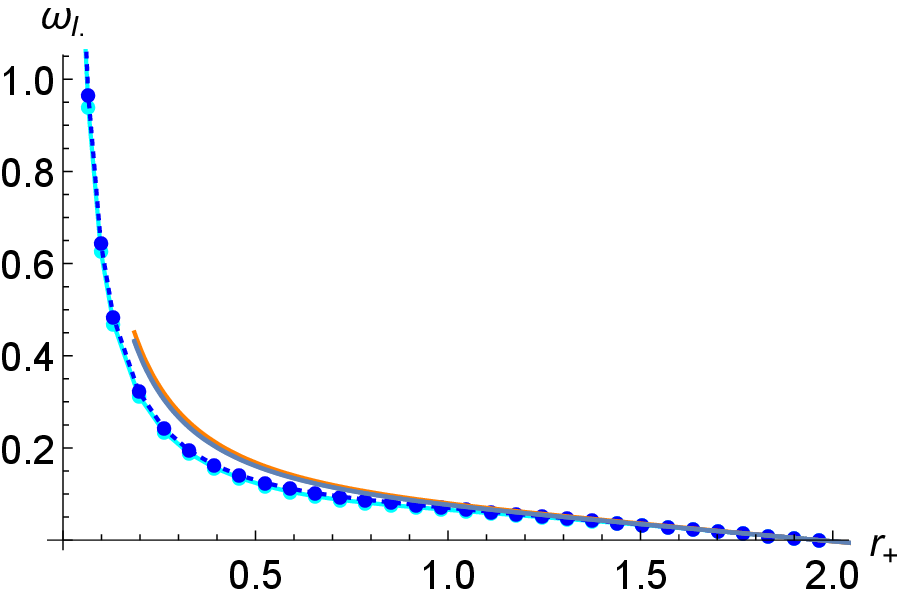}}
$\;\;\;\;\;${\includegraphics[width=7.5cm,keepaspectratio]{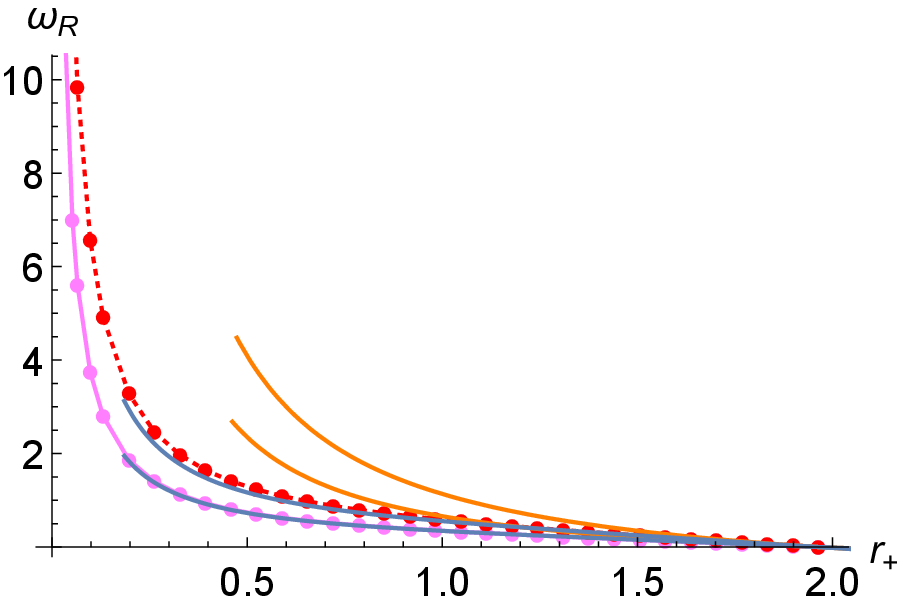}}
\end{center}
\vspace{-0.6cm}
\caption{{\footnotesize The plots of {\it lowest} ($n=0$) $\omega_R$ and $\omega_I$ near $r_+=0$,
corresponding to Figs. 32 and 33. The blue lines represent the best-fit curves
near the limit, given by (\ref{omega_singular_fit}).
}}
\label{figIV}
\end{figure}

\begin{figure}[!htbp]
\begin{center}
{\includegraphics[width=7.5cm,keepaspectratio]{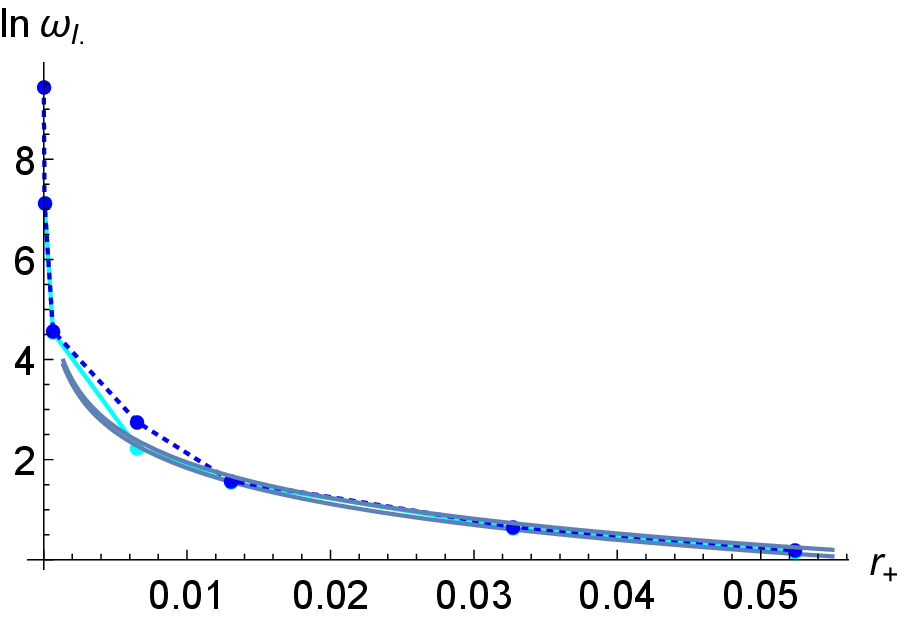}}
$\;\;\;\;\;${\includegraphics[width=7.5cm,keepaspectratio]{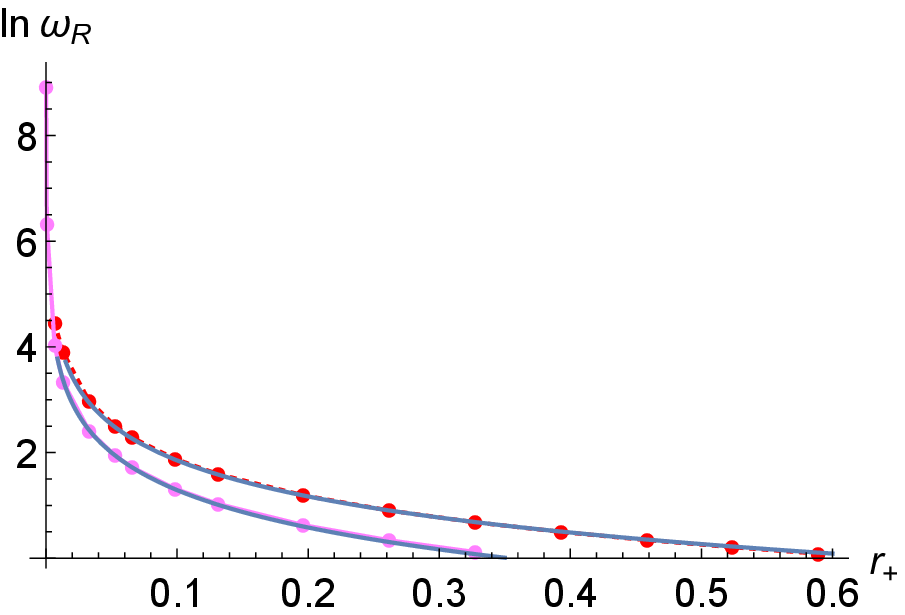}}
\end{center}
\vspace{-0.6cm}
\caption{{\footnotesize The logarithmic plots of {\it lowest} ($n=0$) $\omega_R$ and $\omega_I$ near $r_+=0$,
corresponding to Fig. 34. The blue lines represent the best-fit curves
near the limit, given by (\ref{omega_singular_fit}).
}}
\label{figIV}
\end{figure}

Before finishing this section, we note that the Nariai limit is also
associated with the large-$l$ limit,
\bear
U_1^{(0)} &= & \frac{\kappa_+ ( r_{++}-r_+) }{2 r_+^3} \left[ l (l+1) r_+ + 2 \sqrt{l (l+1)} Q \left(1+\f{Q^2}{\beta^2 r_+^4}\right)^{-1/4}  \right ]+ {\cal O} (l^{0}) ,
\label{V1_0_largeL}\\
U_2^{(0)} &= &\frac{\kappa_+ ( r_{++}-r_+) }{2 r_+^3} \left[ l (l+1) r_+ -2 \sqrt{l (l+1)} Q \left(1+\f{Q^2}{\beta^2 r_+^4}\right)^{-1/4}  \right ] + {\cal O} (l^{0}),
\label{V2_0_largeL}
\eear
showing, at the leading order, the merge of $U_1^{(0)}$ and $U_2^{(0)}$ and the linear dependence $\om_R \approx \sigma l $ with $\sigma=\sqrt{ \kappa_+ ( r_{++}-r_+)/2 r_+^2 }$ as observed in Figs. 19, 21, and 23. which may be compared with the RN case in \cite{Ferr:1984} or the Sch-dS case in \cite{Zhid:2003}. On the other hand, the asymptotic approach of $\om_I$ to a limiting value corresponds to $\om_I=\kappa_+ (n+1/2)$ in (\ref{omega_exact}) \footnote{Of course, these two behaviors may not be directly compared with those of Figs 19, 21, and 23 because of different proportional coefficients depending on whether it is near the Nariai limit or not. Actually, we have $\omega=- 0.01763 i+0.12083~ l ,~- 0.02278 i+0.10775 ~l,~ - 0.02089 i+0.08954~ l$ near the Nariai limit, while $\omega=-0.03055 i+0.1092~ l,~ - 0.03349 i+0.1006 ~l,~ - 0.03345 i+ 0.0910~ l$ for the cases of Figs 19, 21, and 23, which correspond to $\epsilon=0.99091, ~0.70872,~0.57611$, respectively. It is rather surprising that we have rough agreements already even beyond the Nariai limit.}.

We also remark that Figs. 34 and 35 show, as $r_+ \rightarrow 0$, the divergent $\omega$ as \beq
\omega_I \approx r_+^{-\bar{a}} e^{-\bar{b}},~~\omega_R \approx r_+^{-\bar{c}} e^{-\bar{d}}
\eeq
with $(\bar{a},~ \bar{b},~ \bar{c},~ \bar{d}) \approx
(1.0239 ,~2.7715 ,~0.9834 ,~0.4133 ),~(1.0356 ,~2.9366 ,~1.0215 ,~1.0623 )$ for $U_1,~U_2$, respectively. This is consistent with the similar behavior in Fig. 16. It is interesting to note that the similar divergence can be also seen from the $r_+ \rightarrow 0$ limit in the Nariai limit formula (\ref{omega_exact}) with a quite close exponent $\bar{a}=1$ for $\omega_I$, while somewhat different exponent $\bar{c}=3/2$ for $\omega_R$, even though the point-like horizon limit $r_+ \rightarrow 0$ would be far beyond the Nariai limit.

\section{Discussion}

We have studied, using the Schutz-Iyer-Will's WKB method, QNMs for the {\it axial}
gravitational perturbations of electrically charged black holes in
EBI gravity with a positive cosmological constant. We have found that all the axial perturbations are {\it stable}, {\it i.e.} ``non-negative" $\om_I$,  including the {\it flat} ($\Lambda=0$) as well as {\it dS} ($\Lambda>0$) cases, for all cases where
the WKB method applies,
correcting some errors in the literature \cite{Fernando:2004pc}.

It is also found that there are cases where
the conventional WKB method does not apply, like the three-turning-points problems for the lower-$l$ cases in
Figs. 18 - 23, and a more generalized formalism is necessary for
studying their QNMs and stabilities \cite{Galt:1991}.
Regarding the degenerate horizons, where the WKB formula can be marginally
applicable,
our results seem to be consistent with
stability \cite{Chandrasekhar,Kokk:1988,Mell:1989} \footnote{This may be contrasted with the ``horizon
instability" of axisymmetric extremal horizons \cite{Aretakis:2012ei}. It would be
a challenging problem to study the horizon instability of axisymmetric extremal
horizons within WKB approaches or in a more generalized approach \cite{Galt:1991}.}.

On the other hand, we may note in our case that there are actually three types of degenerate horizons with the vanishing Hawking temperature. The first is the Nariai-type horizons, where the (outer) black hole horizons, regardless of whether it is the {\it RN}-type or the {\it Sch}-type, coincide with the cosmological horizon. In this case, we find that the QNMs are completely ``frozen", {\it i.e.}, $ \om \approx 0$ and this would indicate the solution as the
{\it final state}
of the system, {\it i.e.}, no further evolution
\footnote{This may be compared with the similar frozen modes for
{\it heavy-mass} scalar perturbations in a wormhole geometry \cite{Kim:2018ang}.}.
This is the case of $\beta=\beta_{\rm min}$ in Fig. 7, $Q=Q_{\rm min}$ in
Fig. 9, $M=M_{\rm max}$ in Figs. 11, 13, 15, and $\Lambda=\Lambda_{\rm max}$
in Figs. 25, 27, 29. The second is the usual extremal black holes, where
the inner (Cauchy) horizon coincides with the outer (event) horizon,
regardless of the existence of the cosmological horizon. In this case,
there are non-vanishing $\om$'s and they are stable, {\it i.e.}, $\om_I>0$,
as noted above. This is the case of $\beta=\beta_{\rm max}$ in Fig. 7,
$Q=Q_{\rm max}$ in Fig. 9, $M=M_{\rm min}$ in Fig. 13, and
$\Lambda=\Lambda_{\rm min}$ in Figs. 25, 27, 29. One peculiar thing of
our EBI system is that there is another type of extremal black holes, which
is the third case of degenerate horizon, where the black hole horizons
merge at the origin $r_+=0$, {\it i.e.,} the ``point-like" horizon.
In this case, the QNMs are quite {\it long-lived}
with large decay time $\tau \sim 1/\om_I$, which are close to the quasi-resonance modes (QRMs) with $\om_I=0$ \cite{Ohas:2004}. This is the case of $M=M_{\rm min}$ in Figs. 13, 15 and this is a genuine effect
of the {\it non-GR} branch
$\beta Q = 1/2$, which does not have the GR limit \cite{Kim:2016pky}.

We have studied the exact solutions near the Nariai limit and found, for the imaginary frequency parts, good agreements even far beyond the limit. This may indicate the accuracy of our WKB approach. We have also obtained the similar large-$l$ behaviors as observed in the numerical computations, though this needs not be the same as the large-$l$ limit of the far beyond Nariai limit, due to non-commutativity of the two limits.

Several further remarks are in order. First, it is known
that the polar
and axial
perturbations in Einstein gravity are related due to a symmetry,
associated with the
parity \cite{Chandrasekhar}. The similar relations for EBI gravity are expected
to exist due to the same parity symmetry in the equations of motion, but its
rigorous proof would be still an interesting open
problem. Second, it is known that the electric and magnetic duality exists in the background solutions \cite{Garc:1984}. But its validity in the perturbations does not seem to be quite evident and it would be desirable to consider this problem in more details. Third, the relation between QNMs and Hawking temperature in de Sitter space is an important probe of $dS/CFT$-correspondence \cite{Park:1998,Stro:2001}. It would be interesting to find a way to compute QNMs from the appropriate CFT at the spatial infinity \cite{Abda:2002}. Finally, we note that our results for higher-$l$, {\it i.e.}, $l=(1), 2, 3, \cdots$ with $n=0$ show the high real-to-imaginary frequency ratio $\om_R/\om_I$, which ranges about $5 \sim 10$, so that the WKB approximations can be reasonably accurate.

%

\section*{Acknowledgements}
This work was supported by Basic Science Research Program through the National Research Foundation of Korea (NRF) funded by the Ministry of Education, Science and Technology (NRF-2019R1F1A1060409 (JYK), NRF-2018R1D1A1B07049451 (COL), NRF-2016R1A2B401304, 2020R1A2C1010372, 2020R1A6A1A03047877 (MIP)).

\section*{Appendix}
The tetrad components of Ricci tensors for the metric in
(\ref{perturbedmetric}) can be computed in a straightforward manner through the form notations, from the knowledge of {\it spin-connection} one-form ${\bf \omega}$ satisfying the torsion-free condition \cite{Chandrasekhar} \footnote{Due to the double-opposite choice of conventions, one from the different choice of the Riemann and Ricci tensors (see the footnote No.1) and the other from the different choice of the signature of the metric, the tetrad components of the Ricci tensor coincide with those of \cite{Chandrasekhar}. }.
Here we present some components which are relevant for our calculations:
\bear
- R_{00} = && e^ {-2 \nu} [ ( \psi + \mu_2 + \mu_3 )_{,0,0} + \psi_{,0} ( \psi - \nu)_{,0} + \mu_{2,0} ( \mu_2 - \nu )_{,0} + \mu_{3,0} ( \mu_3 - \nu )_{,0} ]   \nonumber  \\
                  && - e^{-2 \mu_2} [ \nu_{,2,2} + \nu_{,2} ( \psi + \nu - \mu_2 +\mu_3 )_{,2} ]   \nonumber  \\
                  && - e^{-2 \mu_3} [ \nu_{,3,3} + \nu_{,3} ( \psi + \nu + \mu_2 -\mu_3 )_{,3} ]   \nonumber  \\
                  && + \frac{1}{2}  e^{2 \psi-2 \nu} [ e^{-2 \mu_2 } Q_{20}^2 + e^{-2 \mu_3 } Q_{30}^2 ] ,    \nonumber   \\
- R_{11} = && e^ {-2 \mu_2} [ \psi_{,2,2} + \psi_{,2} ( \psi + \nu + \mu_3 -\mu_2  )_{,2}  ]   \nonumber  \\
                  && + e^{-2 \mu_3} [ \psi_{,3,3} + \psi_{,3} ( \psi + \nu + \mu_2 -\mu_3  )_{,3}   ]                          \nonumber  \\
                  && - e^{-2 \nu} [ \psi_{,0,0} + \psi_{,0} ( \psi - \nu + \mu_2 +\mu_3 )_{,0} ]  - \frac{1}{2}  e^{2 \psi-2 \mu_2 - 2 \mu_3}  Q_{23}^2    \nonumber   \\
                  && + \frac{1}{2}  e^{2 \psi-2 \nu} [ e^{-2 \mu_3 } Q_{30}^2 + e^{-2 \mu_2 } Q_{20}^2 ] ,      \nonumber   \\
- R_{22} = && e^ {-2 \mu_2} [ ( \psi + \nu + \mu_3 )_{,2,2} + \psi_{,2} ( \psi - \mu_2)_{,2} + \mu_{3,2} ( \mu_3 - \mu_2 )_{,2} + \nu_{,2} ( \nu - \mu_2)_{,2} ]   \nonumber  \\
                  && + e^{-2 \mu_3} [ \mu_{2,3,3} + \mu_{2,3} ( \psi + \nu + \mu_2 -\mu_3 )_{,3} ]   \nonumber  \\
                && - e^{-2 \nu} [ \mu_{2,0,0} + \mu_{2,0} ( \psi - \nu + \mu_2 +\mu_3 )_{,0} ]   \nonumber  \\
                  && + \frac{1}{2}  e^{2 \psi-2 \mu_2} [ e^{-2 \mu_3 } Q_{23}^2 -e^{-2 \nu } Q_{20}^2 ] ,    \nonumber   \\
- R_{01} = &&  \frac{1}{2} e^ {-2 \psi - \mu_2 -\mu_3} [ ( e^{3 \psi - \nu -\mu_2 + \mu_3 } Q_{20})_{,2} +  ( e^{3 \psi - \nu -\mu_3 + \mu_2 } Q_{30})_{,3}   ] ,  \nonumber  \\
- R_{12} = &&  \frac{1}{2} e^ {-2 \psi - \nu -\mu_3} [ ( e^{3 \psi + \nu -\mu_2 - \mu_3 } Q_{32})_{,3}  -  ( e^{3 \psi - \nu +\mu_3 - \mu_2 } Q_{02})_{,0}   ] ,  \nonumber    \\
- R_{02} = &&  e^ {-\mu_2- \nu } [ ( \psi + \mu_3  )_{,2,0}  + \psi_{,2}  (  \psi - \mu_2  )_{,0}   + \mu_{3,2} ( \mu_3 - \mu_2 )_{,0} - ( \psi +\mu_3 )_{,0} \nu_{,2}  ]  \nonumber      \\
                 &&  - \frac{1}{2} e^ {2 \psi - \nu -2\mu_3 -\mu_2} Q_{23} Q_{30} ,  \nonumber    \\
- R_{23} = &&  e^ {-\mu_2- \mu_3 } [ ( \psi + \nu  )_{,2,3}  - ( \psi + \nu)_{,2}  \mu_{2,3} - ( \psi + \nu)_{,3}  \mu_{3,2} + \psi_{,2} \psi_{,3}  +\nu_{,2} \nu_{,3}  ]  \nonumber      \\
                 &&  - \frac{1}{2} e^ {2 \psi - 2\nu -\mu_2 -\mu_3} Q_{20} Q_{30} .   \nonumber
\eear
The components $R_{33} , R_{13},$ and $R_{03} $ can be obtained by interchanging the indices 2 and 3 in the components $R_{22} , R_{12},$ and $R_{02} $.

\end{document}